\documentclass{aa}  

\usepackage{array}
\usepackage{graphicx}
\usepackage{multirow}
\usepackage{color}
\usepackage{hyperref}
\usepackage{adjustbox}
\usepackage[capitalize]{cleveref}
\usepackage{txfonts}
\renewcommand{\arraystretch}{1.2}

\usepackage{hyperref}
\begin{document}

   \title{A systematic study of the ultra-fast outflow responses to luminosity variations in active galactic nuclei}

   \titlerunning{A systematic study of UFO response around highly-accreting SMBHs}
   \authorrunning{Yerong Xu et al.}
   \author{Yerong Xu
          \inst{1,2}\fnmsep\thanks{yerong.xu@inaf.it},
          Ciro Pinto
          \inst{1},
          Daniele Rogantini\inst{3,4},
          Didier Barret\inst{5},
          Stefano Bianchi\inst{6},
          Matteo Guainazzi\inst{7},
          Jacobo Ebrero\inst{8},
          William Alston\inst{9},
          Erin Kara\inst{3},
          \and
          Giancarlo Cusumano\inst{1}
          }

   \institute{INAF - IASF Palermo, Via U. La Malfa 153, I-90146 Palermo, Italy
         \and
            Universit\`a degli Studi di Palermo, Dipartimento di Fisica e Chimica, via Archirafi 36, I-90123 Palermo, Italy
        \and
        MIT Kavli Institute for Astrophysics and Space Research, Massachusetts Institute of Technology, Cambridge, MA 02139, USA 
        \and
        Department of Astronomy and Astrophysics, University of Chicago, 5640 S Ellis Avenue, Chicago, IL 60637, USA
        \and Universit\'e de Toulouse, CNRS, IRAP, 9 Avenue du colonel Roche, BP 44346, 31028 Toulouse Cedex 4, France
        \and
        Dipartimento di Matematica e Fisica, Università degli Studi Roma Tre, via della Vasca Navale 84, I-00146 Roma, Italy 
        \and ESA European Space Research and Technology Centre (ESTEC), Keplerlaan 1, 2201 AZ, Noordwĳk, The Netherlands 
        \and
        Telespazio UK for the European Space Agency (ESA), European Space Astronomy Centre (ESAC), Camino Bajo del Castillo, s/n, 28692 Villanueva de la Cañada, Madrid, Spain 
        \and
        Centre for Astrophysics Research, University of Hertfordshire, College Lane, Hatfield AL10 9AB, UK  
             }

   \date{Received XXX; accepted XXX}

 
  \abstract
   {Ultra-fast outflows (UFOs) have been revealed in a large number of active galactic nuclei (AGN) in the past two decades. Their extreme velocities and high ionization states make them a promising candidate for AGN feedback on the evolution of the host galaxy. However, their exact underlying driving mechanism is not yet fully understood.}
   {Given that the variability of UFOs may be used to distinguish among different launching mechanisms, we aim to search for and characterize the responses of the UFO properties to the variable irradiating luminosity. }
   {We performed a high-resolution time- and flux-resolved spectroscopy of archival \textit{XMM-Newton} observations on six highly-accreting narrow-line Seyfert 1 (NLS1) galaxies, selected by UFO detection and sufficient exposure times. The state-of-the-art methods of the blind Gaussian line scan and photoionization model scan are used to identify UFO solutions. We search for ionized winds and investigate structure of ionized winds and their responses to the luminosity variations. The location, density, and kinetic energy of UFOs are estimated as well.}
   {The powerful photoionization model scan reveals three previously unreported UFOs in RE J1034+396, PG 1244+026 and I ZW 1 with a detection significance above $3\sigma$, and two new warm absorbers (WAs) in RE J1034+396. 5 out of 6 (83\%) AGN in our sample hosts multi-phase ionized winds, where outflows in I ZW 1 are energy-conserved. The relatively low-ionization entrained UFOs are discovered in 4 (66\%) AGN of our sample, supporting the shocked outflow interpretation for ionized winds in AGN. We notice that 2 out of 7 (28\%) UFOs in our sample seem to respond to the radiation field and 3 (43\%) UFOs hint at a radiatively accelerated nature, requiring further observations. Combined with published works, we do not find any correlations between UFO responses and AGN properties except for a tentative ($\sim1.8\sigma$) anti-correlation between the UFO acceleration and the Eddington ratio, to be confirmed by further observations and an enlarged sample. The kinetic energy of UFOs, mostly detected in soft X-rays, is found to have a large uncertainty. We, therefore, cannot conclude whether soft X-ray UFOs have sufficient energy to drive the AGN feedback, although they are very promising based on some reasonable assumptions. The primary UFO in I ZW 1 (detected in the hard X-ray) is the only case in our sample to possess conclusively sufficient energy to affect the host galaxy.}
   {}
   \maketitle
%

\section{Introduction}\label{sec:intro}
Ultra-fast outflows (UFOs) were first discovered in the X-ray spectra of active galactic nuclei (AGN) through blueshifted Fe \textsc{xxv/xxvi} absorption lines around 7\,keV \citep[e.g.][]{2002Chartas,2003Chartas,2003Pounds,2003Reeves}. In the past two decades, a number of high column density ($N_\mathrm{H}>10^{22}\mathrm{cm}^{-2}$), highly ionized ($\log\xi>3$) ultra-fast ($v\geq 10000$\,km/s or 0.03c) outflows have been found in both nearby and high-z AGN with a detection rate around 30\% \citep[e.g.][]{2010Tombesi,2012Patrick,2013Gofford,2020Igo,2021Chartas,2023Matzeu}. The ultra-fast velocity suggests an origin from the innermost regions around the central supermassive black holes (SMBHs). Such extreme physical properties can result in a large amount of kinetic power, possibly sufficient to match the theoretical predictions of effective AGN feedback models \citep[e.g.][]{2005DiMatteo,2010Hopkins}. 
UFOs, therefore, are expected to play a crucial role in regulating the growth of the SMBHs and the bulges of the host galaxies \citep[e.g.][and references therein]{2012Fabian}, offering a promising explanation for the observed well-known AGN-galaxy relation \citep[e,g. $M_\mathrm{BH}\mbox{--}\sigma$,][and references therein]{2013Kormendy}.

UFOs were not only detected in the hard X-ray band but also resolved in the soft X-ray band, thanks to the high-resolution grating instruments, i.e., the Reflection Grating Spectrometer \citep[RGS,][]{2001denHerder} onboard \textit{XMM-Newton} \citep{2001Jansen} and the High Energy Transmission Gratings \citep[HETG,][]{2005Canizares} onboard \textit{Chandra} \citep{2002Weisskopf}. The high-resolution capability can distinguish UFOs from the slow and moderately ionized outflows (the so-called warm absorbers, WAs) in the soft X-ray band. Compared with the Fe \textsc{xxv} and \textsc{xxvi} features, which are usually unresolved by the current CCD detectors and have the potential degeneracy with the X-ray reprocessing emission \citep{2011Gallo,2013Gallo,2015Zoghbi,2022Parker}, UFOs in soft X-rays can be identified through multiple ion transitions \citep[e.g., O \textsc{vii} and O \textsc{viii},][]{2015Longinotti,2016Pounds,2018Pinto,2022Xu}, providing more convincing evidence for the UFO existence. Recently, some atypical UFOs have been detected in the soft X-rays, as they share a similar velocity range but have a lower column density ($N_\mathrm{H}\leq10^{22}\,\mathrm{cm}^{-2}$) and a lower ionization state ($\log\xi\leq4$) than the highly ionized iron K absorbers, revealing the multi-phase origin of UFOs \citep[e.g.,][]{2016Reeves,2019Serafinelli,2021Krongold,2023Xu}.

In spite of the large number of UFOs detected in AGN, the exact wind-driven mechanism has not been well understood so far. The radiatively-driven \citep[e.g.,][]{2000Proga,2010Sim,2016Hagino} and magnetically-driven \citep[MHD, e.g.,][]{2004Kato,2010Fukumura,2015Fukumura} mechanisms were proposed to accelerate winds to relativistic speeds. The former scenario is naturally expected in high-accretion systems, such as narrow-line Seyfert 1 (NLS1) galaxies, which host the low-mass and highly accreting SMBHs \citep[e.g.,][]{2008Komossa}, and quasars, while magnetically-driven UFOs are anticipated in low-accretion systems. The radiation explanation is supported by the correlation between UFO velocity and source luminosity noticed in PDS 456 \citep{2017Matzeu}, IRAS 13224-3809 \citep{2018Pinto}, and Mrk 1044 \citep{2023Xu}. However, the exact behavior of UFOs is complex. For example in 1H 0707-495, an anti-correlation between the UFO velocity and source flux was observed. It was explained by the slim inner accretion flow at super-Eddington states, extending the wind launching radii outwards, resulting in a lower velocity \citep{2021Xu}, although alternative scenarios invoking MHD might be possible. Motivated by such complexity, we aim to perform a systematic study of UFO response to the source variability to better understand their nature and launching mechanisms.

In this study, we will show the detailed high-resolution spectroscopic analysis of a sample of UFOs in 6 nearby NLS1 galaxies, observed by \textit{XMM-Newton} space telescope, mainly focusing on the RGS data. In Sec. \ref{sec:selection} and Sec. \ref{sec:reduction}, we will list the sample selection criteria and our data reduction processes respectively. The adopted analysis methods are described in Sec. \ref{sec:methods}. Our results are summarized in Sec. \ref{sec:results}, where we find previously unreported UFOs in three AGN, and further study the relationship between the UFO properties and the source. We discuss our results and compare them with the literature in Sec. \ref{sec:discussion}. Finally, in Sec. \ref{sec:conclusion} we draw our conclusions and point out prospects.

\section{Sample Selection}\label{sec:selection}
The systems of interest in our study are AGN with UFOs. We thus explored the archival \textit{XMM-Newton} dataset, in particular RGS spectra, which possess both high spectral resolution and sufficient photons due to its large effective area in the soft X-ray band. Our selection is based on the following considerations:
\begin{enumerate}[(1)]
    \item A UFO has to be detected in the system. 
    Since we are also interested in discovering new UFOs in AGN, the search is not limited to the UFO-reported AGN, but rather is extended to all Seyfert galaxies and quasars with an \textit{XMM-Newton} exposure time of $>50\,$ks, resulting in a sample of 307 targets. The long exposure time is necessary for RGS to detect and resolve narrow lines \citep[e.g. see Fig. 11 in][]{2018Kosec}.
    
    \item The UFO absorption features in the soft X-ray band might be heavily contaminated by the transient obscuration event or the persistent torus obscuration. Thus only AGN with a neutral column density $\log{N_\mathrm{H}/\mathrm{cm}^{-2}}<22$ are selected, leading to the sample size down to 179.
    \item To detect the potential UFO response to the source variability, we need enough counts to perform the time-/flux-resolved spectroscopy. In the initial stage, we adopted the product of the averaged RGS flux at $15\AA$ and the total exposure time as a probe for the number of soft X-ray counts, which is easily accessed through the \textit{XMM-Newton} Science Archive (\href{https://nxsa.esac.esa.int/nxsa-web/\#search}{XSA}). The exact number of counts is obtained by extracting the RGS spectra (listed in column 6 of Tab.\ref{tab:sources}). The threshold is set at half the soft counts of PDS 456,  which is the target with the least soft X-ray counts among four previously analyzed sources, i.e. 50000, to ensure statistics. The resulting sample size becomes 29.
    \item The detection significance of UFO is a key quantity for the strict constraints on parameters, which are indispensable for the discovery of UFO response. Therefore, we selected out sources with a UFO detection of $\Delta \rm C\mbox{-}stat>30$ for additional 4 degrees of freedom (d.o.f.), i.e.$>4.6\sigma$, in the stacked spectrum to ensure constraints on UFO parameters in the flux-/time-resolved spectra. This filter cannot be directly applied before performing the physical modeling, so we modified the size of our sample during the analysis and ended up with a number of 6.
\end{enumerate}

Our final sample consists of 10 AGN, including 4 previously studied AGN, all of which are nearby NLS1 galaxies. Their basic information is shown in Tab. \ref{tab:sources}, ordered by the accumulated RGS counts. The black hole mass ($M_\mathrm{BH}$) and the bolometric luminosity ($L_\mathrm{bol}$) are obtained from the literature, while the latter is also estimated in this work (see Sec. \ref{subsec:xabs}). The corresponding Eddington ratio is calculated by $\lambda_\mathrm{Edd}\equiv L_\mathrm{bol}/L_\mathrm{Edd}$, where $L_\mathrm{Edd}\equiv4\pi GM_\mathrm{BH}m_\mathrm{p}c/\sigma_\mathrm{T}$ is the Eddington luminosity. All of them are high-/super-Eddington AGN, expected to launch outflows \citep{2007Ohsuga}. The disk inclination angles are obtained based on the reflection spectroscopy results from the literature and this work, except for RE J1034+396, which exhibits no discernible reflection features in the spectra. For this particular case, the value is loosely constrained by the simulations \citep{2014Hu}.
\begin{table*}[htbp]
\renewcommand{\arraystretch}{1.3}
\caption{Table of the 6 sources in this sample plus 4 previously studied sources, including the source name (1), AGN type (2), redshift (3), black hole mass (4), total on-axis \textit{XMM-Newton} exposure time without solar flare correction (5), total counts in the RGS band with solar flare correction (6), bolometric luminosity (7), Eddington ratio (8) and disk inclination angle (9). See more details in Sec. \ref{sec:selection}.}
\smallskip
\centering
\begin{tabular}{ccccccccc}
\hline
Source & Type & Redshift & $\log(M_\mathrm{BH}/M_\odot)$ & Exposure &  RGS Counts & $L_\mathrm{bol}$ & $\lambda_\mathrm{Edd}$ & $i$ \\
& & z &  & (ks) & ($10^{5}$) &  ($10^{44}\,\mathrm{erg/s}$) & ($\equiv L_\mathrm{bol}/L_\mathrm{Edd}$) & (deg)\\
(1)&(2) & (3) & (4) &(5)&(6)&(7)&(8) & (9)\\
\hline
\multicolumn{9}{c}{This work}\\
\hline
1H 1934-063 & NLS1 & 0.0102 & $6.46^{+0.20}_{-0.20}$\tablefootmark{a,b} &492& 4.2  & 1.73\tablefootmark{m,n} & $0.47\pm0.22$ & $42^{+14}_{-2}$\tablefootmark{n}\\
RE J1034+396 & NLS1 & 0.042 & $6.40^{+0.2}_{-0.4}$\tablefootmark{c} &1417& 3.7 &   $2.7\mbox{--}5.4$\tablefootmark{n,o} & $1.29\pm0.43$ & $30^{+28}_{-28}$\tablefootmark{w} \\
PG 1244+026 & NLS1  &0.048 &$7.11^{+0.14}_{-0.41}$\tablefootmark{b,d,e} &764& 2.8 & $14\mbox{--}17$\tablefootmark{n,p} & $1.04\pm0.59$ & $31^{+1}_{-1}$\tablefootmark{n}\\
PG 1211+143 & NLS1/quasar  &0.0809& $8.16^{+0.12}_{-0.15}$\tablefootmark{f} &876& 2.3 &  $31\mbox{--}53$\tablefootmark{n,q} & $0.23\pm0.08$ & $30^{+1}_{-1}$\tablefootmark{n}\\
I ZW 1 & NLS1  &0.061& $7.45^{+0.08}_{-0.12}$\tablefootmark{g} &528& 1.8 &  $22\mbox{--}39$\tablefootmark{n,r} & $0.88\pm0.21$ & $39^{+4}_{-2}$\tablefootmark{n}\\
IRAS 17020+4544 & NLS1 & 0.0604& $6.72^{+0.10}_{-0.10}$\tablefootmark{b,h} &209& 0.8 &  $7.8\mbox{--}9.9$\tablefootmark{n,s} & $1.33\pm0.31$ & $55^{+4}_{-4}$\tablefootmark{n} \\
\hline
\multicolumn{9}{c}{Published work}\\
\hline
Mrk 1044 & NLS1  &  0.016& $6.45^{+0.12}_{-0.13}$\tablefootmark{i} & 733 &  9.8 & $1.3\mbox{--}1.6$\tablefootmark{t} & $0.40\pm0.12$ & $34^{+1}_{-2}$\tablefootmark{x}\\ 
1H 0707-495 & NLS1 &0.0406& $6.30^{+0.30}_{-0.30}$\tablefootmark{j}  &1385& 4.1 &  $2\mbox{--}55$\tablefootmark{u} & $9.05\pm5.43$ & $43^{+2}_{-2}$\tablefootmark{y}\\
IRAS 13224-3809 & NLS1  &0.0658& $6.54^{+0.41}_{-0.08}$\tablefootmark{k} &2131& 3.3 &  $4\mbox{--}13$\tablefootmark{v} & $1.13\pm0.58$ & $59^{+1}_{-1}$\tablefootmark{y}  \\
PDS 456 & quasar  &0.184& $9.3^{+0.4}_{-0.4}$\tablefootmark{l} &1146& 1.1 &  1800\tablefootmark{l} & $0.72\pm0.66$ & $65^{+2}_{-2}$\tablefootmark{l}\\
\hline
\end{tabular}
\tablefoot{References: \tablefoottext{a}{\citet{2008Malizia}}
\tablefoottext{b}{\citet{2012Ponti}}
\tablefoottext{c}{\citet{2010Bian}}
\tablefoottext{d}{\citet{2014Kara}}
\tablefoottext{e}{\citet{2008Marconi}}
\tablefoottext{f}{\citet{2004Peterson}}
\tablefoottext{g}{\citet{2006Vestergaard}}
\tablefoottext{h}{\citet{2001Wang}}
\tablefoottext{i}{\citet{2015Du}}
\tablefoottext{j}{\citet{2016Done}}
\tablefoottext{k}{\citet{2015Chiang}}
\tablefoottext{l}{\citet{2009Reeves}}
\tablefoottext{m}{\citet{2022Xu}}
\tablefoottext{n}{This work}
\tablefoottext{o}{\citet{2016Czerny}}
\tablefoottext{p}{\citet{2013Jin}}
\tablefoottext{q}{\citet{2018Kriss}}
\tablefoottext{r}{\citet{2004Porquet}}
\tablefoottext{s}{\citet{2020Gonzalez}}
\tablefoottext{t}{\citet{2022Husemann}}
\tablefoottext{u}{\citet{2021Xu}}
\tablefoottext{v}{\citet{2018Buisson}}
\tablefoottext{w}{\citet{2014Hu}}
\tablefoottext{x}{\citet{2023Xu}}
\tablefoottext{y}{\citet{2018Parker}}
}
\label{tab:sources}
\vspace{-3mm}
\end{table*}

\section{Data Reduction}\label{sec:reduction}
The analyzed observations of the sources in our sample are shown in Tab. \ref{tab:obs}. We discarded the observations with less than 20 ks exposure time. Specifically for RE J1034+396, we also excluded several observations prior to 2020 that had a similar flux level with the extensive campaigns in 2020 and 2021. This was to avoid potential influence from the long-term continuum variation on flux-resolved spectroscopy. The data sets were reduced following the standard SAS threads with the \textit{XMM-Newton} Science Analysis System (SAS v20.0.0) and calibration files available by September 2022. In this work, apart from the RGS data, we also use EPIC-pn \citep[a CCD-based instrument,][]{2001Struder} and Optical Monitor \citep[OM,][]{2001Mason} data to help constrain the spectral energy distribution (SED) of the AGN. 

Briefly, we reduced the EPIC-pn data using \textsc{epproc} and filtered the time intervals affected by the background flares, which show the count rates larger than 0.5 counts/sec in the $10\mbox{--}12$\,keV range. The source and background spectra were extracted from a circular region with a radius of 30 arcsec centered on and offset but near the source respectively. The pile-up effect was examined by task \textsc{epatplot}. Only one observation of RE J1034+396 (0506440101) and four observations of PG 1211+143 (0112610101, 05020501(2)01, 0745110301) are affected by pile-up. Therefore, an annulus region with an inner radius of 32 arcsec and an outer radius of 45 arcsec was applied to extract the source spectrum of RE J1034+396, while an annulus with an inner radius of 10 arcsec and an outer radius of 30 arcsec was adopted for the PG 1211+143 source spectra. The RGS data were processed by the \textsc{rgsproc} package with a flare filter of 0.3 counts/sec. The first-order RGS spectra were extracted from a cross-dispersion region of 1 arcmin width and the background spectra were extracted from photons beyond 98\% of the source point-spread function as default. We only used the good time intervals (GTI) common to both RGS 1 and 2 and combined their spectra for the high signal-to-noise (S/N). The OM data were reduced with \textsc{omichain} and stacked into a time-averaged spectrum for each source, since they are relatively stable and much less variable than the X-ray flux, at most 25\% variations in our sample. The response files are retrieved from the ESA webpage\footnote{https://www.cosmos.esa.int/web/xmm-newton/om-response-files}. 

\begin{figure*}[htbp]
    \centering
	\includegraphics[width=\columnwidth]{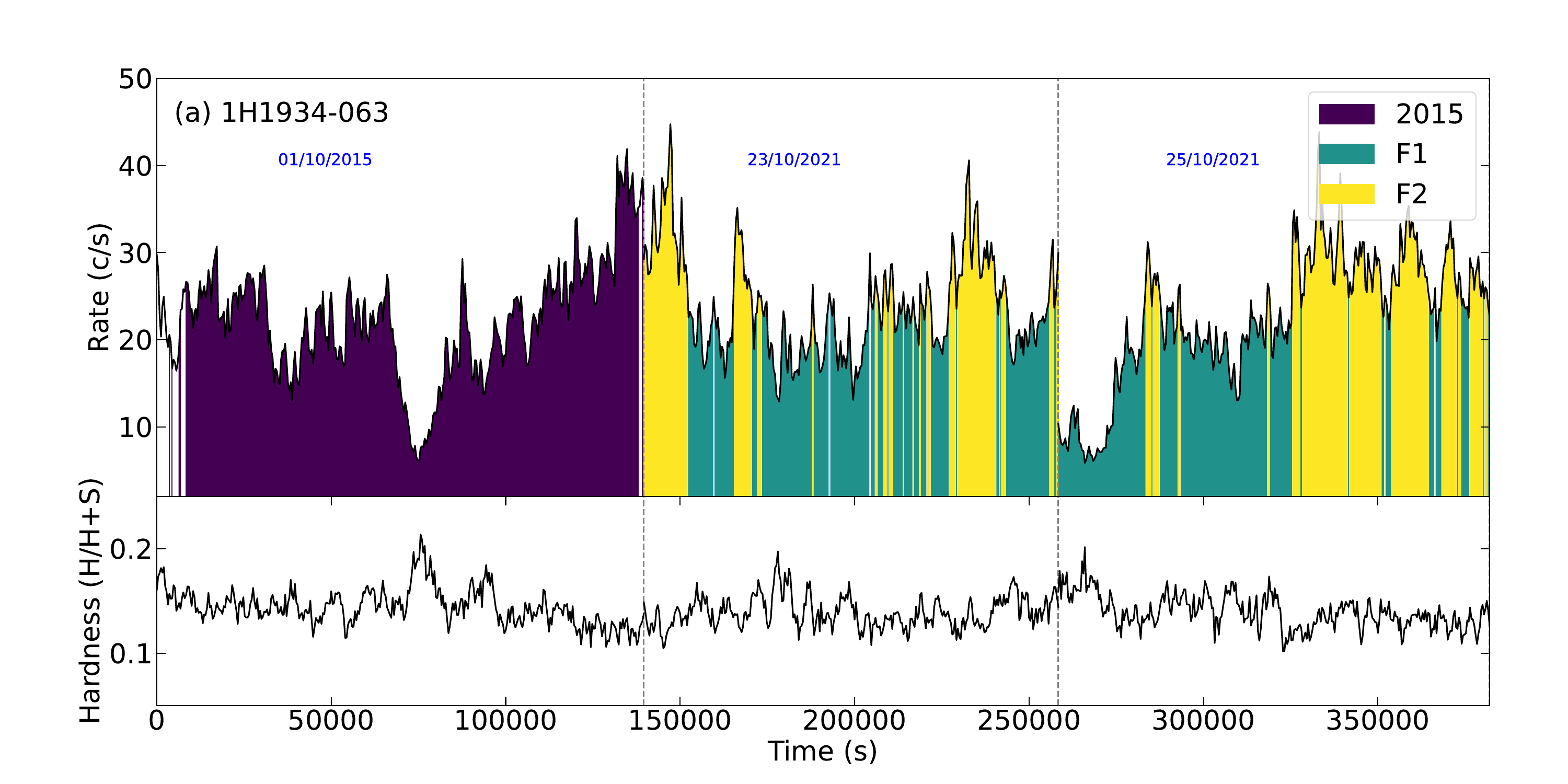}
	\includegraphics[width=\columnwidth]{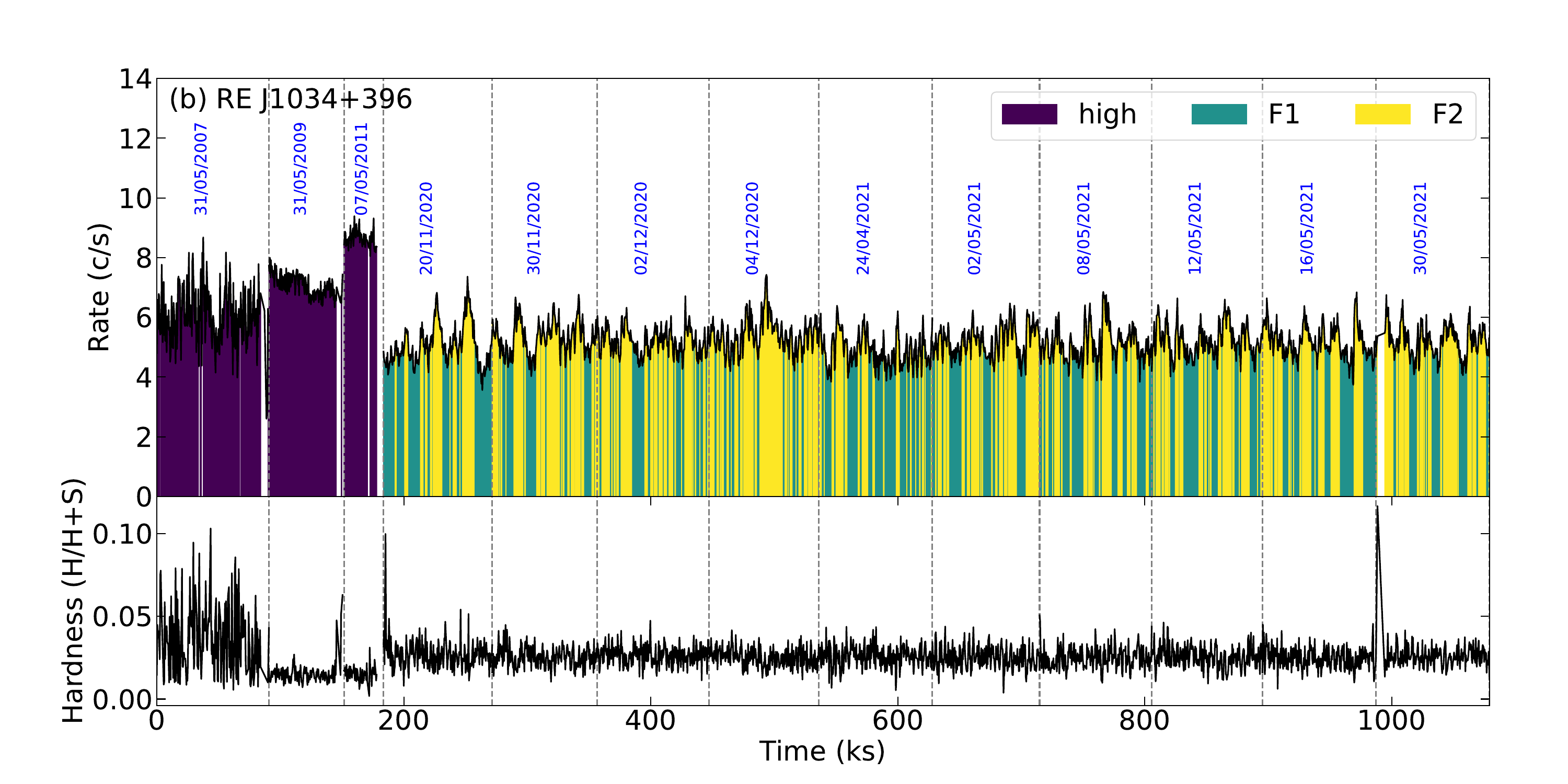}
	\includegraphics[width=\columnwidth]{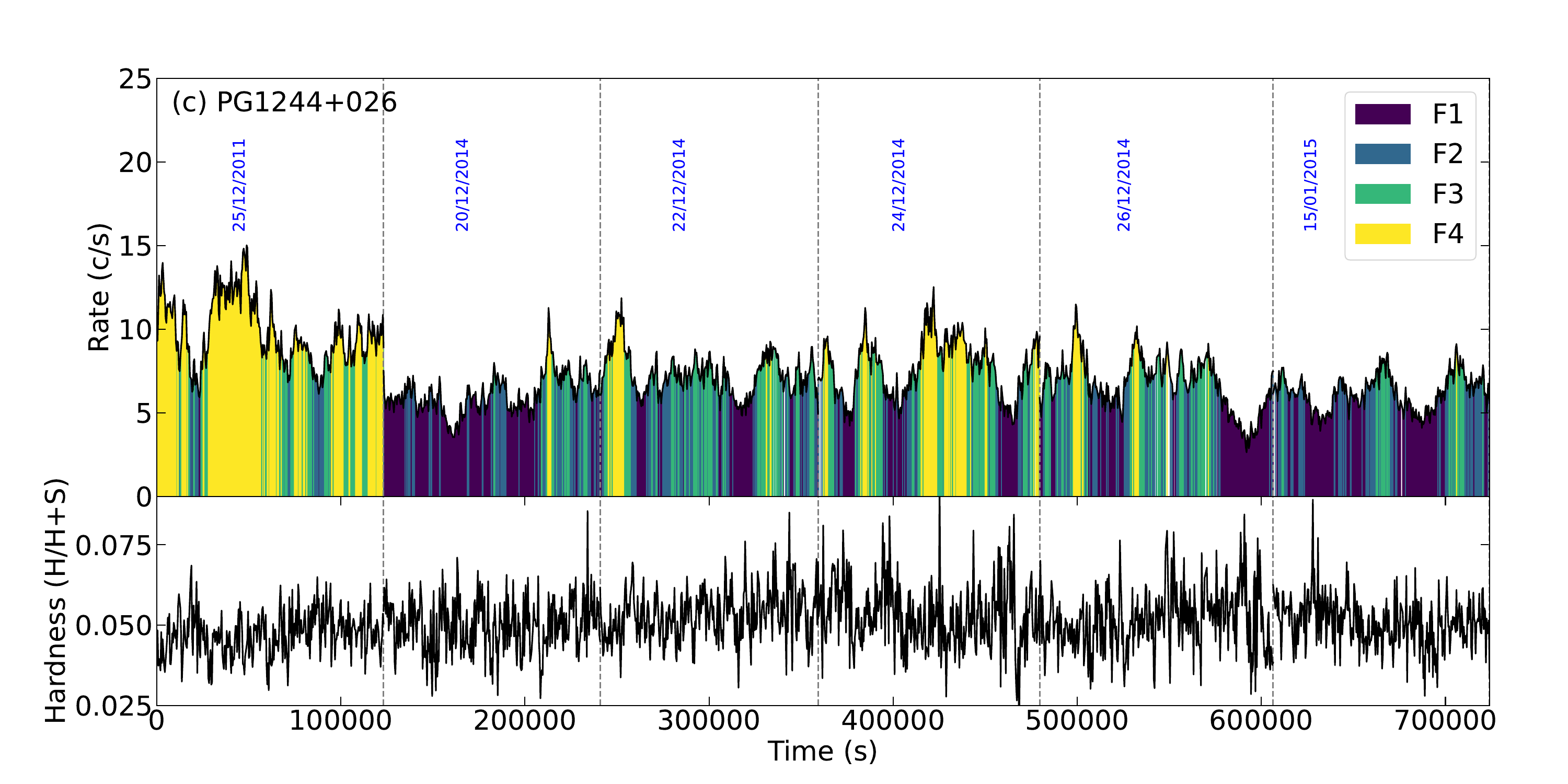}
	\includegraphics[width=\columnwidth]{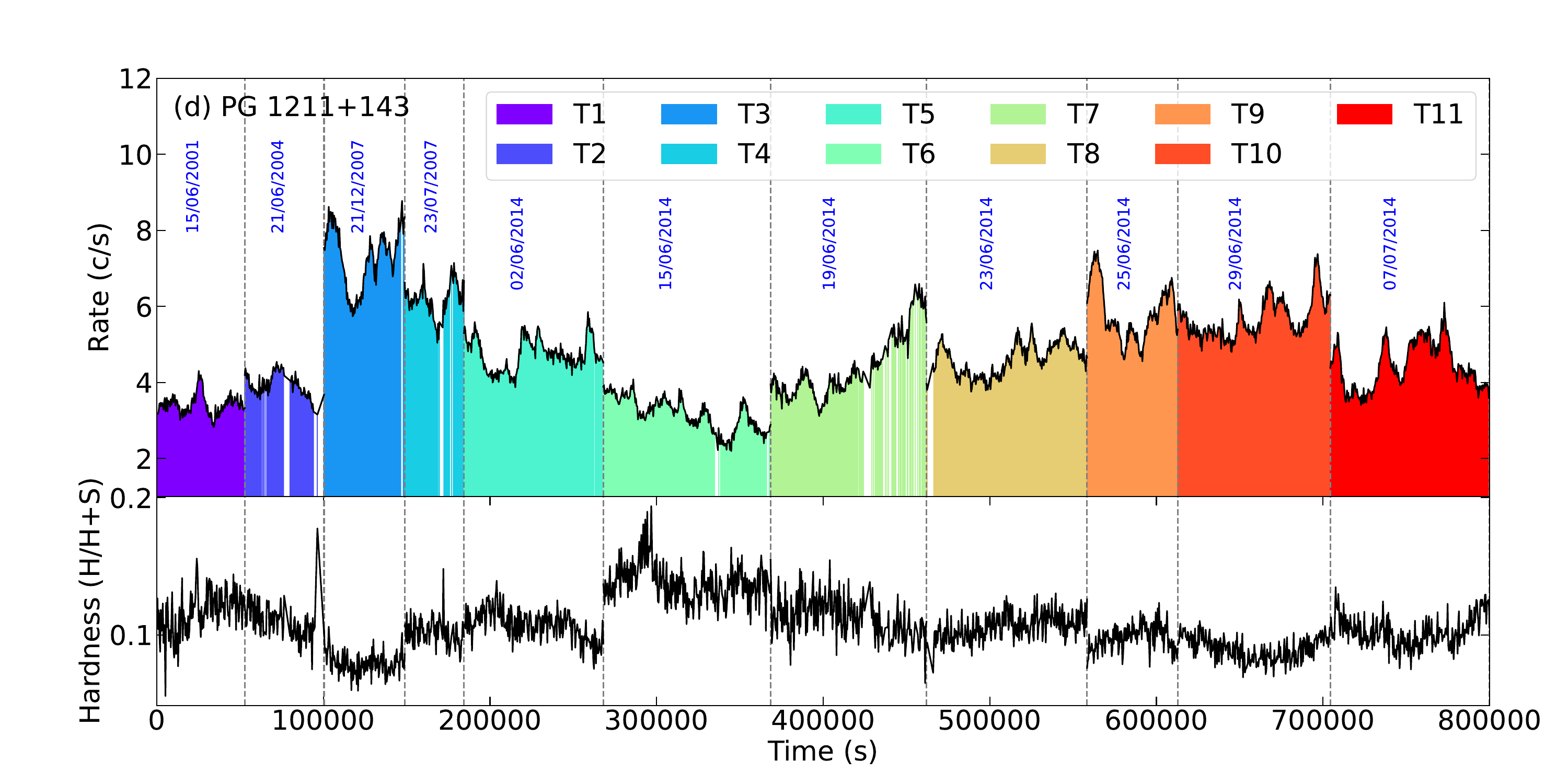}
	\includegraphics[width=\columnwidth]{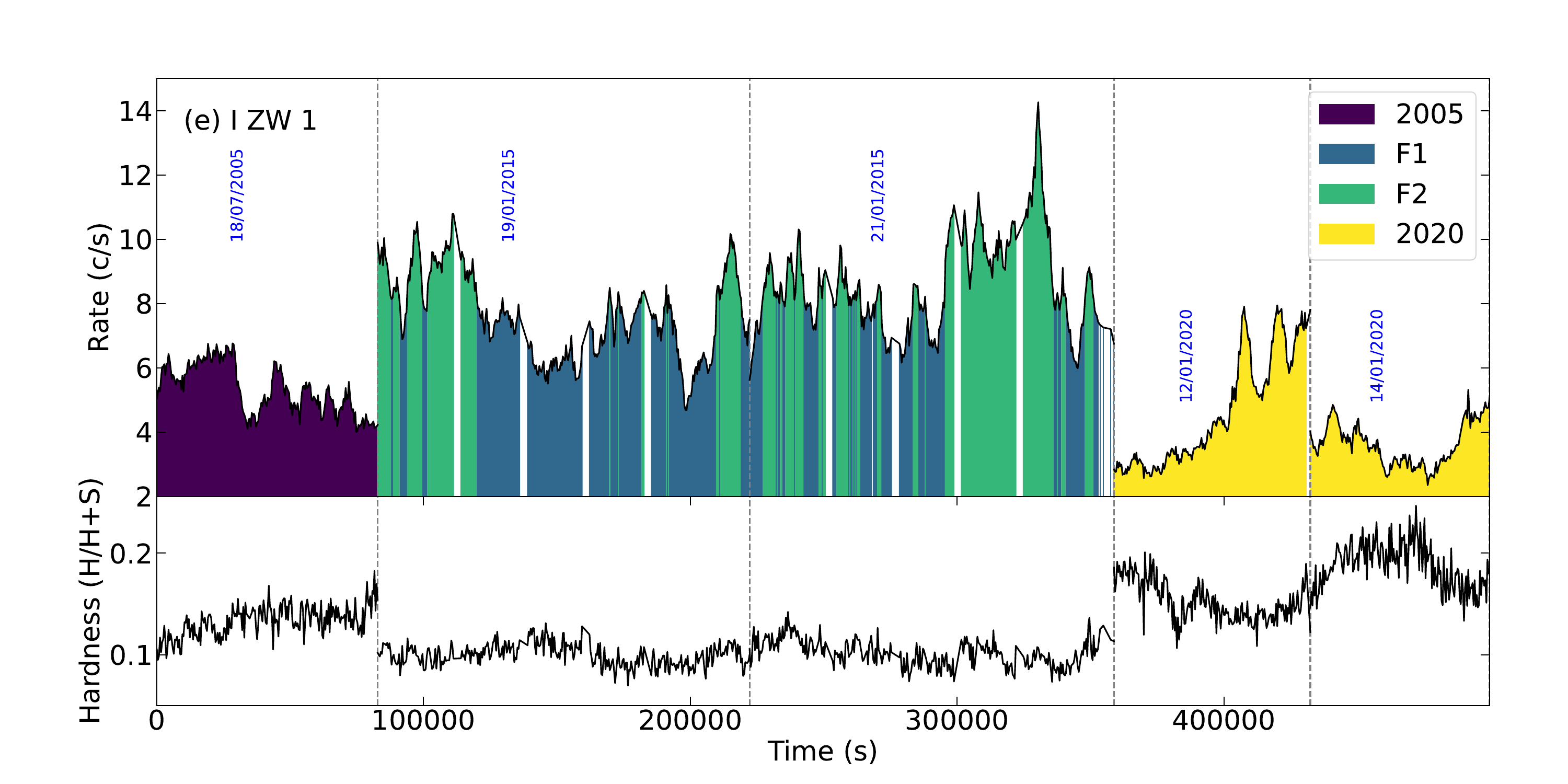}
	\includegraphics[width=\columnwidth]{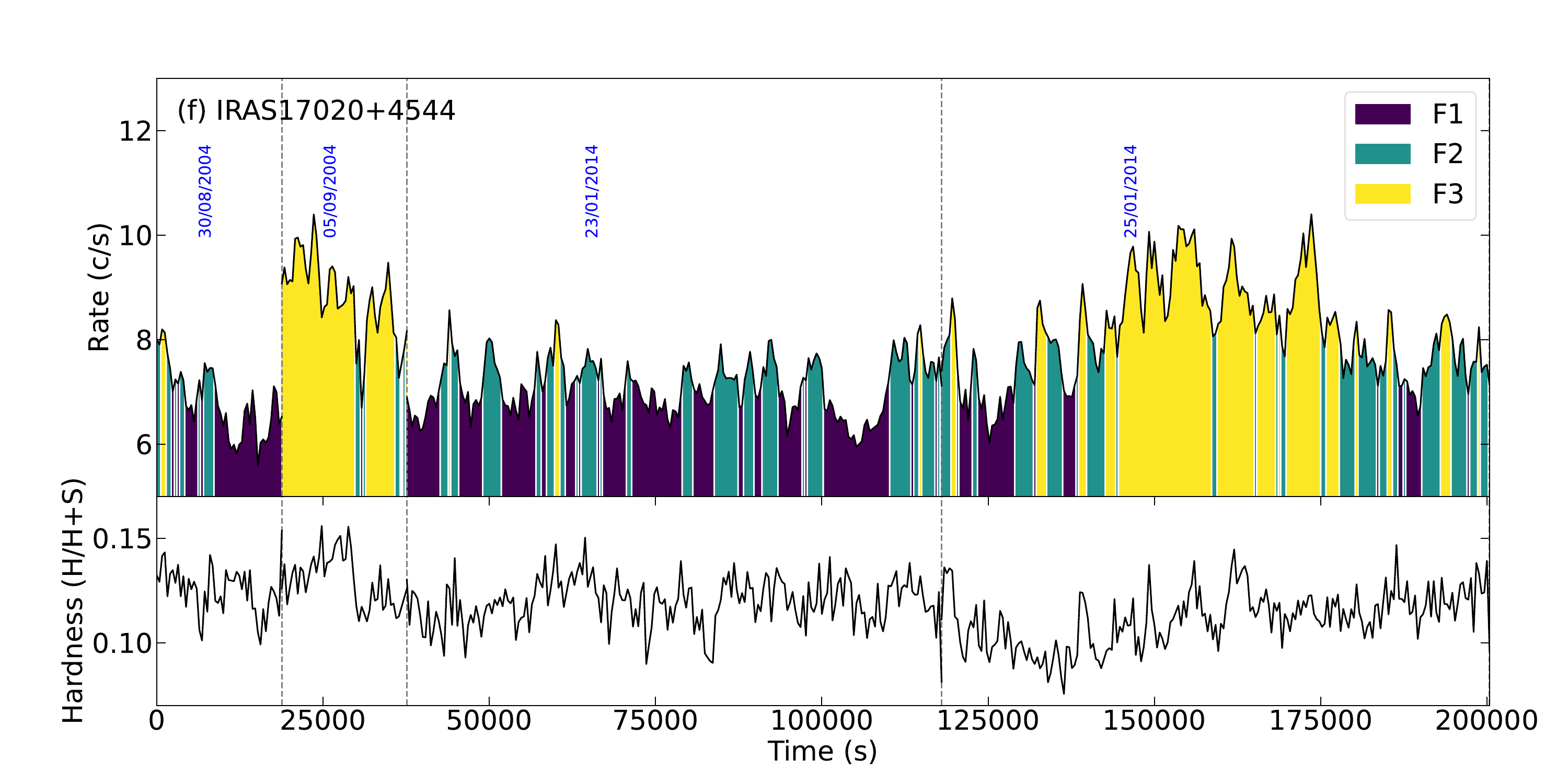}
    \caption{The EPIC-pn ($0.3\mbox{--}10$\,keV) light curve (\textit{upper}) and the corresponding hardness ratio (\textit{lower}) of the observations of 1H 1934-063 (\textit{a}), RE J1034+396 (\textit{b}), PG 1244+026 (\textit{c}), PG 1211+143 (\textit{d}), I ZW 1 (\textit{e}) and IRAS 17020+4544 (\textit{f}). The individual and flux-/time-resolved spectra of each source stacked by different approaches (listed in Tab. \ref{tab:obs}) are marked in different colors with labels. }
    \label{fig:lc}
\end{figure*}

To investigate the dependence of UFOs on source variability, we adopted different stacking approaches for the EPIC-pn and RGS spectra of each object and generated the time-averaged, time-resolved, and flux-resolved spectra. The stacking approaches and corresponding labels are listed in Tab. \ref{tab:obs}. The EPIC-pn ($0.3\mbox{--}10$\,keV) light curves and corresponding hardness ratios ($\mathrm{HR}=\mathrm{H}/\mathrm{H+S}$, H: $2\mbox{--}10$\,keV; S: $0.3\mbox{--}2$\,keV) are extracted with the \textsc{epiclccorr} task and shown in Fig. \ref{fig:lc}. For the flux-resolved approach, we equally divided the light curves into several flux levels (the number shown in Tab. \ref{tab:obs}), which makes the number of counts of each level comparable ($>10^{4}$ counts). The GTI files were created with the \textsc{tabgitgen} package and the flux-resolved spectra were subsequently extracted and stacked. For each source, except for 1H 1934-063, a time-averaged spectrum from all observations is also extracted and labeled as `avg', while the averaged spectrum of 1H 1934-063 results from stacking two new observations in 2021 and refers to as `2021'. The EPIC-pn and RGS spectra were separately stacked with \textsc{epicspeccombine} and \textsc{rgscombine} and grouped to over-sample the instrumental resolution at least by a factor of 3. 

\begin{table*}
\centering
\caption{Table of the 6 sources analyzed in this work. The second column lists the stacking approach according to the observations. The number in the brackets means the number of the generated spectra after stacking. The individual \textit{XMM-Newton} observations are shown in the third column. For clarity, the corresponding labels of the stacked and individual spectra are listed in the fourth column, where the time-resolved spectra are referred to as T1\dots T11 chronologically and the flux-resolved spectra are referred to as F1\dots F4, from the faintest to brightest state. For each source, except for 1H 1934-063, the time-averaged spectrum from all observations is also generated and referred to as the `avg' spectrum, while for 1H 1934-063, the two new observations (Obs. ID: 08910101(2)01) are stacked and named as `2021'.
}
\begin{tabular}{lccccc}
\hline
\hline
Sources & Stacking Approach & Obs. ID & Labels \\
\hline
\multirow{2}{*}{1H 1934-063} & individual (1) & 0761870201 & 2015 \\
            & flux-resolved (2) & 0891010101, 0891010201  & F1, F2\\
\hline
\multirow{4}{*}{RE J1034+396} & time-averaged (1) & 0506440101, 0561580201, 0675440301 & High\\
            & \multirow{3}{*}{flux-resolved (2)} & 0865010101, 0865011001, 0865011101,  & \multirow{3}{*}{F1, F2}\\ 
            & & 0865011201, 0865011301, 0865011401,  \\
            & & 0865011501, 0865011601, 0865011701, 0865011801 \\
\hline
\multirow{2}{*}{PG 1244+026} & \multirow{2}{*}{flux-resolved (4)} & 0675320101, 0744440101, 0744440201 & \multirow{2}{*}{F1, F2, F3, F4}\\
            &   &0744440301, 0744440401, 0744440501 \\
\hline
\multirow{3}{*}{PG 1211+143} & \multirow{3}{*}{time-resolved  (11)} & 0112610101, 0208020101, 0502050101, 0502050201 & T1, T2, T3, T4\\
            & & 0745110101, 0745110201, 0745110301, 0745110401 & T5, T6, T7, T8 \\ 
            & & 0745110501, 0745110601, 0745110701 & T9, T10, T11 \\
\hline
\multirow{3}{*}{I ZW 1}      & individual (1) & 0300470101 & 2005\\
            & flux-resolved (2) & 0743050301, 0743050801 & F1, F2\\
            & time-averaged (1) & 0851990101, 0851990201 & 2020\\
\hline
IRAS 17020+4544 & flux-resolved  (3) &0206860101, 0206860201, 0721220101, 0721220301 & F1, F2, F3\\
\hline
\end{tabular}
\label{tab:obs}
\vspace{-3mm}
\end{table*}
\section{Methods}\label{sec:methods}
In this section, we introduce our analysis methods, including the spectral modeling (see Sec. \ref{subsec:contiuum}), Gaussian line scan (see Sec. \ref{subsec:gaussian}), and the photoionization model scan (see Sec. \ref{subsec:xabs}) to visualize the UFO absorption lines in each spectrum, identify the best-fit solution for UFOs and obtain the UFO properties.

\subsection{Continuum Modeling}\label{subsec:contiuum}
The X-ray data analysis software XSPEC \citep[v12.12.1,][]{1996Arnaud} is used for broadband-band spectral analysis of the EPIC-pn and RGS data. We consider the RGS spectra between $0.4\mbox{--}1.77$\,keV, and the EPIC-pn spectra only between $1.77\mbox{--}10$\,keV (except for RE J1034+396 and PG 1244+026 between $1.77\mbox{--}8$\,keV due to the background domination above 8\,keV) in our analysis, due to the influence of the lower resolution but higher count rate of EPIC-pn on the detection of atomic features. The instrumental differences are taken into account by adopting a variable cross-calibration factor \texttt{constant}. We use the C$\mbox{--}$stat \citep{1979Cash} statistics and estimate the uncertainties of all parameters at the 90\% confidence level (i.e. $\Delta\mathrm{C}\mbox{--}\mathrm{stat}=2.71$) unless explicitly stated. In this paper, the luminosities are calculated by the \texttt{cflux} model with the assumption of $H_\mathrm{0}=70$\,km/s/Mpc, $\Omega_\mathrm{\Lambda}=0.73$ and $\Omega_\mathrm{M}=0.27$. We use \texttt{tbabs} in XSPEC to model the Galactic absorption with Galactic column densities provided by \citet{2016HI4PI}, with the exception of 1H 1934-063. For this particular case, we accounted for the relatively heavy Galactic absorption ($\sim10^{21}\,\mathrm{cm}^{-2}$) using the high-resolution photoabsorption model \texttt{ismabs} instead. We adopt the solar abundance calculated by \citet{2009Lodders} to keep consistent with the subsequently used photoionization model in Sec. \ref{subsec:xabs}. The redshift of AGN is taken into account by \texttt{zashift} in XSPEC. In general, the broadband X-ray continuum of AGN consists of a primary power-law component from hot Comptonization \citep[the hot corona,][]{1993Haardt}, a soft excess, and an X-ray reprocessing of the primary irradiation \citep{2005Ross}, displayed by fluorescent lines (especially Fe K emission). The origin of the soft excess is still being debated, either relativistically blurred high-density reflection or warm Comptonization \citep[e.g.,][]{2018Petrucci,2019Garc,2020Middei,2020Petrucci,2021XuSE}. In this paper, since we are only interested in the atomic features instead of the origin of the broadband continuum, we adopt the model \texttt{diskbb}, characterized by the temperature at the inner accretion disk, to phenomenologically account for the soft excess for the sake of simplicity. The primary and reprocessing emissions are explained by a flavor of the relativistic reflection model \citep[RELXILL v1.4.3,][]{2014Garc}, which includes a hot Comptonization continuum (\texttt{relxillCp}). This model is characterized by the temperature $kT_\mathrm{e}$ and the emissivity index $q$ of the hot corona, the spectral slope $\Gamma$, the spin of the black hole $a_\star$, and several disk properties (inner $R_\mathrm{in}$ and outer $R_\mathrm{out}$ radius, inclination angle $i$, ionization state $\log\xi$, and iron abundance $A_\mathrm{Fe}$).

Therefore, the general continuum model for most targets in this work is:
\begin{center}
    \texttt{constant}*\texttt{tbabs}*\texttt{zashift}*\texttt{cflux}*(\texttt{diskbb}+\texttt{relxillCp}),
\end{center}
except for RE J1034+396 and IRAS 17020+4544. Due to the lack of discernible reflection features in the spectra of RE J1034+396, we adopted the best model combination from \citet{2021Jin}, which assumes the warm Comptonization explanation for the soft excess and consists of three Comptonization components for the hot, intermediate, and warm Comptonization emission. In IRAS 17020+4544, a single reflection component cannot explain the relativistically broadened Fe K emission. Therefore, in accordance with the best-fit continuum model from \citet{2020Gonzalez}, we included an additional \texttt{laor} component, which models the disk line affected by strong gravity, with disk parameters (the inclination angle, inner and outer disk radius) linked to those of \texttt{relxillCp}, while the line energy, emissivity index, and normalization are left free. For the remaining targets, our general continuum model also aligns with the best combination in prior studies on those sources  \cite[e.g.][]{2014Kara,2016LobbanRef,2022Xu,2022Wilkins}. 
The long-term invariants, such as the black hole spin, disk inclination angle, and iron abundance, were linked together across the averaged, time- and flux-resolved spectra of the same source during the analysis. Moreover, we have tested a few different continuum models, such as changing the flavors of RELXILL and replacing the phenomenological model \texttt{diskbb} with a warm corona or a high-density relativistic reflection model. We found that the choice of the continuum model does not affect the following results on the line modeling and detection of UFOs. Details of continuum parameters are unrelated to our goal and thus are not shown in this work. Only inclination angles derived from the reflection model are listed in Tab.\ref{tab:sources} for further analysis in Sec. \ref{subsec:incl}.

\subsection{Gaussian Line Scan}\label{subsec:gaussian}
The blind Gaussian line scan over the spectra, which have been modeled by the continuum components, is a straightforward way to visualize the atomic features on the continua. The method is to fit an additional Gaussian line upon the continuum model with a logarithmic grid of energy steps over $0.4\mbox{--}10$\,keV and record the statistical improvement ($\Delta\mathrm{C}\mbox{--}\mathrm{stat}$) at each step. The energy centroid and the line width are fixed at each step, while the normalization is free to be negative or positive. To identify both the narrow and broad atomic features, we assumed the line widths at $\sigma_v$ of 500, 1500, 4500\,km/s, and the corresponding numbers of the energy steps at 2000, 700, 300, respectively, to maintain the balance between the computational cost and the resolving power of instruments. The recorded statistical improvements provide a rough estimate of the detection significance of individual lines in a single trial (i.e. in the unit of $\sigma$), in the form of the square root of $\Delta\mathrm{C}\mbox{--}\mathrm{stat}$ times the sign of the normalization \citep{1979Cash}. This analysis approach is performed for every spectrum extracted from the observations of the sources in our sample. The results are shown in \cref{app:fig:gaussian-1h1934-063,app:fig:gaussian-rej1034+396,app:fig:gaussian-pg1244+026,app:fig:gaussian-pg1211+143,app:fig:gaussian-izw1,app:fig:gaussian-iras17020+4544}, where the centroid and the uncertainty of prominent UFO absorption features discovered in the following photoionization modeling are respectively highlighted by the vertical dashed \textit{red} lines and shaded regions. The details of the line scan results over each spectrum of individual sources are presented in Appendix \ref{app:sec:obs+spectral}. 

\subsection{Photoionization Model Scan}\label{subsec:xabs}
To obtain the physical properties of UFOs, we employ the physical photoionization model, \texttt{pion}, in the SPEX package \citep{1996Kaastra}. This model self-consistently calculates the transmission and emission spectra of ionized gas in photoionization equilibrium and only needs to provide the SED of the radiation field. To implement the \texttt{pion} model in XSPEC, we adopted the code used in \citet{2019Parker} to construct the tabulated XSPEC version of \texttt{pion}. In this work, we only make use of the absorption component of \texttt{pion} (solid angle $\Omega=0$ and covering fraction $C_\mathrm{F}=1$), i.e., equals to the \texttt{xabs} code in SPEX, and the re-built model is thus named as \texttt{xabs\_xs}. The \texttt{xabs} model is characterized by four main parameters: column density $N_\mathrm{H}$, ionization parameter $\xi$, turbulence velocity $\sigma_v$ and the line-of-sight (LOS) velocity of gas $v_\mathrm{LOS}$.

\begin{figure}[htbp]
    \centering
	\includegraphics[width=\columnwidth]{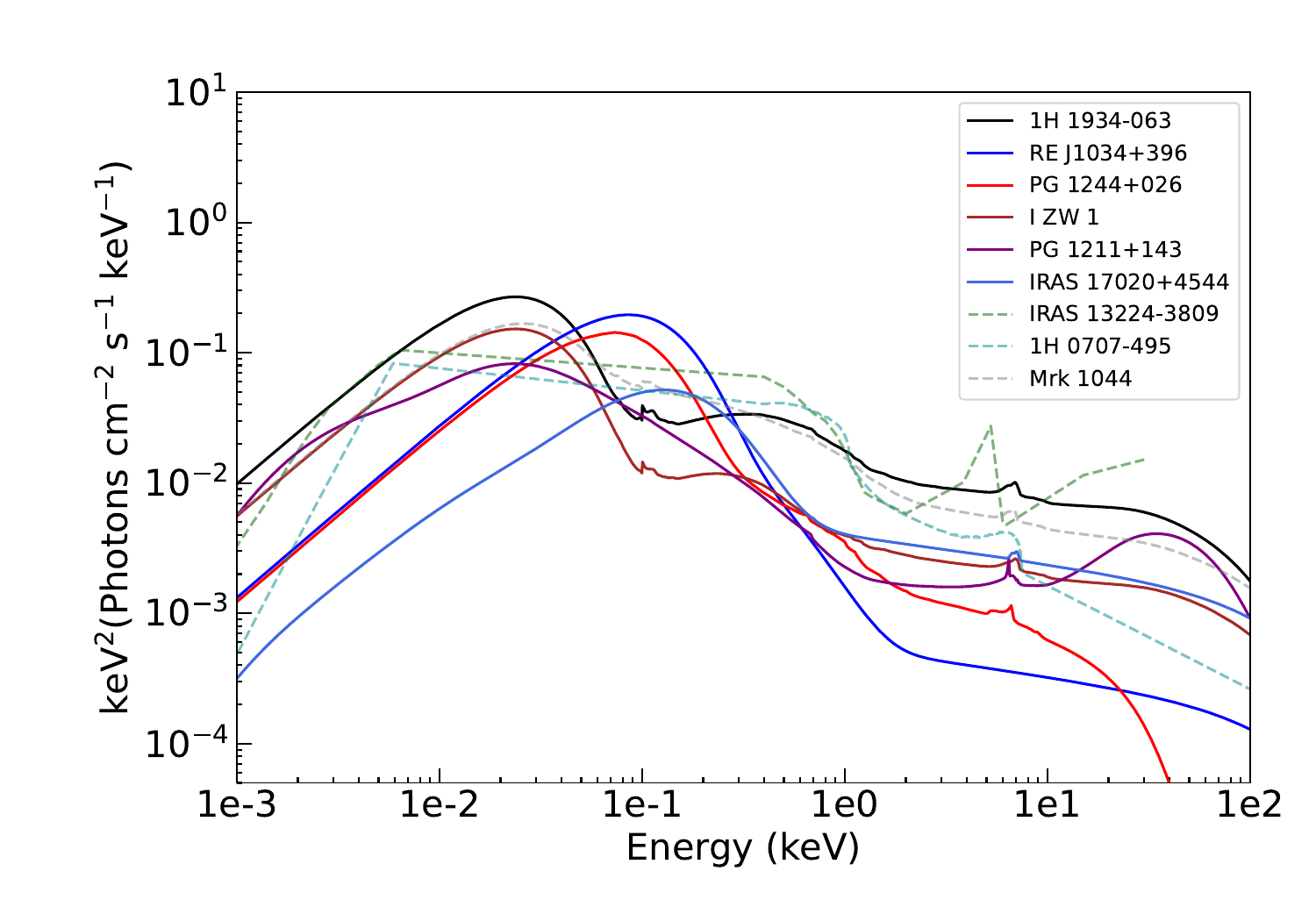}
    \caption{The averaged SED of sources in our sample compared with 1H 0707-495 \citep{2021Xu}, IRAS 13224-3809 \citep{2018Jiang}, and Mrk 1044 \citep{2023Xu}. 
    }
    \label{fig:sed}
\end{figure}
The intrinsic SEDs of individual sources are derived from the UV, constrained by OM spectra, to the hard X-ray band (from 1\,eV to 10\,keV). We adopted the AGN SED model, \texttt{AGNSLIM} \citep{2018Kubota,2019Kubota}, sharing the same temperature between the disk and the seed photon of the warm Comptonization, only for SED calculations. This model allows us to directly measure the Eddington ratio, listed in Tab.\ref{tab:sources}. The relativistic reflection component was included as well and fixed at the best fit derived from the continuum modeling in Sec. \ref{subsec:contiuum} without the primary continuum in the model. The Galactic extinction is also considered by the \texttt{redden} model according to \citet{2011Schlafly}. Nonetheless, for RE J1034+396, PG 1244+026, and I ZW 1, the UV/optical data cannot be well explained by \texttt{AGNSLIM}. In these cases, we employed the phenomenological model \texttt{diskbb} for the UV/optical spectra. Specifically, the characterized disk temperatures of these sources are $30^{+16}_{-15}$, $28^{+17}_{-15}$, and $12^{+10}_{-9}$\,eV respectively. We roughly estimated their Eddington ratios by calculating the bolometric luminosity ($10^{-3}\mbox{--}10^{3}$\,keV) predicted by the model. The intrinsic time-averaged SEDs of each source are shown in Fig. \ref{fig:sed}. Due to the concerns about the potential impact of the loosely constrained UV/optical measurements on photoionization modeling, we investigated and confirmed that variations in the disk temperature within the uncertainties do not significantly affect the ionization balance of the plasma, and the conclusions of this work.

Ideally, given the SED of the radiation field, we can obtain the best-fit parameters of \texttt{xabs\_xs} by directly fitting the spectra. However, the fits to the photoionization model are easy to fall in the local minima of the parameter space due to the fact that the same absorption features (if without enough statistics) can be comparably explained by different solutions with different ionization states and speeds. In addition, some weak UFO solutions might be overlooked if there is no obviously significant single absorption line but multiple moderately significant absorption features in the spectra. Therefore, we launch a systematic scan over a multi-dimension grid of the parameters ($\log\xi,z_\mathrm{LOS},\sigma_v$) of \texttt{xabs\_xs} upon the continuum model for each spectrum. This method is capable of identifying the globally best-fit solution and showing the locations of all potential solutions in the parameter space, which might reveal the multiphase outflows as long as the other solutions do not degenerate with the primary solution. The grids consist of the ionization parameter $\log\xi$, ranging from 0 to 5 with a step of $\Delta\log\xi=0.1$, the turbulent velocity $\sigma_v$ ($\sigma_v=500,1500,4500$\,km/s) and the LOS velocity $v_\mathrm{LOS}$, from -105000\,km/s to 0 with an increment, like Gaussian line scan, depending on the choice of $\sigma_v$ ($c\Delta z_\mathrm{LOS}=500,700,1500$\,km/s for $\sigma_v=500,1500,4500$\,km/s respectively). In the scan, the column density $N_\mathrm{H}$ and the continuum parameters are left free to vary. The C-stat improvement is recorded at each grid to reveal the detection significance of the absorber.

\begin{figure*}
\centering
    \includegraphics[width=\columnwidth, trim={20 0 20 10}]{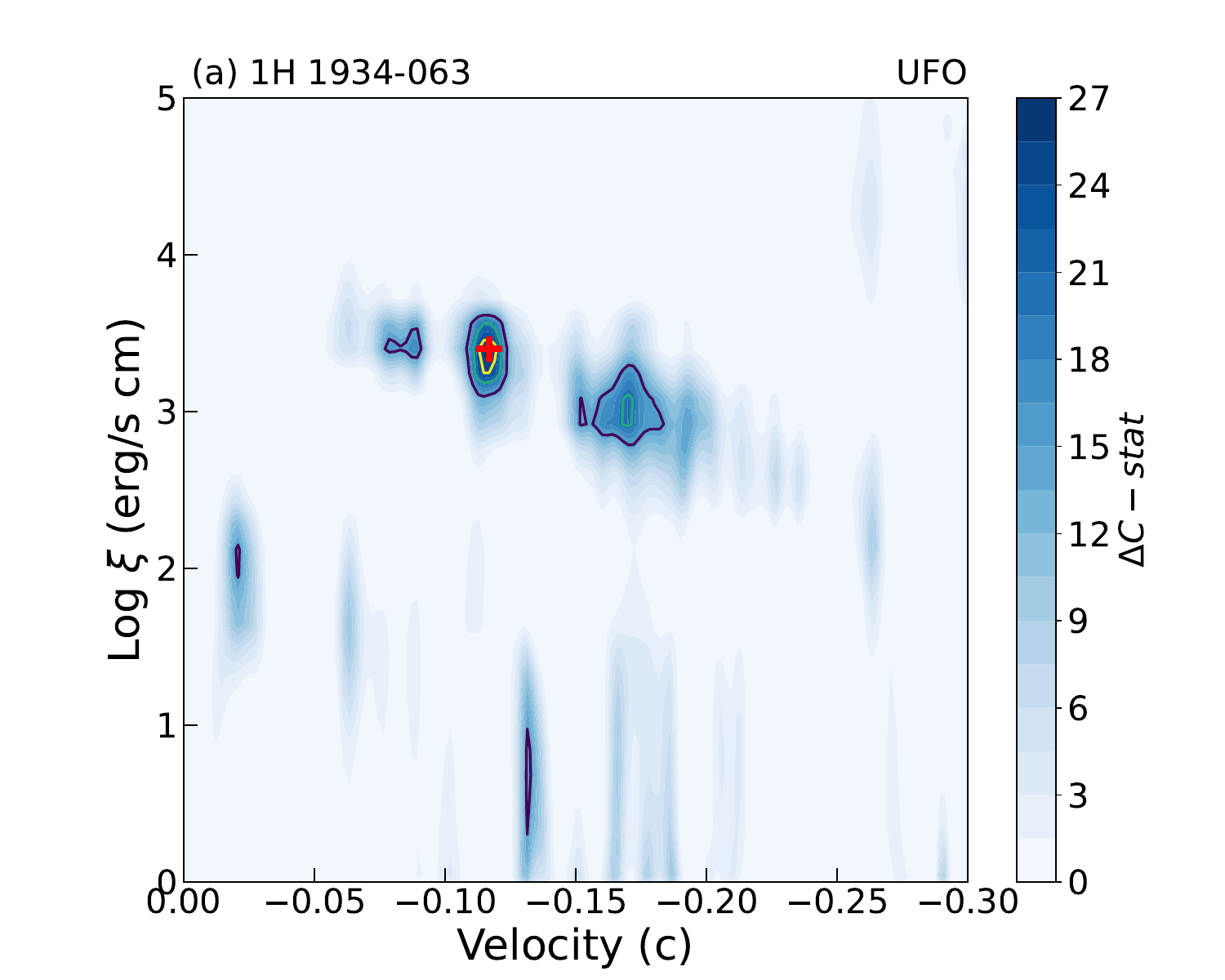}
    \includegraphics[width=\columnwidth, trim={20 0 20 10}]{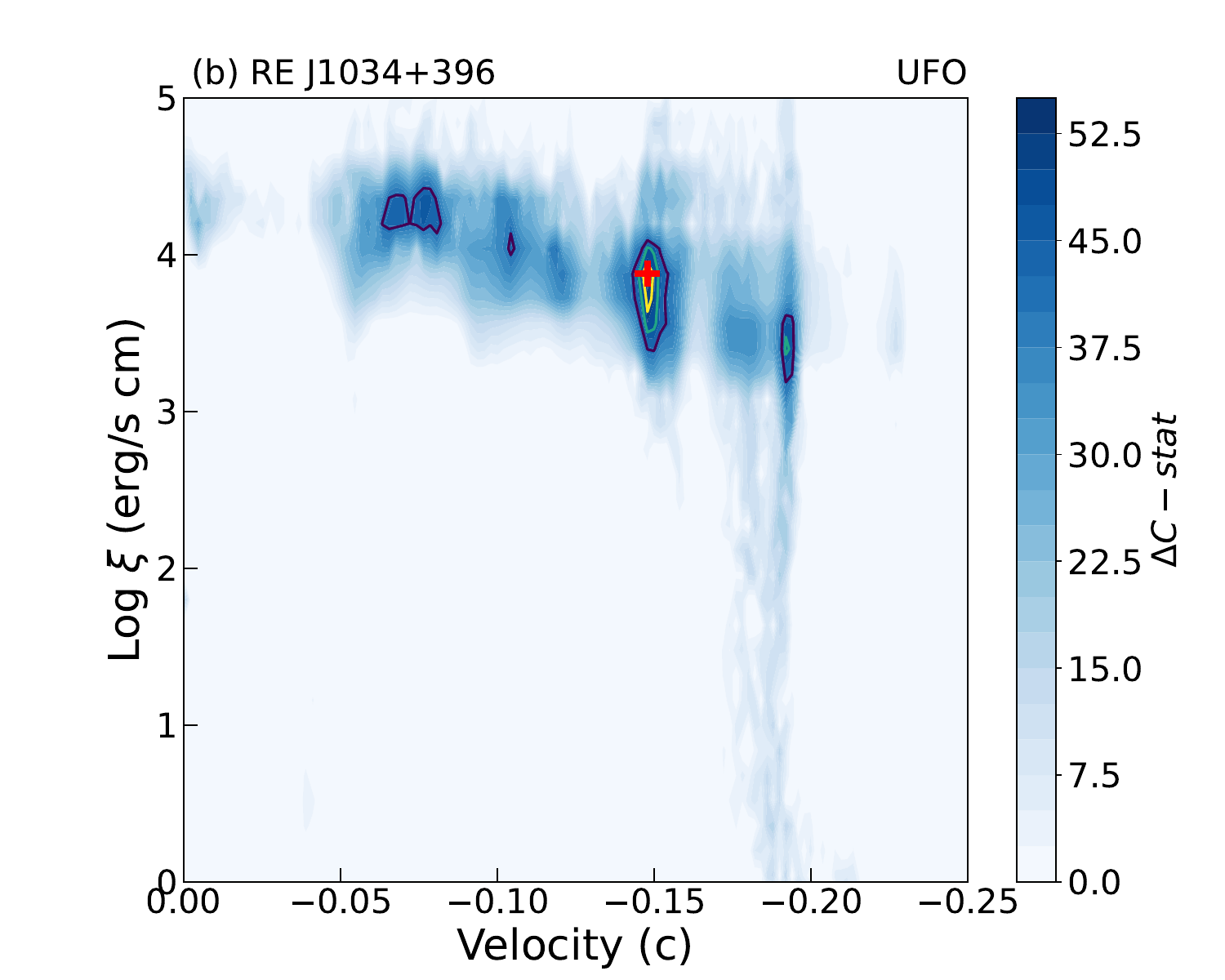}
    \includegraphics[width=\columnwidth, trim={20 0 20 10}]{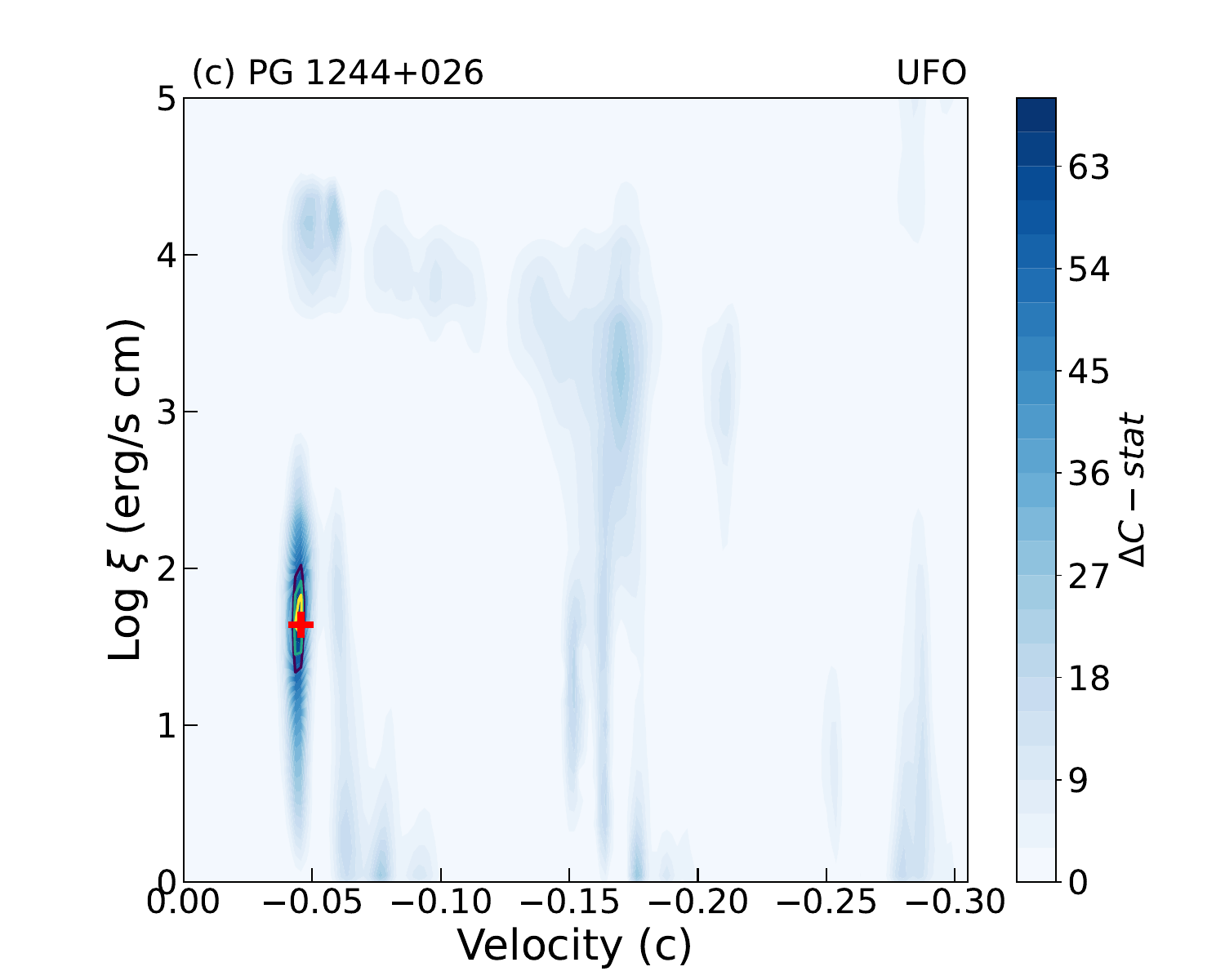}
    \includegraphics[width=\columnwidth, trim={20 0 20 10}]{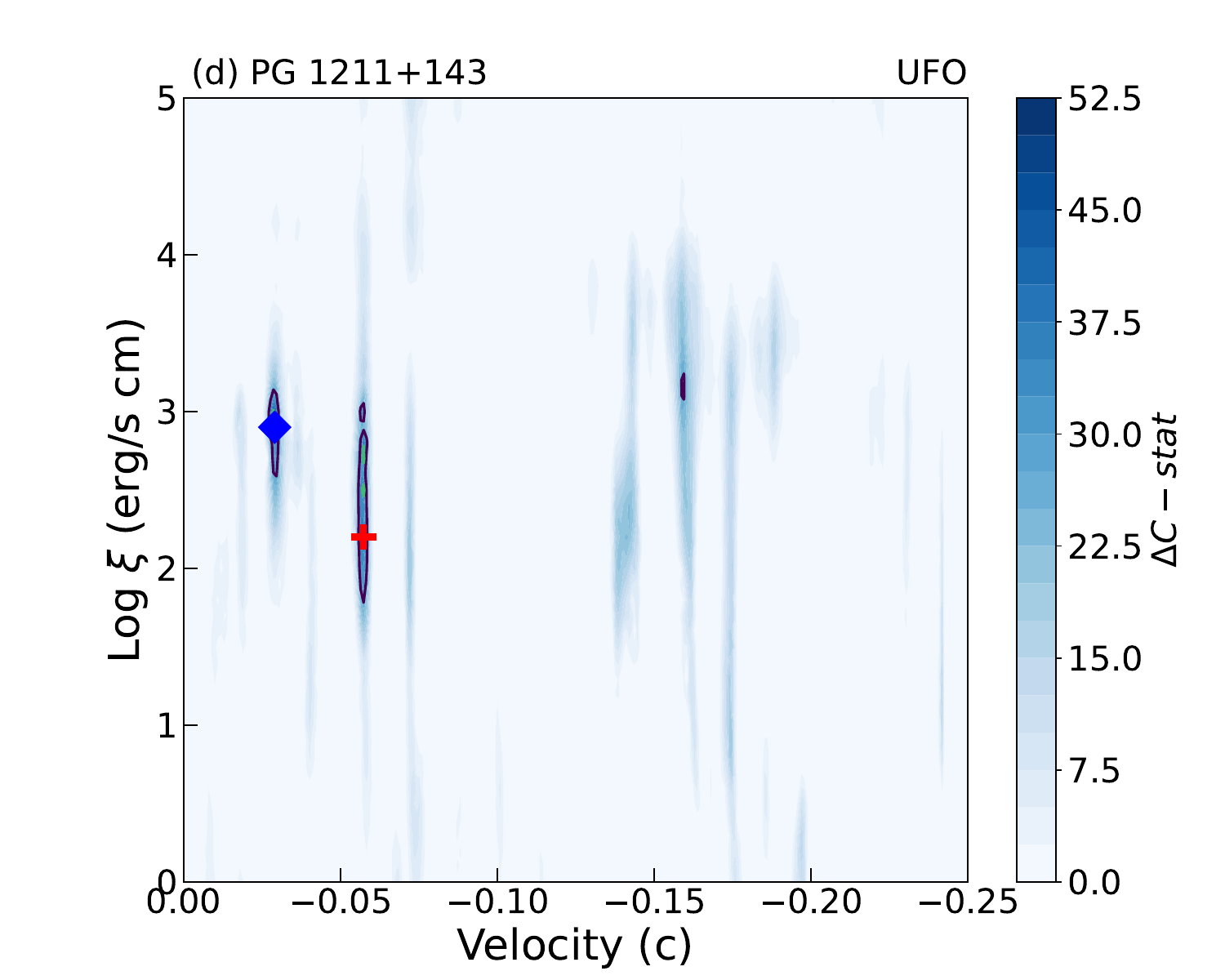}
    \includegraphics[width=\columnwidth, trim={20 0 20 10}]{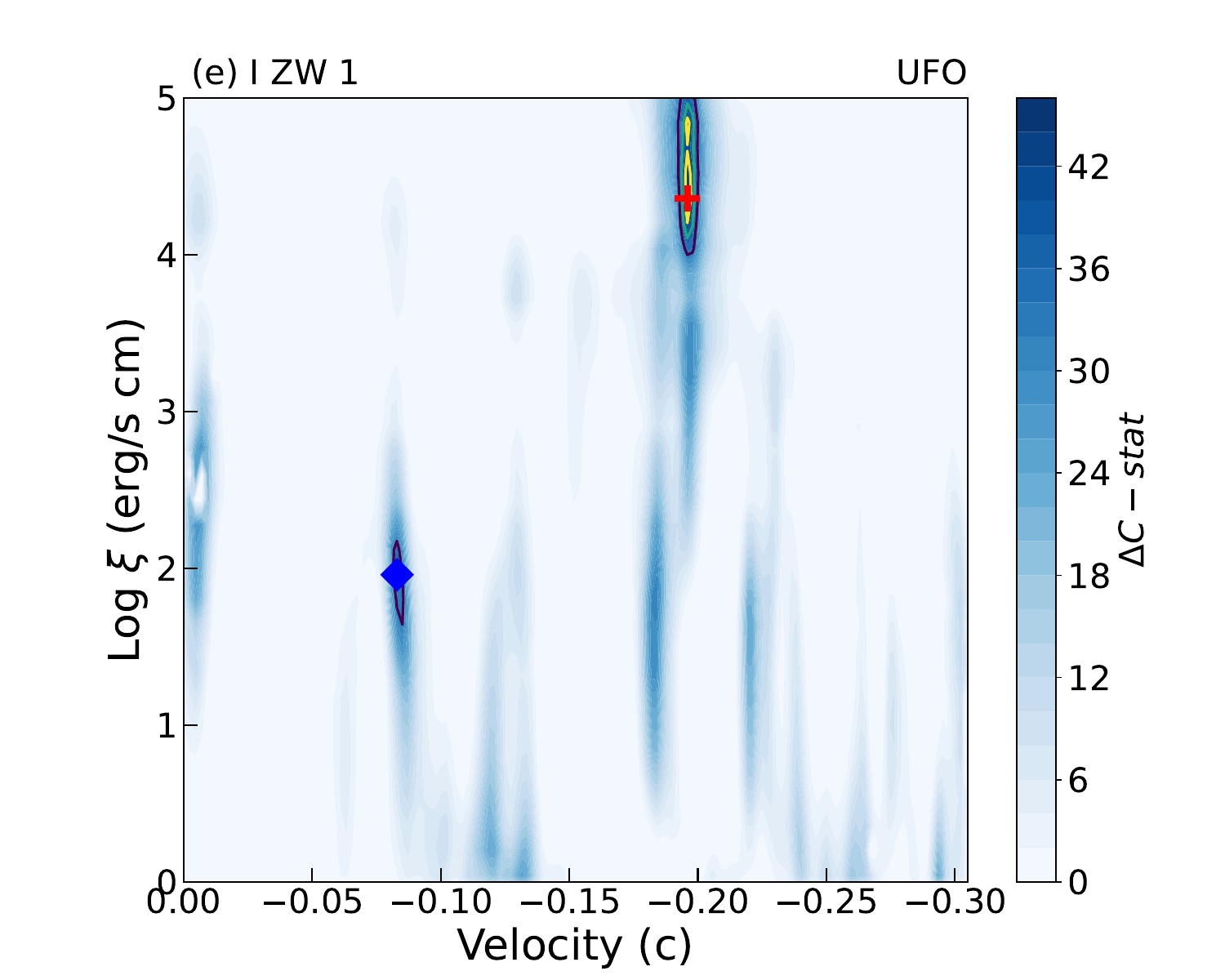}
    \includegraphics[width=\columnwidth, trim={20 0 20 10}]{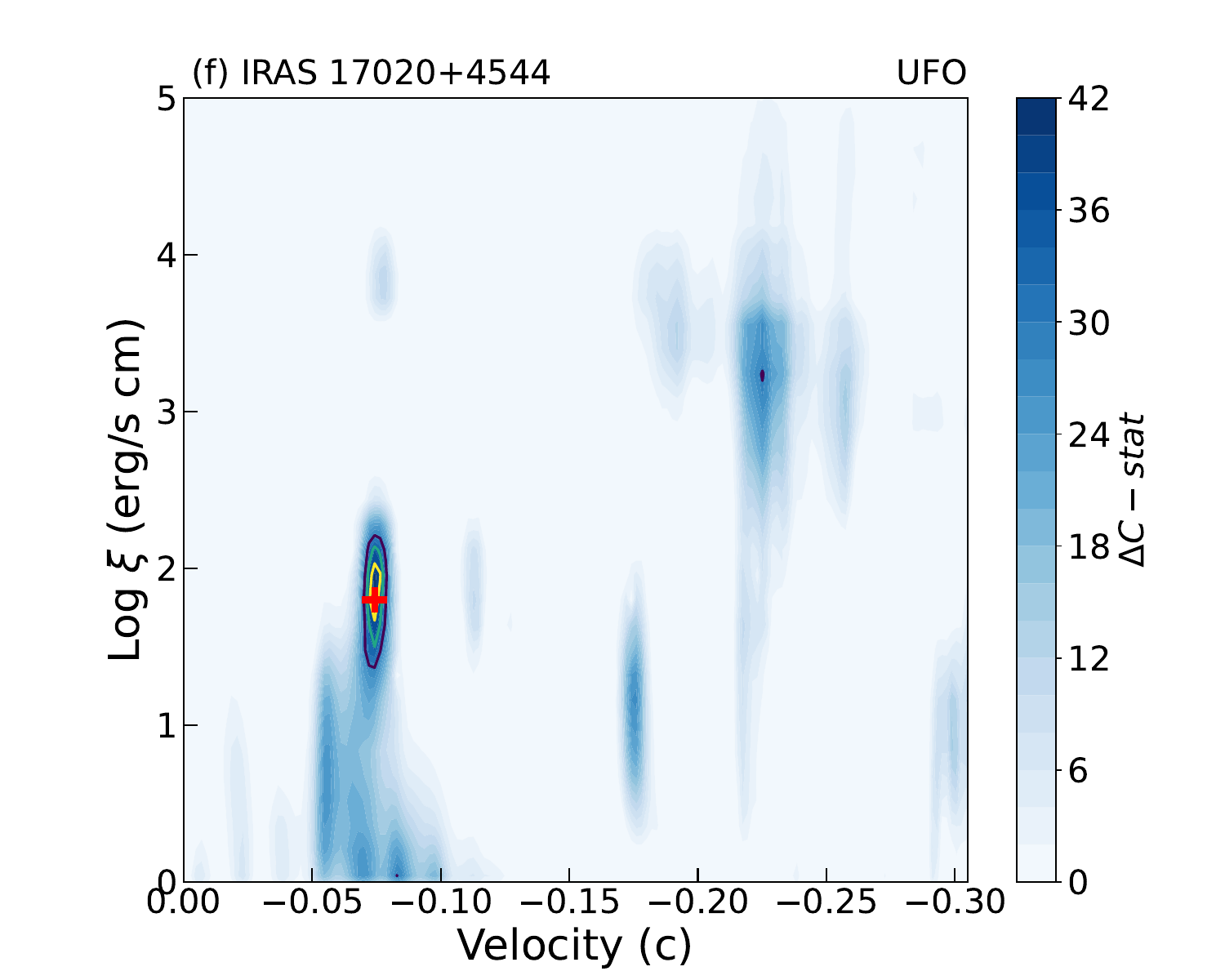}
    \caption{Photoionization absorption model search for the time-averaged spectra of 1H 1934-063 (\textit{a}), RE J1034+396 (\textit{b}), PG 1244+026 (\textit{c}), PG 1211+143 (\textit{d}), I ZW 1 (\textit{e}) and IRAS 17020+4544 (\textit{f}) over the baseline model (continuum or plus WAs). The color illustrates the statistical improvement after adding an absorption component with a line width of 1500\,km/s (except PG 1211+143 with $\sigma_v=500$\,km/s). The \textit{purple}, \textit{green}, and \textit{yellow} contour lines correspond to 99.73\%, 95.4\%, and 68.3\% confidence level. The solution with the most significant detection is marked by a \textit{red cross} and the secondary detection is marked by a \textit{blue diamond}, while it may not be the final solution, as another potential solution can be the most significant detection during the spectral modeling when the line width $\sigma_v$ is free to vary.}
    \label{fig:absorption_scan}
\end{figure*}

The photoionization model scan is performed on every spectrum in our sample. If the secondary solutions have a significance above $3\sigma$ in a single trial, i.e. $\Delta$C-stat/d.o.f.$=16.25/4$ (the real significance of several solutions will be measured by simulations in Sec.\ref{subsec:xabsfits}), we will iteratively include an additional absorption component into the model and re-perform the scan upon the new baseline model until the significance of adding a further absorption component to the model falls below $3\sigma$. For clarity, we only present the scan results detecting UFOs across the time-averaged spectrum of each target in Fig.\ref{fig:absorption_scan}, while the results detecting WAs, such as 1H 1934-063 \citep{2022Xu}, RE J1034+396 (firstly detected), I ZW 1 \citep{2007Costantini,2022Rogantini}, and IRAS 17020+4544 \citep{2018Sanfrutos} are not shown. We exhibit results with a line width of 1500\,km/s with the exception of PG 1211+143 ($\sigma_v=$500\,km/s) due to the consistency among scan results with different line widths. The velocity on the X-axis is the relativistically corrected velocity according to the equation: $v/c=\sqrt{(1+v_\mathrm{LOS}/c)/(1-v_\mathrm{LOS}/c)}-1$. The grid with the strongest detection is marked by a red cross but it may not be the final result, because during the direct spectral modeling, the line width is free to vary and the best fit may fall in another solution. But the scan plots at least provide a reference map for the globally best-fit solution of UFOs. The second UFO solution is marked by a blue diamond (i.e. in PG 1211+143 and I ZW 1). The \textit{purple}, \textit{green}, and \textit{yellow} contour lines represent 3, 2, and 1$\sigma$ uncertainty (i.e. 99.73\%, 95.4\% and 68.3\% confidence level).

\section{Results}\label{sec:results}
In this section, we present the results of the photoionization modeling (Section.\ref{subsec:xabsfits}) and the tentative dependence of UFO detection on viewing angles (Section.\ref{subsec:incl}).

\subsection{Photoionization Modeling}\label{subsec:xabsfits}
According to the scan results in Fig.\ref{fig:absorption_scan}, we directly fit the photoionization absorption model with a free turbulent velocity to each spectrum in our sample with the initial values obtained from the scan results. Conservatively, we also test other peaks in the scan plots in case their significance is larger than those obtained from the scan with a fixed line width, e.g. the best-fit UFO in RE J1034+396. The final best-fit parameters of UFOs are listed in Tab.\ref{tab:xabsfits}, while those of WAs are listed in Tab.\ref{app:tab:WAfits}. The single trial significance is estimated from the $\chi^2$ test (similar to the C-statistics for a large number sample \citep{1979Cash}). Only the absorbers with a significance above $3\sigma$ (i.e. $\Delta\mathrm{C\mbox{--}stat/d.o.f.}>16/4$) are taken into account, which explains why some previously reported UFOs are not included in our results, e.g. 4 UFOs in IRAS 17020+4544 but only 1 UFO with $\Delta\mathrm{C\mbox{--}stat/d.o.f.}>16/4$ \citep{2018Sanfrutos}. Despite the presence of numerous solutions in the scan plots, multiple UFOs are not detected in our results, except for PG 1211+143 and I ZW 1. This is because the solutions partially degenerate and account for the same line features, making the secondary solutions insufficiently apparent once the primary UFO is considered. We have also estimated the potential influence of the strong emission lines in some spectra on the modeling of the absorption component. We compared the fits before and after including a photoionization emission component (calculated from \texttt{pion} in SPEX) and found it does not significantly affect the fits and our conclusions, perhaps due to the narrow line widths of the emission lines. The descriptions of WAs and UFOs in individual sources are presented in Appendix.\ref{app:sec:obs+spectral}.

\begin{table*}[!htbp]
\renewcommand{\arraystretch}{1.25} 
\centering
\caption{Table of the best-fit UFO parameters derived from extracted spectra of targets in this work. The uncertainties of parameters are estimated at the 90\% confidence level ($\Delta \mathrm{C-stat}=2.71$).
}
\begin{tabular}{c|cccccc|c}
\hline
\hline
 Sources & \multicolumn{6}{c|}{UFO}   & Luminosity \\
\hline
       & $\log\xi$ & $N_\mathrm{H}$  & $\sigma_\mathrm{v}$  & $v_\mathrm{LOS}$ & $\Delta\rm C\mbox{-}stat$ & single trial 
       & $L_\mathrm{0.4\mbox{--}10\,keV}$ \\
       & $\mathrm{erg\,cm\,s^{-1}}$ &($10^{21}$\,cm$^{-2}$) &  (km/s) &  (km/s) &  & significance ($\sigma$)
       & ($10^{43}$\,erg/s) \\
\hline
1H 1934-063 &&&&&&&\\
\cline{2-7}
2015 & $1.60^{+0.20}_{-0.18}$ &  $0.05^{+0.02}_{-0.01}$ & $900^{+700}_{-500}$ & $-24000^{+300}_{-300}$ & 30 & 4.6 &   $1.74^{+0.02}_{-0.01}$ \\
2021 & $3.31^{+0.14}_{-0.10}$ & $2.4^{+1.0}_{-0.6}$ & $2200^{+4500}_{-1400}$ & $-37000^{+1500}_{-1200}$ & 32& 4.7&  $1.73^{+0.02}_{-0.02}$\\
F1 & $3.32^{+0.28}_{-0.12}$ & $2.3^{+2.2}_{-0.8}$ &  $700^{+1400}_{-600}$ &  $-35800^{+1200}_{-1500}$ & 20 &3.5&$1.44^{+0.02}_{-0.01}$\\
F2 & $3.32^{+0.18}_{-0.09}$ & $2.1^{+1.8}_{-0.6}$ & $300^{+1300}_{-300}$ & $-38000^{+600}_{-600}$ & 17 & 3.1&  $2.22^{+0.02}_{-0.02}$\\
\hline
RE J1034+396 &&&&&&\\
\cline{2-7}
avg & $3.38^{+0.11}_{-0.11}$ &  $2.7^{+2.9}_{-1.0}$ & $<100$ &  $-63500^{+300}_{-300}$ & 75 &7.9  &   $3.8^{+0.1}_{-0.1}$\\
F1 &  $3.26^{+0.22}_{-0.18}$ & $1.8^{+2.3}_{-1.3}$ & $200^{+400}_{-100}$ & $-63300^{+300}_{-300}$& 28&4.4&$3.3^{+0.1}_{-0.2}$  \\
F2 & $3.43^{+0.19}_{-0.14}$ & $2.9^{+3.3}_{-2.0}$ &  $<100$ &  $-63700^{+300}_{-300}$ & 34&5.0 & $4.0^{+0.2}_{-0.2}$\\
high & $3.60^{+0.35}_{-0.13}$ & $9.1^{+6.7}_{-5.3}$ & $<100$ & $-48000^{+1000}_{-1400}$ & 30 &4.6&   $4.2^{+0.1}_{-0.1}$ \\

\hline
PG 1244+026 &&&&&&\\
\cline{2-7}
avg & $1.42^{+0.12}_{-0.18}$ &  $0.15^{+0.08}_{-0.05}$ & $100^{+100}_{-50}$ &  $-13500^{+300}_{-300}$ & 114&$>8$ &   $7.0^{+0.1}_{-0.1}$ \\
F1 &  $1.59^{+0.19}_{-0.26}$ & $0.23^{+0.07}_{-0.10}$  & $<100$  & $-13200^{+300}_{-300}$& 32&4.7 & $5.4^{+0.1}_{-0.1}$ \\
F2 & $1.46^{+0.33}_{-0.19}$ & $0.25^{+0.07}_{-0.10}$ &  $100^{+100}_{-50}$ &  $-13700^{+300}_{-300}$ & 53 &6.5& $7.1^{+0.1}_{-0.1}$\\
F3 & $1.3^{+0.4}_{-0.2}$ & $0.12^{+0.10}_{-0.06}$ & $100^{+200}_{-50}$ & $-13600^{+300}_{-300}$ & 22&3.7 &  $8.3^{+0.1}_{-0.1}$  \\
F4 & $1.3^{+0.4}_{-0.5}$ & $0.04^{+0.02}_{-0.02}$ & $300^{+1000}_{-200}$ & $-13500^{+600}_{-300}$ & 18&3.2 &  $10.7^{+0.2}_{-0.2}$\\
\hline
PG 1211+143 &&&&&&\\
\cline{2-7}
\multirow{2}{*}{avg} & $1.79^{+0.09}_{-0.09}$ &  $0.42^{+0.07}_{-0.06}$ & $<100$ &  $-17600^{+300}_{-300}$ & 126&$>8$ &   \multirow{2}{*}{$19.9^{+0.2}_{-0.2}$} \\
 &  $2.25^{+0.18}_{-0.14}$ & $0.15^{+0.18}_{-0.07}$  & $3200^{+2600}_{-900}$  & $-10800^{+900}_{-1200}$ & 48&6.1 &  \\
\cline{2-7}
\multirow{2}{*}{T1$^\star$} &  $2.33^{+0.09}_{-0.09}$ & $2.7^{+1.2}_{-0.7}$  & $100^{+100}_{-50}$  & $-20700^{+300}_{-300}$ & 57 &6.8 & \multirow{2}{*}{$33^{+12}_{-9}$ } \\
 &  $5.0^{+0.2}_{-0.3}$ & $600^{+800}_{-400}$  & $400^{+400}_{-300}$  & $-27300^{+900}_{-300}$ & 24 &3.9 &  \\
\cline{2-7}
T3 & $3.03^{+0.19}_{-0.18}$ & $2.6^{+2.0}_{-0.9}$ & $200^{+500}_{-100}$ & $-12800^{+300}_{-300}$ & 69&7.6 &  $28.3^{+0.6}_{-0.6}$ \\
T4 & $1.63^{+0.32}_{-0.31}$ & $0.6^{+0.5}_{-0.3}$ & $<100$  & $-17800^{+600}_{-600}$ & 22&3.7 &  $27.0^{+0.6}_{-0.6}$\\
T5 & $4.5^{+0.4}_{-0.1}$  & $1300^{+600}_{-300}$ & $<100$ & $-51900^{+600}_{-600}$ & 23 &3.8 &  $57^{+11}_{-11}$\\
T6 & $1.53^{+0.21}_{-0.19}$ & $2.3^{+0.7}_{-0.8}$  & $<100$ & $-17800^{+300}_{-300}$ & 100&$>8$ &  $15.7^{+0.4}_{-0.3}$\\
T7 & $2.22^{+0.19}_{-0.21}$ & $0.7^{+0.5}_{-0.4}$ & $200^{+100}_{-100}$ & $-17600^{+300}_{-300}$ & 73&7.8 &  $18.5^{+0.3}_{-0.3}$\\
T8 & $3.33^{+0.19}_{-0.29}$ & $20^{+30}_{-10}$ &  $<100$  &  $-41200^{+900}_{-900}$ & 17 &3.1 &  $18.6^{+0.8}_{-0.7}$\\
T9 & $4.17^{+0.17}_{-0.15}$ & $406^{+180}_{-200}$ & $<100$ & $-76300^{+300}_{-600}$ & 17 &3.1 &  $35^{+6}_{-2}$\\
T10 & $2.17^{+0.19}_{-0.19}$ & $0.4^{+0.3}_{-0.2}$ & $100^{+100}_{-50}$ & $-10100^{+300}_{-300}$ & 23&3.8 &  $25^{+3}_{-7}$  \\
 \cline{2-7}
\multirow{2}{*}{T11$^\star$} & $4.19^{+0.31}_{-0.09}$  & $110^{+260}_{-60}$ & $100^{+500}_{-100}$  & $-72900^{+600}_{-600}$ & 23&3.8 & \multirow{2}{*}{ $19^{+1}_{-1}$ } \\
 & $2.47^{+0.42}_{-0.36}$  & $0.7^{+0.4}_{-0.2}$ & $8400^{+1400}_{-1500}$  & $-11000^{+1300}_{-1200}$ & 20&3.5 &   \\
  \cline{2-7}
\hline
I ZW 1 &&&&&&\\
\cline{2-7}
\multirow{2}{*}{avg$^\star$} & $4.80^{+0.11}_{-0.12}$ &  $1500^{+100}_{-400}$ & $<100$ &  $-64200^{+300}_{-300}$ & 56&6.7 &   \multirow{2}{*}{$72^{+16}_{-12}$}  \\
 & $2.26^{+0.12}_{-0.17}$ &  $0.14^{+0.12}_{-0.09}$ & $2000^{+700}_{-800}$ &  $-25800^{+600}_{-600}$ & 48&6.1 &    \\
 \cline{2-7}
\multirow{2}{*}{F1$^\star$} & $4.54^{+0.32}_{-0.10}$ & $1500^{+1000}_{-900}$ &  $<100$ &  $-64000^{+300}_{-300}$ & 51 &6.4& \multirow{2}{*}{$71^{+19}_{-21}$} \\
 & $2.06^{+0.20}_{-0.12}$ & $0.15^{+0.13}_{-0.08}$ & $2500^{+700}_{-900}$ & $-26100^{+600}_{-600}$ & 23&3.8 & \\
 \cline{2-7}
\multirow{2}{*}{F2$^\star$} & $4.50^{+0.10}_{-0.08}$ & $1500^{+500}_{-600}$  &  $<100$ &  $-65400^{+300}_{-300}$ & 19&3.3 & \multirow{2}{*}{$87^{+23}_{-22}$} \\
 & $2.69^{+0.19}_{-0.27}$ & $0.36^{+0.23}_{-0.21}$ & $1500^{+600}_{-600}$  & $-25800^{+600}_{-600}$ & 29 &4.5 &   \\
 \cline{2-7}
2020 & $3.76^{+0.21}_{-0.11}$ & $7^{+4}_{-4}$ & $600^{+900}_{-500}$ & $-41300^{+600}_{-600}$ & 18 &3.2 &  $12.6^{+0.3}_{-0.3}$ \\
\hline
IRAS 17020+4544 &&&&&&\\
\cline{2-7}
avg & $1.9^{+0.1}_{-0.2}$ &  $0.30^{+0.07}_{-0.06}$ & $2500^{+500}_{-500}$ &  $-23200^{+600}_{-600}$ & 49 &6.2 &   $22.6^{+0.5}_{-0.5}$ \\
F1 &  $1.5^{+0.4}_{-0.2}$ & $0.19^{+0.08}_{-0.10}$  & $3000^{+2000}_{-600}$  & $-23100^{+1500}_{-1500}$  & 17 &3.1& $20.5^{+0.2}_{-0.8}$ \\
F2 & $1.8^{+0.2}_{-0.2}$ & $0.26^{+0.07}_{-0.05}$ & $1700^{+900}_{-900}$ &  $-23100^{+600}_{-600}$ & 19 &3.3 & $22.6^{+0.6}_{-0.6}$ \\
F3 & $1.9^{+0.2}_{-0.2}$ & $0.33^{+0.12}_{-0.09}$ & $2000^{+800}_{-800}$ & $-23200^{+600}_{-600}$ & 22 &3.7 &  $28.7^{+1.0}_{-0.8}$  \\
\hline
\hline
\end{tabular}
\label{tab:xabsfits}
\begin{flushleft}
{$^{\star}$ A secondary UFO is detected in this spectrum.}
\end{flushleft}
\end{table*}

\begin{figure}
\centering
    \includegraphics[width=\columnwidth, trim={20 0 20 10}]{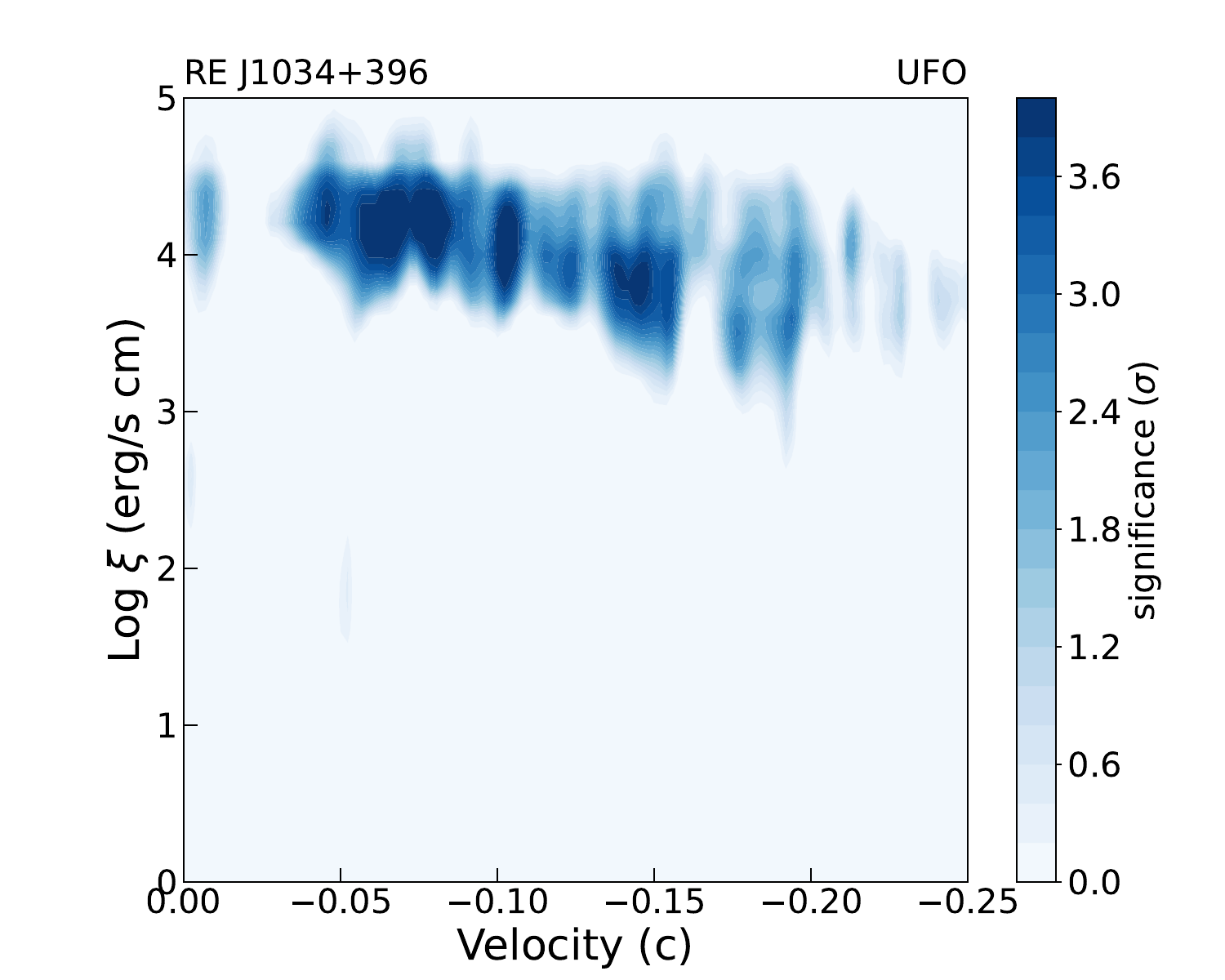}
    \includegraphics[width=\columnwidth, trim={20 0 20 10}]{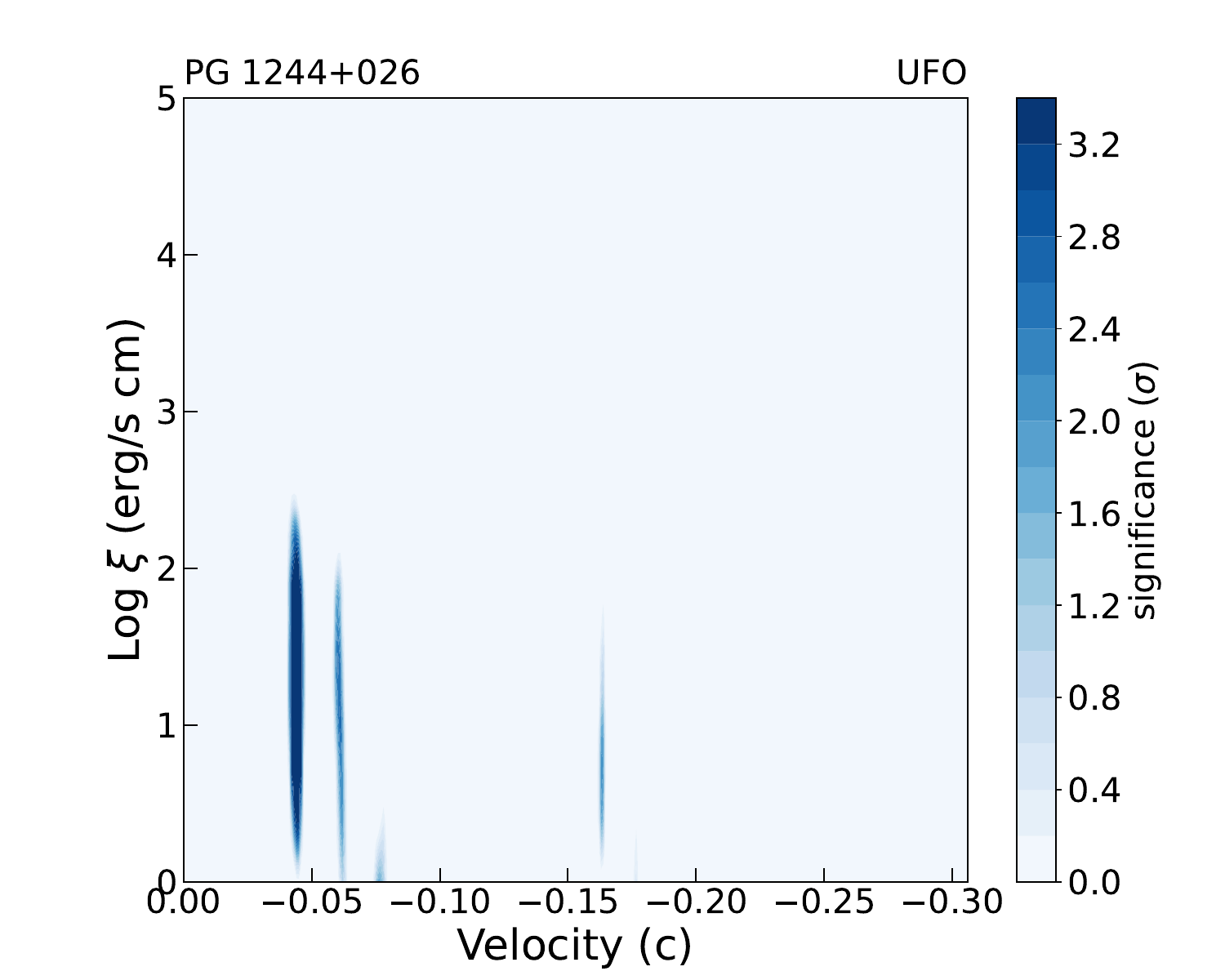}
    \includegraphics[width=\columnwidth, trim={20 0 20 10}]{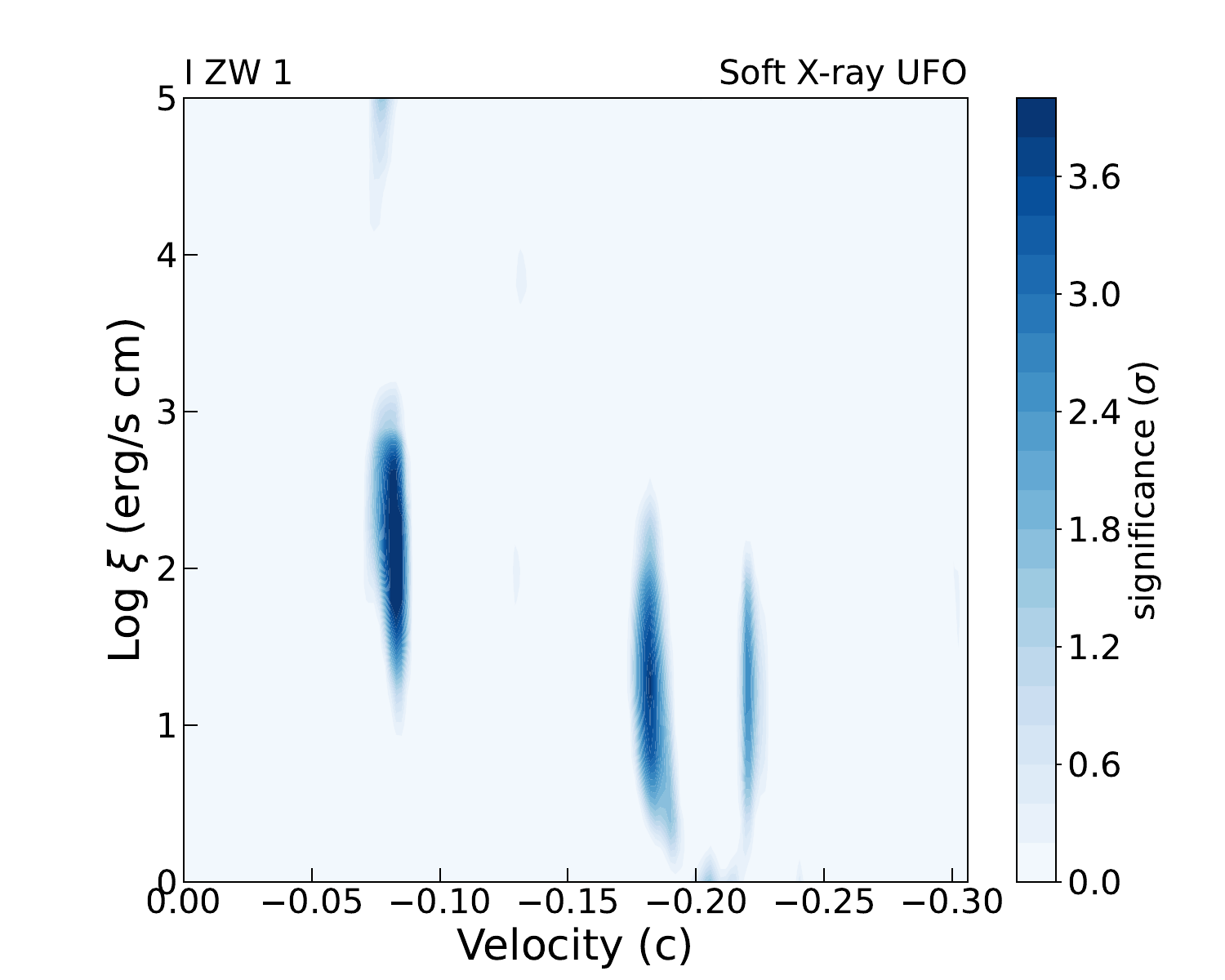}
    \caption{The Monte Carlo (MC) simulations of photoionization absorption model search for the time-averaged spectra of RE J1034+396 (\textit{top}), PG 1244+026 (\textit{middle}), and I ZW 1 (\textit{bottom}) over the baseline model (continuum or plus several absorbers). The color illustrates the detection significance of the newly discovered UFOs considering the look-elsewhere effect. The upper limit of the significance is $3.89\sigma$ as the number of simulated spectra is 10000.}
    \label{fig:absorption_scan_MC}
\end{figure}

Notably, UFOs in RE J1034+396 and PG 1244+026, and the second UFO in I ZW 1 (low ionization), referred as to `I ZW 1-2', are reported for the first time in this work. Two WAs in RE J1034+396 are also for the first time detected. Importantly, these detections are not merely a result of including new and previously unanalyzed observations of these sources but are primarily attributed to the effectiveness of our methodology. To estimate the detection significance of newly discovered UFOs, we performed the Monte Carlo (MC) simulations. For each case, we simulated 10,000 spectra based on the best-fit model before including the newly detected UFO component (i.e. continuum+3WAs for RE J1034+396 and I ZW 1, and continuum model for PG 1244+026). The photoionization scan described in Sec. \ref{subsec:xabs} was performed on each simulated spectra. For each searched grid $i$, we estimate the significance by comparing the $\Delta$C-$\mathrm{stat}_{\mathrm{real},i}$ of an additional UFO absorption for the real spectrum with the $\Delta$C-$\mathrm{stat}_{\mathrm{MC}}$ of an additional UFO absorption within \textit{the whole searched parameter space} for simulated spectra. This step takes into account the \textit{look-elsewhere effect}. The ratio between the number of simulations yielding occurrences of $\Delta$C-$\mathrm{stat}_\mathrm{MC}>\Delta$C-$\mathrm{stat}_{\mathrm{real},i}$ and the total number of MC simulations is the $p$-value of the grid $i$.
The collected significance maps are shown in Fig.\ref{fig:absorption_scan_MC}. The significance of UFO in PG 1244+026 is $\sim3.3\sigma$ (smaller than the single trial significance in Tab.\ref{tab:xabsfits}) and those in RE J1034+396 and I ZW 1 are $\geq3.89\sigma$ (limited by the number of simulations), confirming the presence of newly discovered UFOs.

\subsection{Inclination-dependent UFO detection?}\label{subsec:incl}
\begin{figure}[htbp]
    \centering
	\includegraphics[width=\columnwidth]{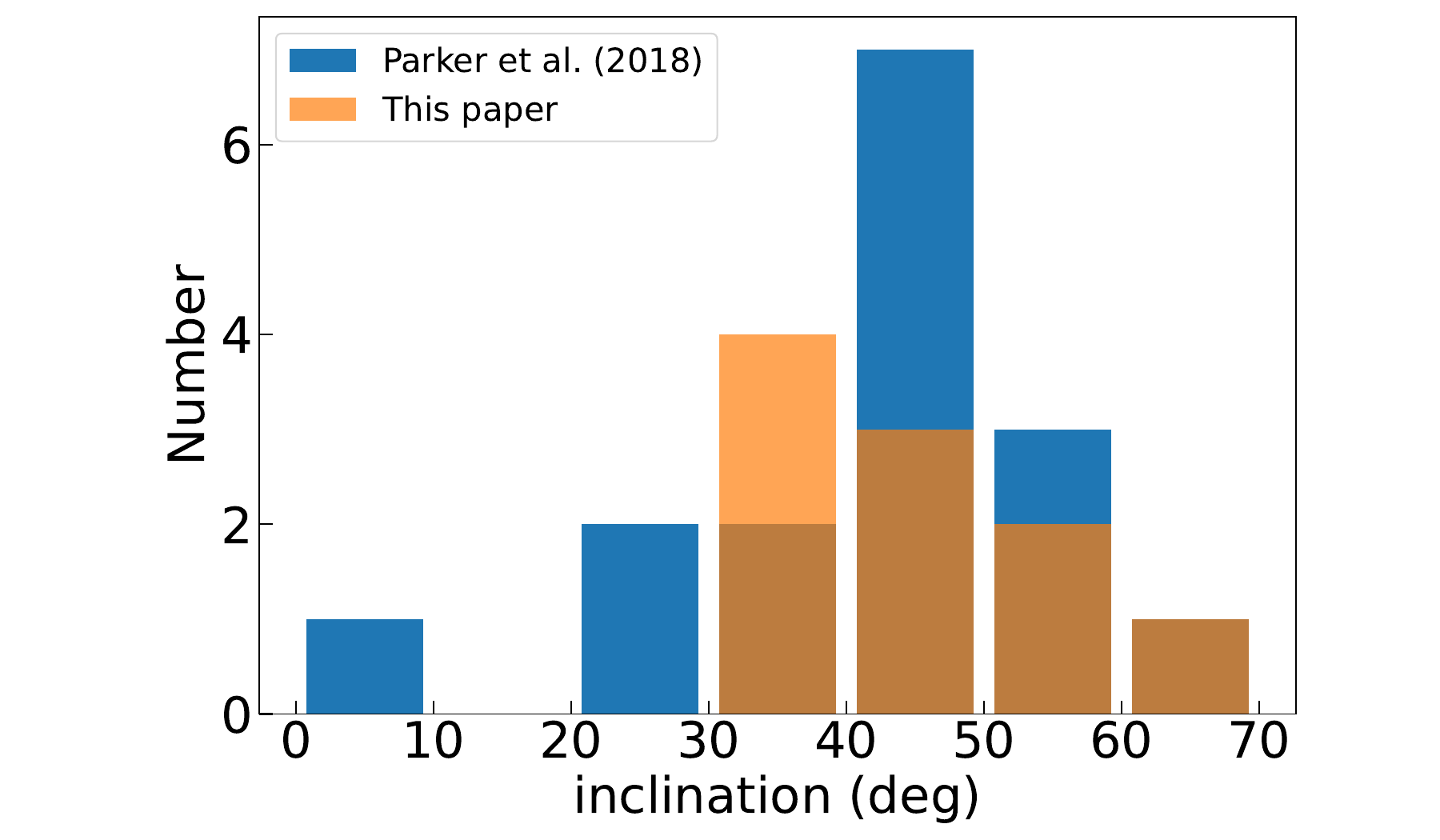}
    \caption{The distribution of disk inclination angle of sources in our sample plus four studied sources compared with the sample of \citet{2018Parker}. Most UFOs in the sample are detected with a disk inclination angle between $30\mbox{--}60^\circ$.
    }
    \label{fig:incl_histogram}
\end{figure}

By employing the relativistic reflection model for modeling the Fe K emission, we can constrain the inclination angle of the accretion disk. It is observed that most UFOs in our sample are detected in AGN with $i>30^\circ$. Combined with the sample of \citet{2018Parker}, most of which are Type 1 AGN, we expand the population size to 26 AGN, including UFO detection and inclination angles derived from relativistic reflection spectroscopy, shown in Fig.\ref{fig:incl_histogram}. UFO detections are strongly concentrated at $30\mbox{--}60^\circ$, suggesting UFOs are detectable only within specific viewing angles. However, this phenomenon most likely results from selection effects. At low inclinations, our LOS intersects directly the innermost region of the accretion disk, where the plasma may be fully ionized and undetectable, even if a UFO exists \citep{2018Pinto}. In the edge-on scenario, the nucleus is heavily obscured by the dusty torus (typically Type 2 AGN, e.g. \citep{2024Marin}), preventing the detection of soft X-ray UFOs. In cases that are not Compton-thick with a column density of $N_\mathrm{H}<10^{24}\,\mathrm{cm}^{-2}$, UFOs are detectable in hard X-rays, while UFOs in Type 2 AGN, which have a similar detection fraction as that in Type 1 AGN \citep{2010Tombesi}, are not equally included in the sample, leading to the lack of UFO detection at high inclinations. Another caveat is the degeneracy between the inclination angle and other parameters of the reflection model (e.g. Fe abundance $A_\mathrm{Fe}$ and ionization parameter $\log\xi$), which may affect the values of the inclination angle \citep{2014Reynolds}. Consequently, due to the biases mentioned above, the hypothesis of inclination-dependent UFO detection is tentative, awaiting an enhanced sample size of AGN with detections of both UFOs and relativistic reflection, as well as model-independent measurements of inclination angles.

\section{Discussion}\label{sec:discussion}
By performing the state-of-the-art photoionization model scan over archival \textit{XMM-Newton} spectra of 7 AGN, we obtain parameters of ionized outflows and find three previously unreported UFOs in RE J1034+396, PG 1244+026, and I ZW 1-2 and two new WAs in RE J1034+396. In this section, we discuss the multiphased outflows in our sample (Sec. \ref{subsec:multiphase}), potential correlations between UFO parameters and X-ray luminosities as well as their relations with intrinsic AGN properties (Sec. \ref{subsec:evolution}), the energetics of UFOs (Sec. \ref{subsec:energetics}), and the capability of future missions to improve our understanding of UFOs (Sec. \ref{subsec:athena}).

\subsection{Multi-phase outflows}\label{subsec:multiphase}
One of the most relevant features of our analysis is the simultaneous detection of multi-phase outflows. Except for PG 1244+026, all AGN in our sample host multi-phase outflows (UFOs and WAs). Multi-phase UFOs are observed in two of them, i.e. PG 1211+143 and I ZW 1. UFOs in PG 1211+143 have already been found multi-phase in \citet{2016Pounds2UFO,2018Reeves,2023Pounds}, while these papers do not mention some of the UFO components that we present in this paper (e.g. UFOs in T5, T8, and T9) as they did not perform the photoionization model scan like what we did. Thanks to this powerful method, UFOs in I ZW 1 are for the first time discovered to have multiple phases in this work.

\begin{figure}
\centering
    \includegraphics[width=\columnwidth, trim={20 0 20 10}]{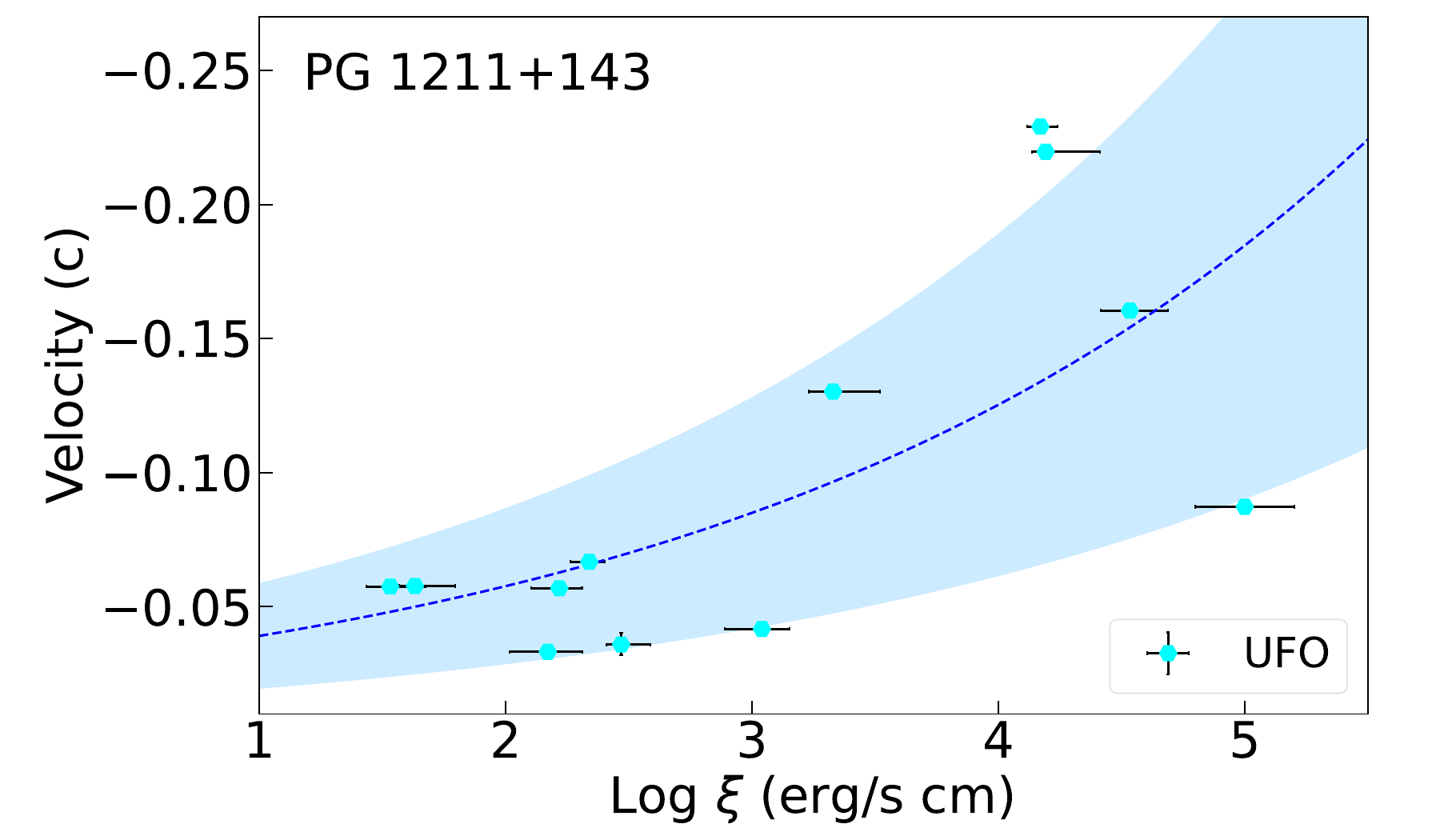}
    \includegraphics[width=\columnwidth, trim={20 0 20 10}]{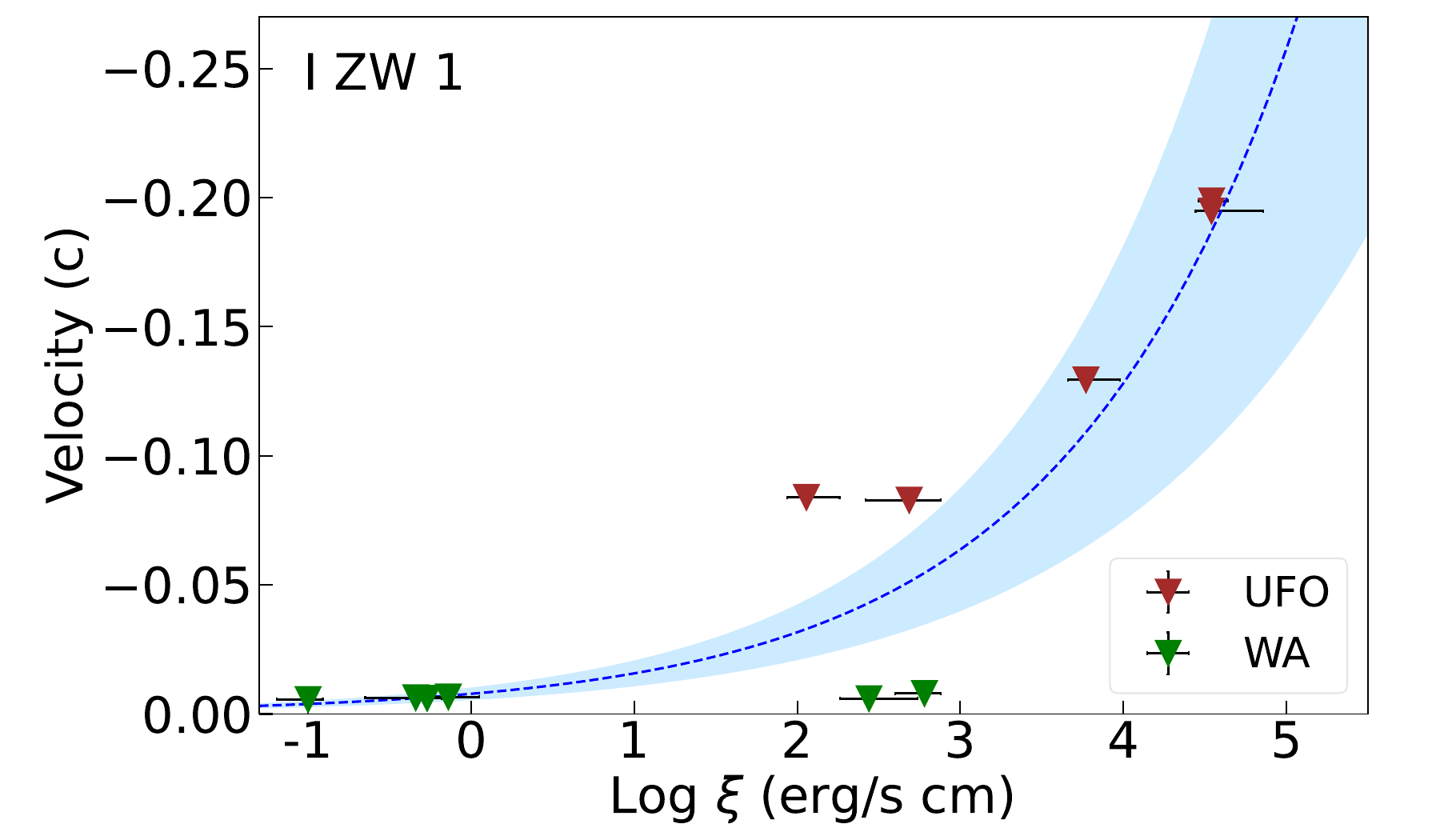}
    \caption{The velocity of ionized winds versus ionization parameters in PG 1211+143 (\textit{top}) and I ZW 1 (\textit{bottom}). The blue line shows the powerlaw regression fitted to data with $1\sigma$ uncertainty shaded.}
    \label{fig:multiphase-logxi-zv}
\end{figure}

To investigate the structure of ionized winds in these two AGN, we plot the velocity of outflows versus ionization parameters in Fig.\ref{fig:multiphase-logxi-zv} and fit them with a powerlaw function. We include detected WAs into the plot of I ZW 1, while the WA in PG 1211+143 is detected in only one observation, so it is discarded. The Spearman's rank correlation coefficient $\rho_\mathrm{sp}$ between $\xi$ and $v$ are $r=-0.64$ in PG 1211+143 and -0.81 in I ZW 1 with $p$-values of 0.02 and $0.003$ respectively, suggesting strong correlations. Here we perform the powerlaw fit for following physical-motivated discussions. The fit for PG 1211+143 provides
\begin{equation}\label{eq:multiphase}
    v/c=(-0.026\pm0.013)\xi^{(0.17\pm0.06)},
\end{equation}
and the fit for I ZW 1 yields
\begin{equation}
    v/c=(-0.008\pm0.002)\xi^{(0.31\pm0.03)}.
\end{equation}
A similar correlation was also observed or hinted at other AGN, e.g. NGC 4051 \citep{2013Pounds}, IRAS 13224-3809 \citep{2018Pinto} and 1H 0707-495 \citep{2021Xu}, compatible with the scenario that the UFO launched from the inner and hotter accretion disk requires a larger velocity to escape the local Keplerian velocity.

According to the definition of the kinetic energy of outflows $L_\mathrm{kin}\propto v_w^3/\xi$ (see details in Eq.\ref{eq:kinetic} in Sec. \ref{subsec:energetics}), $v^3/\xi=\mathrm{constant}$ (i.e. $v\propto \xi^{1/3}$) implies an energy-conserved outflow. Therefore, our result suggests that the kinetic energy of UFOs in I ZW 1 is conserved during the propagation from the inner accretion disk to large scales if they originate from a stratified wind. However, in PG 1211+143 and the power index $(0.17\pm0.06)$ is inconsistent with either energy-conserved ($v\propto \xi^{1/3}$) or momentum-conserved ($v\propto \xi^{1/2}$) outflows, where the momentum of outflow is expressed by $\dot{P}_w\sim \dot{M}_\mathrm{out}v_w\propto v_w^2/\xi$. The possible reason is that outflows in PG 1211+143 do not originate from a stratified wind but from random disk instability, indicated by rapidly variable UFO parameters \citep{2016Pounds}.

In most AGN (except PG 1244+026) in our sample, we detect UFO and WA components with similar best-fit ionization parameters (see Tab.\ref{tab:xabsfits} and Tab.\ref{app:tab:WAfits}). These components may be a signature of entrained UFOs (E-UFOs) as proposed by \citet{2013Pounds} and \citet{2019Serafinelli}. These authors suggested a scenario whereby WAs are produced in the shock where UFOs collide with the surrounding medium. In this case, the interstellar medium (ISM) is entrained by UFOs launched from the inner disk, leading to E-UFOs, which keep the low ionization state of ISM with the ultra-fast velocities from UFOs. Our results support this interpretation for these UFO/WA `pairs’. An intriguing case is RE J1034+396, where a WA is detected with a very high ionization parameter \citep[$\log\xi\sim3.6$,][]{2024Zhou}), in agreement with the UFO detected at all epochs, although it is still compatible with the shocked outflow mode. We observe that the ratio of velocities in these UFO/WA component pairs with a similar ionization parameter is larger than the value predicted by shock theory \citep[$\sim4$][]{2013Pounds}, possibly because the momentum is lost during the shock.

\subsection{The evolution of UFO properties across different X-ray luminosities}\label{subsec:evolution}

To further investigate the dependence of the wind properties on the source luminosity, we plot the best-fitting column density, ionization parameter, and velocity of UFOs at their $1\sigma$ uncertainties versus the unabsorbed luminosity between $0.4\mbox{--}10\,$keV in \ref{app:fig:UFOs-lum} and fit them with a powerlaw function. Only 4 UFOs are fitted after excluding UFOs with distinct parameters from others (to avoid tracking different phases) and with only two measurements (see details in Appendix.\ref{app:sec:correlation}). The fitted power indexes, referred to as $\Gamma_\mathrm{N_H}$, $\Gamma_\mathrm{\log\xi}$ and $\Gamma_\mathrm{velocity}$, are listed in Tab.\ref{tab:fits}, which also include results from previous archival studies, where only $\Gamma_\mathrm{velocity}$ was calculated for PDS 456.

\begin{table}[htbp]
\setlength{\tabcolsep}{3pt}
\centering
\caption{Table of the fitted power indexes ($\mathrm{parameter}=L_\mathrm{X}^{\Gamma}$) of the column density, ionization parameter, and velocity of the photoionized plasma in our sample (plus the published results) with their $1\,\sigma$ uncertainties versus the unabsorbed luminosity between $0.4\mbox{--}10\,$keV. 
}
\begin{tabular}{lccc}
\hline
\hline
Names & $\Gamma_{N_H}$ & $\Gamma_{\mathrm{log}\xi}$ & $\Gamma_\mathrm{velocity}$ \\
\hline
\multicolumn{4}{c}{This work}\\
\hline
RE J1034+396 &  $4.61\pm2.90$  & $1.95\pm1.09$  & $-0.09\pm0.31$ \\
PG 1244+026 & $-2.45\pm0.81$ & $-0.85\pm0.87$ & $0.03\pm0.05$ \\
PG 1211+143 &   $6.87\pm2.37$ & $3.91\pm1.94$ & $0.81\pm0.62$ \\
IRAS 17020+4544 & $1.05\pm1.10$  & $1.41\pm2.20$ & $0.017\pm0.022$ \\
\hline
\multicolumn{4}{c}{Published work}\\
\hline
Mrk 1044\tablefootmark{a} & $1.59\pm0.98$  & $1.08\pm0.41$ & $0.39\pm0.16$ \\
1H 0707-495\tablefootmark{b} & $1.25\pm0.51$  & $1.83\pm0.77$ & $-0.15\pm0.05$ \\
IRAS 13224-3809\tablefootmark{c} & $-0.08\pm0.04$  & $0.28\pm0.11$ & $0.05\pm0.02$ \\
PDS 456\tablefootmark{d}  &  &   & $0.24\pm0.03$ \\
\hline
\end{tabular}
\tablefoot{References: \tablefoottext{a}{\citet{2023Xu}}
\tablefoottext{b}{\citet{2021Xu}}
\tablefoottext{c}{\citet{2018Pinto}}
\tablefoottext{d}{\citet{2017Matzeu}}}
\label{tab:fits}
\vspace{-3mm}
\end{table}

Due to the limited measurements, fitted parameters are loosely constrained (see Tab.\ref{tab:fits} and broad shaded regions in Fig.\ref{app:fig:UFOs-lum}), although the variations could still be seen through individual parameters. In our sample, only 2 UFOs (RE J1034+396 and PG 1211+143) exhibit positive $\Gamma_\mathrm{\log\xi}$ within uncertainties, while the remaining UFOs (i.e. 1H 1934-063, PG 1244+026, I ZW 1, and IRAS 17020+4544) are consistent with zero, except for I ZW 1-2, which may have a higher ionization parameter at brighter states. $\Gamma_\mathrm{N_H}$ usually follows the same trend as $\Gamma_\mathrm{\log\xi}$ due to the systematic degeneracy that only outflows with a higher column density can be observable at a higher ionization state. The negative $\Gamma_\mathrm{N_H}$ of PG 1244+026 is explained in Sec. \ref{app:subsec:pg1244+026}. According to the definition of the ionization parameter: $\xi\equiv L_\mathrm{ion}/n_\mathrm{H}R^2$, where $L_\mathrm{ion}$ is the ionizing luminosity calculated between $1\mbox{--}1000\,$Ryd (13.6\,eV$\mbox{--}$13.6\,keV), $n_\mathrm{H}$ is the hydrogen number density, and $R$ is the distance between the materials and the ionizing source, positive $\Gamma_\mathrm{\log\xi}$ suggest a UFO responding to the ionizing luminosity. As a result, we prefer to claim that 2 UFOs in our sample seem to respond to the radiation field (more-ionized-when-brighter) while for the other 5, we do not yet have found evidence, which may be intrinsic or caused by limited measurements.

As for velocity, only 1 UFO (PG 1211+143) exhibits a positive $\Gamma_\mathrm{velocity}$ with a large uncertainty due to a multi-phase nature, and UFOs with only two measurements (1H 1934-063 and I ZW 1) hint positive $\Gamma_\mathrm{velocity}$ but limited by measurements, while the others are consistent with zero. The positive $\Gamma_\mathrm{velocity}$ points towards a radiatively-driven UFO \citep{2003King}, while a zero value may result from the stacked flux-resolved spectra smearing out the lines and trends or an intrinsically stable nature of UFOs.

\begin{figure}[htbp]
    \centering
	\includegraphics[width=\columnwidth]{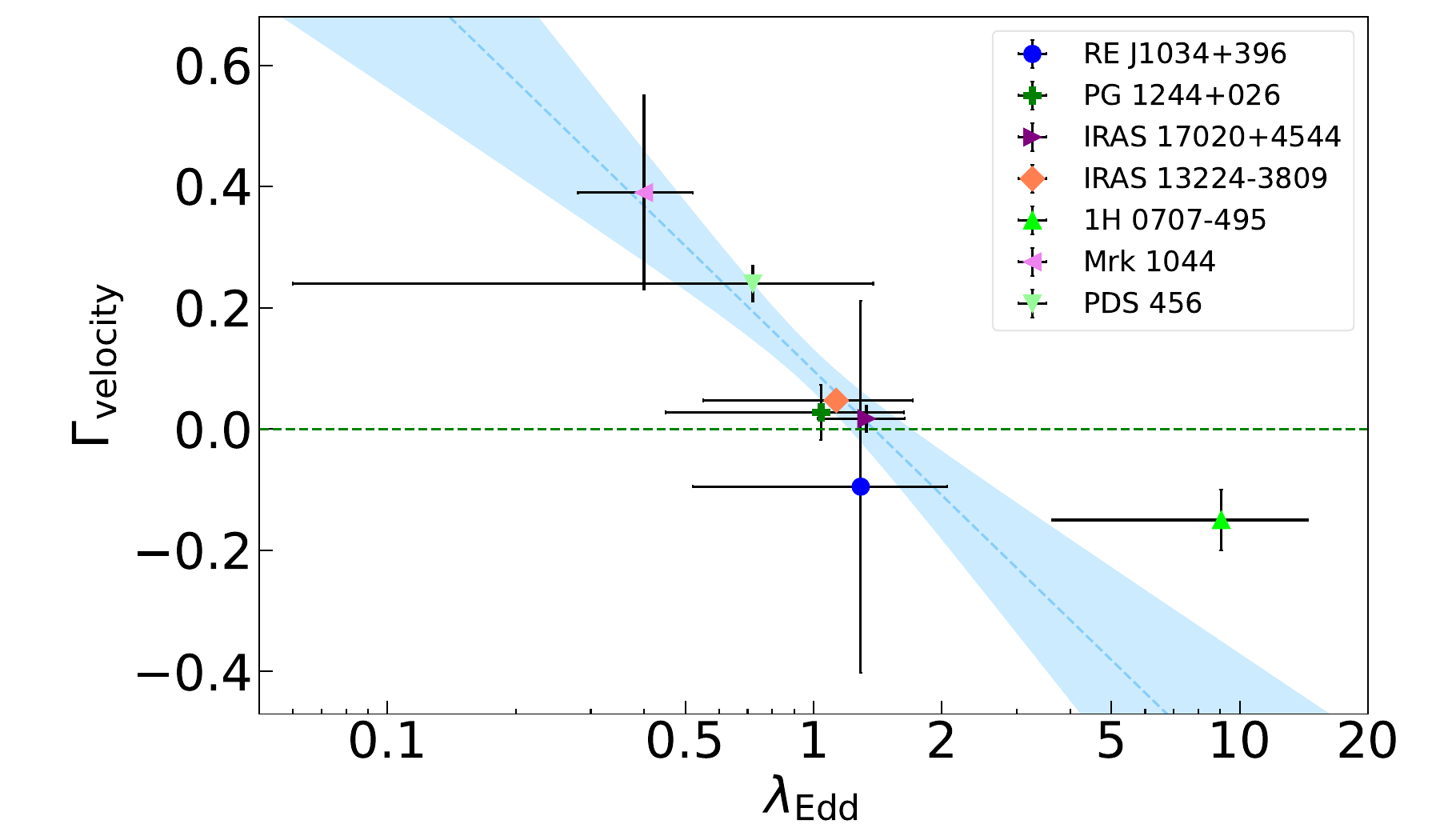}
    \caption{The power indexes of the UFO velocity $\Gamma_\mathrm{velocity}$ as a function of X-ray luminosity versus Eddington ratio $\lambda_\mathrm{Edd}$. The data come from the results listed in Tab.\ref{tab:sources} and \ref{tab:fits}. The horizontal dashed line denotes $\Gamma_\mathrm{velocity}=0$. The blue line shows the linear regression fitted to the data in the linear-log scale with $1\sigma$ uncertainty shaded. PG 1211+143 is excluded from this plot due to multi-phase UFOs.
    }
    \label{fig:velocity_mdot}
\end{figure}

We further tried to explore the potential correlations between the dependence of UFO properties on luminosity and the intrinsic AGN properties (Tab.\ref{tab:sources}), combined with published works, for any indications. The properties consists of the black hole mass $M_\mathrm{BH}$, bolometric luminosity $L_\mathrm{bol}$, Eddington ratio $\lambda_\mathrm{Edd}$, and inclination angle $i$. The plots for each combination are depicted in Fig.\ref{app:fig:slope} combined with AGN in published works. However, we did not find any statistically significant correlations among all pairs of quantities except for a potential correlation between $\lambda_\mathrm{Edd}$ and $\Gamma_\mathrm{velocity}$ (see the bottom plot of panel $c$ of Fig.\ref{app:fig:slope}). After removing PG 1211+143, which is obtained from multi-phase UFOs, we find a tentative anti-correlation with a Pearson correlation coefficient of $r=-0.72$ and a $p$-value of $0.07$, corresponding to $\sim1.8\sigma$ significance. That pair is depicted and fitted in Fig.\ref{fig:velocity_mdot} with the $1\sigma$ uncertainty, yielding 
\begin{equation}
\Gamma_\mathrm{velocity}=(-0.69\pm0.21)\log\lambda_\mathrm{Edd}+(0.10\pm0.04).   
\end{equation}

\begin{figure}
    \centering
    \includegraphics[width=\columnwidth, trim={0 10 0 0}]{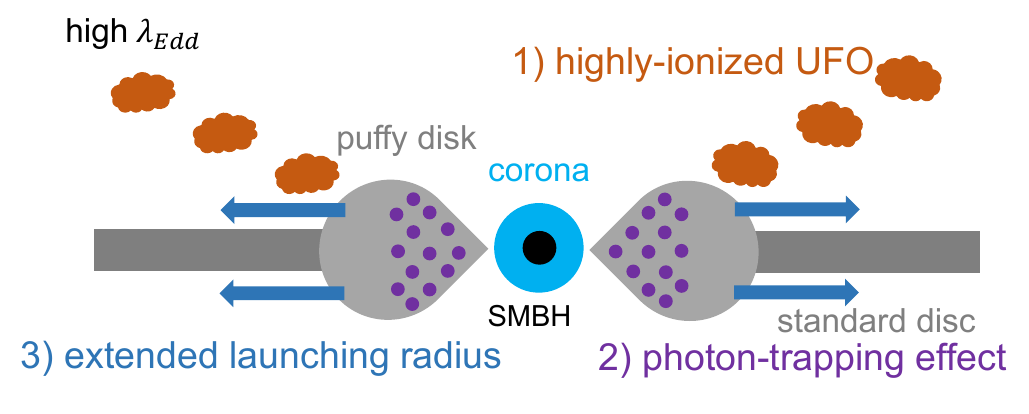}
    \caption{The simplified schemes of AGN with high accretion rates. The SMBH is surrounded by a standard thin disk and a hot corona. UFOs are launched by the radiation field from the accretion disk. The inner part of the accretion disk is slim in the high-accretion regime. Three competitive scenarios are explored to explain a possible $\Gamma_\mathrm{velocity}$ vs $\lambda_\mathrm{Edd}$ anti-correlation (see Sec.\ref{subsec:evolution}).}
    \label{fig:schematic}
\end{figure}

If the tentative ($<2\sigma$) anticorrelation is real, it indicates that in high-/super-Eddington regimes, the radiation acceleration on winds may decrease with increasing Eddington ratios, since the dependence of the velocity on the source luminosity, $\Gamma_\mathrm{velocity}$, represents a form of `radiation acceleration'. In the standard thin disk accretion model \citep{1973Shakura}, an anticipated outcome of the force multiplier is the increasing acceleration of UFOs as the Eddington ratio rises. We consider three possibilities for high-Eddington systems with a slim inner disk \citep[e.g.,][]{1988Abramowicz,2007Ohsuga,2019Jiang,2023Curd} (summarized in Fig.\ref{fig:schematic}) to explain the tentative anticorrelation: 1) In high-Eddington AGN, the spherization radius $R_\mathrm{sp}$  \citep{1973Shakura,2007Poutanen}, where the height and radius of the disk are comparable $H\sim R$ and winds are assumed to be launched from this point, is proportional to the Eddington ratio, $R_\mathrm{sp}\propto R_\mathrm{in}\lambda_\mathrm{Edd}$. The wind launching radius will thus increase as the Eddington ratio rises, coupled with reduced Keplerian velocity, thus leading to a slower launching velocity. The same phenomenon was observed in the population study of UFOs \citep[SUBWAYS,][]{2024Gianolli}. 2) In the slim disk, frequent interactions between matter and photons would delay the liberation of the radiation energy or even trap photons \citep{2002Ohsuga}, resulting in reduced radiation acceleration. 3) The intense radiation field can result in an over-ionization of plasma in the inner part of the accretion disk, leading to a decrease in the cross-section $\sigma$ \citep{1979Rybicki,2007Kallman}, $\sigma\propto\log E/E$, where $E$ is the energy available for the ionization of plasma. The launched winds thus become transparent to photons and experience low radiatively-driven force, dampening the radiation acceleration. 

Nevertheless, we have to caution that the observed trend is not statistically strong ($\sim1.8\sigma$) and just based on a sample of only 7 cases, making it just an indication at the moment. The confirmation of this trend and the discrimination between three possible mechanisms of deceleration require more observations and targets, especially AGN across various accretion rates ($\lambda_{Edd}<0.1$ or $\lambda_{Edd}>1$). Furthermore, theoretical simulations about how the properties of UFOs launched by different mechanisms evolve with the accretion rates should be done in the future for comparison with observational results.

\subsection{Implications for AGN feedback} \label{subsec:energetics}

It is expected that UFOs at small scales can carry out sufficient power to quench or trigger star formation in their hosts and affect the evolution of galaxies \citep[e.g.][]{2005DiMatteo,2010Hopkins,2017Maiolino}. According to simulations, the deposition of kinetic energy larger than 0.5\% of the Eddington luminosity into the ISM is sufficient to produce considerable feedback on the host galaxy. The kinetic power of the outflows can be expressed as: 
\begin{equation}\label{eq:kinetic}
    L_\mathrm{kin}=\frac{1}{2}\dot{M}_\mathrm{out}v_\mathrm{w}^2=2\pi C_\mathrm{V}\Omega R^2\rho v_\mathrm{w}^3 
\end{equation}
where $\dot{M}_\mathrm{out}=4\pi C_\mathrm{V}\Omega R^2\rho v_\mathrm{w}$ is the mass outflow rate; $C_\mathrm{V}$ the volume filling factor; $\Omega$ the solid angle normalized by $4\pi$; $R$ the distance between the ionizing source and outflows; $\rho$ the outflow mass density; $v_\mathrm{w}$ is the wind velocity. The mass density is defined as $\rho=n_\mathrm{H}m_\mathrm{p}\mu$, where $n_\mathrm{H}$ is the hydrogen number density; $m_\mathrm{p}$ the proton mass and $\mu=1.2$ the mean atomic mass assuming solar abundances. The solid angle is assumed at $\Omega/4\pi=0.3$ in this paper, which is determined by the observational UFO detection rate \citep{2010Tombesi} and is consistent with the GR-MHD simulations of radiatively-driven winds in super-Eddington systems \citep{2013Takeuchi}.

We estimate the wind location following the method adopted in \citet{2013Tombesi} and \citet{2015Gofford}. The lower limit on the wind location is constrained by assuming that the UFO velocity is larger than or equal to the escape velocity, $R_\mathrm{w}\geq R_\mathrm{esc}=2GM_\mathrm{BH}/v_\mathrm{w}^2$, while the upper limit is obtained by hypothesizing that the thickness of UFOs is lower than or equal to its maximal distance from the source, $\Delta R\leq R_\mathrm{w}$, combined with the definition of ionization parameter $\xi\equiv L_\mathrm{ion}/n_\mathrm{H}R^2$ and $N_\mathrm{H}=\int^{\infty}_{R_\mathrm{max}}{n(r)dr}$, leading to $R_\mathrm{w}\leq L_\mathrm{ion}/\xi N_\mathrm{H}$. The estimated distance and the corresponding number density are listed in Tab.\ref{tab:kinetic}, according to the definition of the ionization parameter ($\xi\equiv L_\mathrm{ion}/n_\mathrm{H}R^2$). The UFO properties are taken from the results of the stacked spectra, so the secondary UFO of PG 1211+143 in the time-averaged spectrum is also included and denoted as `PG 1211+143-2'.

\begin{table*}[htbp]
\setlength{\tabcolsep}{2pt}
\centering
\caption{Table of the inferred wind parameters in our sample (plus the published results). See details in the content.
}
\begin{adjustbox}{margin=-0.5cm 0cm 0cm 0cm} 
\begin{tabular}{lcccccc}
\hline
\hline
Names & $\log R_w$ & $\log n_\mathrm{H}$ &$\log \dot{M}_\mathrm{out}$ &  $\log L_\mathrm{kin}$ & $[C_\mathrm{V}]^{a}$ & $L_\mathrm{kin}/L_\mathrm{Edd}^{b}$  \\
      & (cm) & ($\mathrm{cm}^{-3}$)  &  (g/s) & (erg/s) & ($10^{-3}$) & \%  \\
\hline
\multicolumn{7}{c}{This work}\\
\hline
1H 1934-063 & $13.8\mbox{--}18.8(17.1)$ & $2.6(6^\star)\mbox{--}12.6$ & $21.6(23.3)\mbox{--}26.6$ &$40.4(42.1)\mbox{--}45.4$ & $9.0\pm2.4$ & $6.8\times10^{-3}(0.4)\mbox{--}7.6\times10^{2}$ [$3.5\pm2.0$]  \\
RE J1034+396 & $13.3\mbox{--}19.9(17.6)$ & $1.6(6^\star)\mbox{--}14.7$ & $21.4(23.6)\mbox{--}27.9$ & $40.6(42.8)\mbox{--}47.2$  & $7.1\pm2.6$ & $1.2\times10^{-2}(2.1)\mbox{--}4.7\times10^{4}$ [$26.0\pm18.6$]  \\
PG 1244+026 &  $15.3\mbox{--}22.8(18.5)$  & $-2.6(6^\star)\mbox{--}12.4$   &  $21.5(25.8)\mbox{--}29.0$  & $39.4(43.7)\mbox{--}47.0$ & $0.7\pm0.2$ & $1.6\times10^{-4}(3.4)\mbox{--}5.6\times10^{3}$ [$1.0\pm0.5$] \\
PG 1211+143 & $16.1\mbox{--}22.2(18.4)$  & $-1.6(6^\star)\mbox{--}10.6$  & $22.9(26.7)\mbox{--}29.0$   & $41.0(44.8)\mbox{--}47.1$  & $1.2\pm0.4$  & $5.8\times10^{-4}(3.6)\mbox{--}7.3\times10^{2}$ [$0.4\pm0.2$] \\
PG 1211+143-2 & $16.5\mbox{--}22.2(18.2)$  & $-2.1(6^\star)\mbox{--}9.3$  & $22.6(26.6)\mbox{--}28.3$ & $40.4(44.4)\mbox{--}46.1$  & $5.6\pm2.1$ & $1.3\times10^{-4}(1.3)\mbox{--}6.2\times10^{1}$ [$0.15\pm0.09$] \\
I ZW 1     & $14.3\mbox{--}16.2(17.2)$  & $7.9(6^\star)\mbox{--}11.8$  &  $25.2(24.2)\mbox{--}27.1$  & $44.4(43.4)\mbox{--}46.3$  & $71.2\mbox{--}26.6$ & $7.0\times10^{0}(0.8)\mbox{--}5.7\times10^{2}$ [$17.8\pm9.0$] \\
I ZW 1-2   & $15.1\mbox{--}22.8(18.5)$ & $-2.7(6^\star)\mbox{--}12.8$ & $21.5(25.8)\mbox{--}29.2$  &  $40.0(44.3)\mbox{--}47.7$ & $0.5\pm0.2$ & $2.8\times10^{-4}(6.0)\mbox{--}1.5\times10^{4}$ [$3.2\pm2.0$] \\
IRAS 17020+4544 &  $14.4\mbox{--}22.8(18.6)$  &  $-2.4(6^\star)\mbox{--}14.4$  & $21.1(25.3)\mbox{--}29.5$  &  $39.6(43.7)\mbox{--}47.9$ & $0.2\pm0.1$  & $5.4\times10^{-4}(8.2)\mbox{--}1.3\times10^{5}$ [$3.6\pm1.7$] \\
\hline
\multicolumn{7}{c}{Published work}\\
\hline
Mrk 1044\tablefootmark{c} &$13.6\mbox{--}15.4$ & $9.0\mbox{--}12.6$  & 24.2  &  43.2 & 7.0  & 4.4 \\
1H 0707-495\tablefootmark{d} & $13.4\mbox{--}14.8$ & $10.7\mbox{--}13.0$ & 24.5  & 43.4 & 8.0 & 13.7 \\
IRAS 13224-3809\tablefootmark{e,f} &  $13.6\mbox{--}14.2$ &  $10\mbox{--}12$ & 24.4 & 43.4  & & 3.8  \\
PDS 456\tablefootmark{g} & $16.2\mbox{--}16.8$ &  $6.5\mbox{--}8.9$ & 27.1 & 46.3 &  & 15.0 \\
\hline
\end{tabular}
\end{adjustbox}
\tablefoot{
The inferred parameters are calculated by assuming a solid angle of $\Omega/4\pi=0.3$ \citep{2010Tombesi,2013Takeuchi}.
\tablefoottext{$\star$}{The lower limit on the density is assumed above the typical density of narrow line region (NLR, $>10^6\,\mathrm{cm}^{-3}$). The corresponding results are shown in parentheses.}
\tablefoottext{a}{The filling factor is calculated by assuming that the outflow rate is comparable to the accretion rate $\dot{M}_\mathrm{out}\sim \dot{M}_\mathrm{acc}$ \citep{2018Kobayashi}.}
\tablefoottext{b}{The value in the bracket is estimated by inputting $C_\mathrm{V}$ into Eq.\ref{eq:kinetic} based on assumption of $\dot{M}_\mathrm{out}\sim \dot{M}_\mathrm{acc}$.}
References: 
\tablefoottext{c}{\citet{2023Xu}}
\tablefoottext{d}{\citet{2021Xu}}
\tablefoottext{e}{\citet{2017Parker}}
\tablefoottext{f}{\citet{2018Pinto}}
\tablefoottext{g}{\citet{2015Nardini}}}
\label{tab:kinetic}
\vspace{-3mm}
\end{table*}

\begin{figure}
    \centering
    \includegraphics[width=\columnwidth, trim={0 0 0 0}]{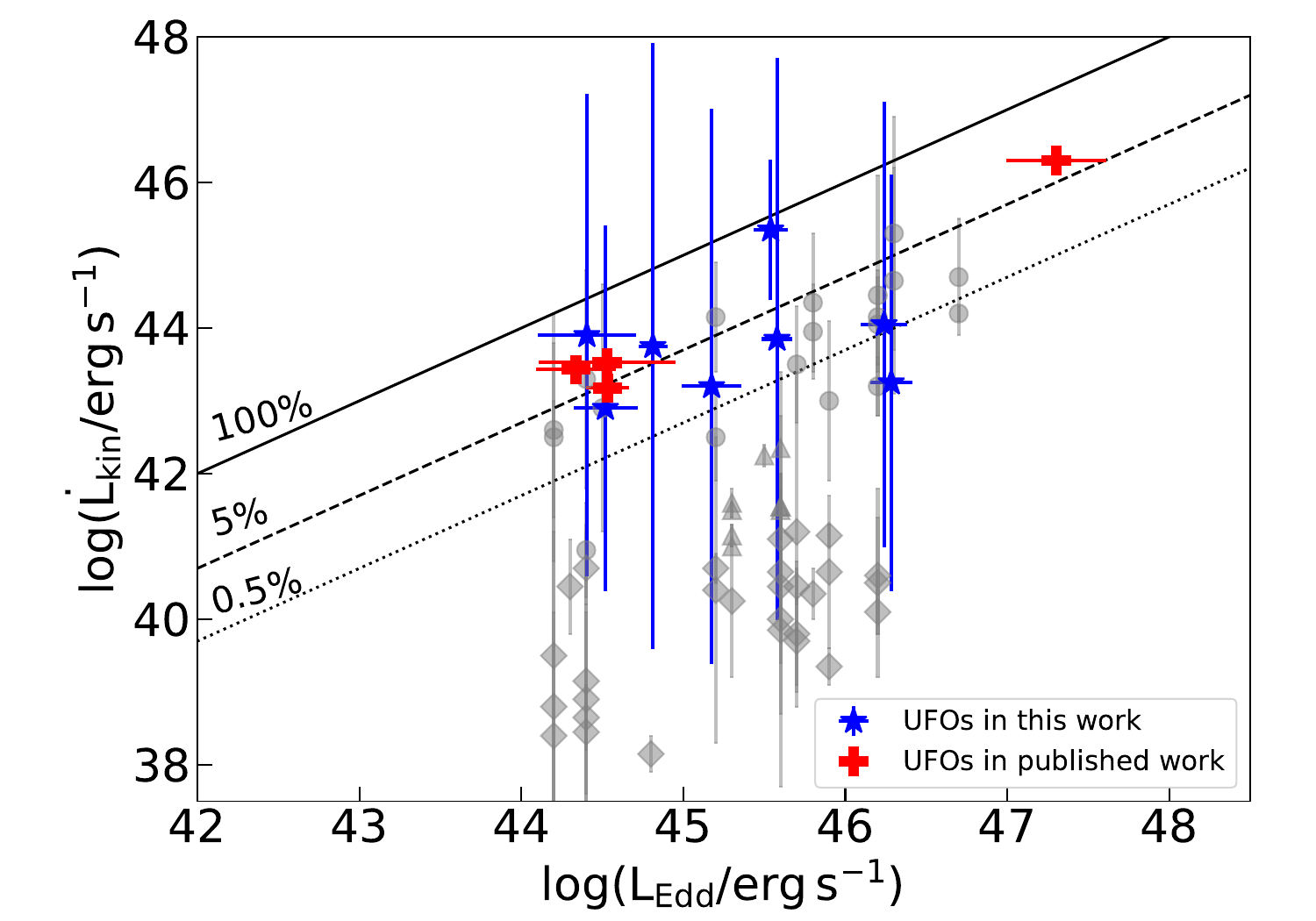}
    \caption{The kinetic luminosity of UFOs in this (\textit{blue stars}, based on the most conservative assumptions) and published (\textit{red crosses}) works versus the Eddington luminosity. The \textit{grey circles}, \textit{triangles}, and \textit{diamonds} correspond to UFOs, non-UFOs, and WAs in \citet{2013Tombesi}, respectively, for comparison. The transverse lines indicate the ratios between the outflow kinetic luminosity and Eddington luminosity of 100 percent (\textit{solid}), 5 percent (\textit{dashed}), and 0.5 percent (\textit{dotted}).}
    \label{fig:kinetic}
\end{figure}

The value of $C_\mathrm{V}$ is difficult to determine because it depends on the ionization and clumpiness of the gas. At low ionization states, the flow is likely to be clumpy, where the column density of the wind can be given by $N_\mathrm{H}\sim C_\mathrm{V}n(R)\Delta R$ with $C_\mathrm{V}<1$ allowing for an inhomogeneous gas, while at high ionization states it can be considered largely smooth with $C_\mathrm{V}\sim1$ and $R\rightarrow R_\mathrm{max}$. Therefore, following \citet{2015Gofford}, we substitute for $R_\mathrm{min}$ and $R_\mathrm{max}$, leading to the upper and lower limits on the mass outflow rate:
\begin{gather}
    \dot{M}_\mathrm{out}^\mathrm{max} \sim 1.2\Omega m_\mathrm{p}L_\mathrm{ion}/\xi v_w \label{eq:mass-rate-max}\\
   \dot{M}_\mathrm{out}^\mathrm{min} \sim 2.4\Omega m_\mathrm{p}GM_\mathrm{BH}N_\mathrm{H} v_w \label{eq:mass-rate-min}
\end{gather}
where in Eq.\ref{eq:mass-rate-max} $n_\mathrm{H}R^2$ is replaced by $L_\mathrm{ion}/\xi$ due to the definition. The estimated mass outflow rate and kinetic power of UFOs in our sample (plus the published results) are listed in Tab.\ref{tab:kinetic}. The ratio between estimated kinetic power and Eddington luminosity is listed as well. For comparison, we plot the kinetic energy versus the Eddington luminosity of UFOs in this (\textit{blue}) and published (\textit{red}) works in Fig.\ref{fig:kinetic}, compared with the sample in \citet{2013Tombesi} (marked by \textit{grey} symbols). We note that the kinetic power of soft X-ray UFOs in our sample has a much larger uncertainty than those in \citet{2013Tombesi}, except for the primary UFO in I ZW 1 (detected in the Fe K region). It is because UFOs detected in soft X-rays usually have smaller velocities and column densities compared with those of hard X-ray UFOs \citep[e.g.][]{2019Serafinelli,2021Laha}. Given such large uncertainties of individual soft X-ray UFOs, we cannot conclude that UFOs in our sample have sufficient kinetic energy to affect the host galaxy, although it is still promising, requiring better constraints on wind parameters.

We note that most lower limits on the density of UFOs in Tab.\ref{tab:kinetic} are unreasonably small, resulting from the relatively low column densities. Theoretically, UFOs are expected to originate from the nuclear region of AGN, i.e. within the narrow line region (NLR), which has typical densities of $10^{3}\mbox{--}10^{6}\,\mathrm{cm}^{-3}$ \citep[e.g.][]{1990Netzer}. Therefore, we take $n_\mathrm{H}=10^{6}\,\mathrm{cm}^{-3}$ as the lower limit and infer the corresponding wind parameters, shown in the parentheses in Tab.\ref{tab:kinetic}. Under this assumption, we can further constrain the lower bounds on the kinetic energy of UFOs, most of which in this work surpasses the theoretical threshold ($>0.5\%L_\mathrm{Edd}$), indicating UFOs have significant impacts on the evolution of host galaxies.

Alternatively, we try to further constrain the wind parameters by assuming that the outflow mass rate is comparable to the accretion rate $\dot{M}_\mathrm{out}\sim\dot{M}_\mathrm{acc}$, which is reasonable in high-accretion systems, the filling factor can be estimated through Eq. 23 in \citep{2018Kobayashi}. The estimated filling factors are between $10^{-4}\mbox{--}10^{-2}$ listed in Tab.\ref{tab:kinetic}, averagely larger than those estimated for WAs, $10^{-2}\mbox{--}10^{-6}$ \citep{2005Blustin}, as UFOs are expected to be closer to SMBHs and less porous than WAs. The corresponding kinetic energy of UFOs is listed in the bracket of the last column. Most of them are above the theoretical criterion except for UFOs in PG 1211+143, consistent with what was found in \citet{2018Danehkar}, although the total kinetic power of two UFOs reaches the threshold.

Based on the above investigations from two different but both reasonable assumptions, our results suggest that UFOs detected in \textit{soft X-rays} also have sufficient power to affect the surrounding medium and host galaxy, complementing the conclusion made from only hard X-ray UFOs \citep{2013Tombesi}. Within our sample, the primary UFO in I ZW 1 is the only case with conclusively sufficient power to affect the host galaxy, which is consistent with the estimation from \citet{2019Reeves} ($15\mbox{--}25\%$). Additionally, the location of this UFO is tightly constrained, $\log R_w\sim14.3\mbox{--}16.2\,\mathrm{cm}$, indicating its origin within the inner accretion disk with a high density $>10^{8}\,\mathrm{cm}^{-3}$. Other UFOs are promisingly powerful enough to build effective AGN feedback while requiring further stricter constraints, in need of future observatories, e.g. \textit{ATHENA} \citep{2013Nandra}, and new methods, e.g. time-dependent photoionization models \citep{2022Rogantinitpho,2023Luminari} and resolved density-sensitive lines \citep{2017Mao}.

\subsection{\textit{ATHENA}/X-IFU simulations}\label{subsec:athena}
The future mission, \textit{ATHENA} \citep{2013Nandra}, with unprecedented spectral resolution and large effective areas, will resolve the systematic limitations of the flux-resolved spectroscopy and tightly constrain the nature of UFOs. The X-ray Integral Field Unit \citep[X-IFU,][]{2018Barret,2023Barret} onboard \textit{ATHENA} will collect sufficient photons for spectroscopy within a short timescale with well-resolved absorption lines, thus avoiding the risk of spectral broadening due to the stacked flux-resolved spectra. This will enable us to trace the UFO responses to the source variability and provide tight constraints on their launching mechanisms.

\begin{figure}
    \centering
    \includegraphics[width=\columnwidth, trim={10 10 20 10}]{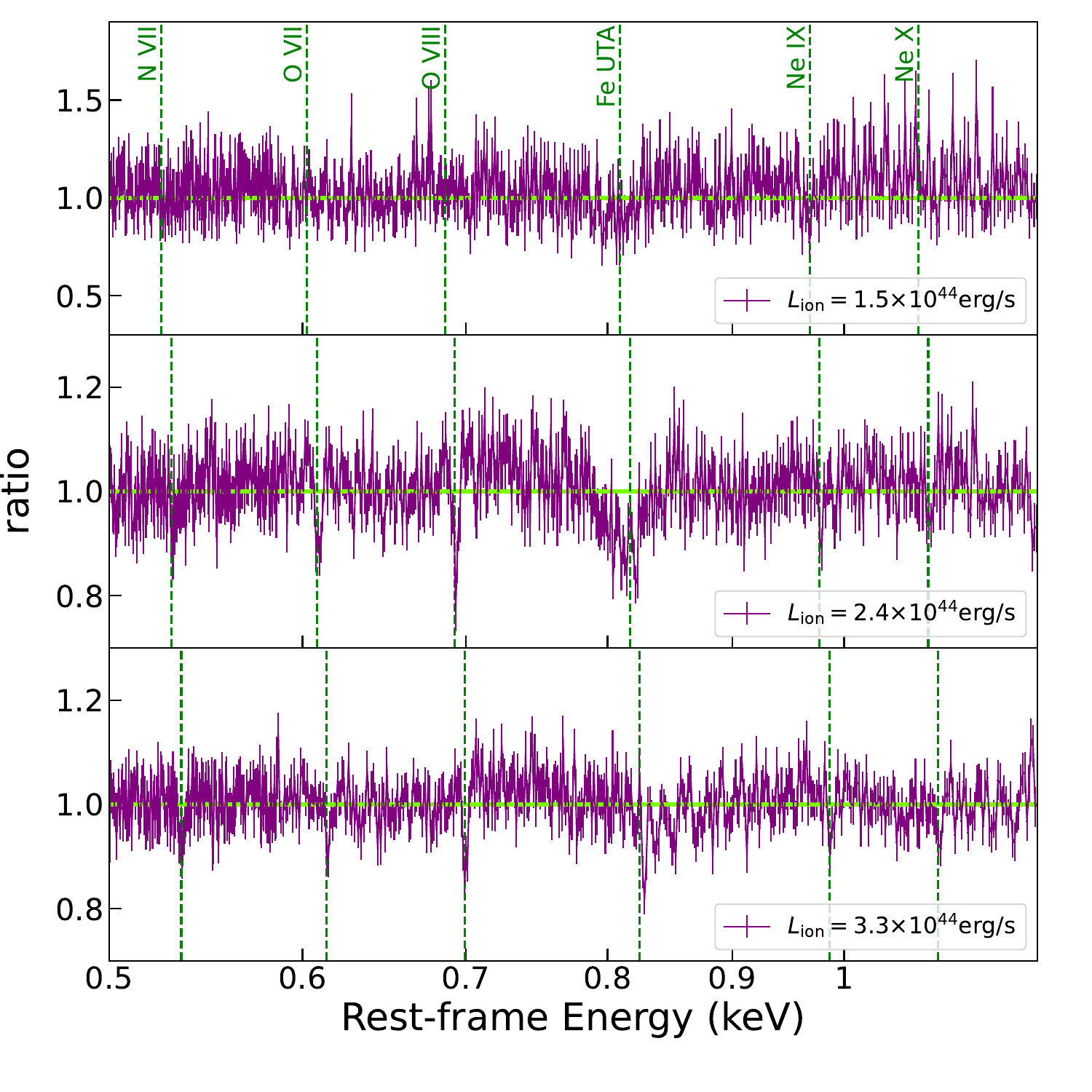}
    \caption{The data/model ratio for the \textit{ATHENA}/X-IFU spectra of PG 1211+143 at low (\textit{top}), middle (\textit{middle}), and high (\textit{bottom}) states with an exposure time of 10\,ks. The simulation of spectra is based on the best-fit continuum model, with the UFO ionization state predicted by the fit (see Tab.\ref{tab:fits}) and normalization adapted to reproduce the luminosity indicated in the legends. The UFO velocity is assumed to follow the trend of I ZW 1, instead of the constant trend of PG 1211+143, to simulate the change in UFO velocity. The vertical dashed green lines present the blueshifted UFO absorption features at different states. }
    \label{fig:athena}
\end{figure}

We utilize the \texttt{fakeit} tool within XSPEC to create simulated \textit{ATHENA}/X-IFU spectra of, for example, PG 1211+143 at different luminosities with new response files (released on 12/07/2023)\footnote{http://x-ifu.irap.omp.eu/resources/for-the-community}. The aim is to illustrate the corresponding UFO variations. We hypothesize that the source brightness linearly increases from its dimmest ($L_\mathrm{ion}=1.5\times10^{44}$erg/s) to its brightest ($L_\mathrm{ion}=3.3\times10^{44}$erg/s) state within 30\,ks, achieved by adjusting only the normalization of \texttt{diskbb} and \texttt{relxillCp} components. The UFO ionization state is assumed to evolve based on our fit results found in Tab.\ref{tab:fits}, while the velocity is assumed to increase, instead of being constant, to simulate the change in UFO velocity. The observation is divided into three 10-ks segments, which is the typical variability timescale of UFOs \citep{2017ParkerPCA,2023Reeves}. In Fig.\ref{fig:athena}, we depict the corresponding data/model ratios with respect to the broadband continuum model. The blueshifted UFO features are indicated by vertical dashed green lines. The significance of UFO detection in each spectrum is at least larger than 5$\sigma$ ($\Delta\mathrm{C-stat/d.o.f.}=53\mbox{--}376/4$), whereas in the simulated 100-ks \textit{XMM-Newton} spectra, the UFO in the faintest state is undetectable ($\Delta\mathrm{C-stat}=6$). Additionally, the constraints on UFO parameters by \textit{ATHENA}/X-IFU with only a 10\,ks exposure are a factor of $1\mbox{--}4$ stricter than those by \textit{XMM-Newton} with a $100$\,ks exposure time, (e.g. $\Delta\log\xi_{XMM}=0.06$ versus $\Delta\log\xi_\mathrm{X-IFU}=0.03$) \citep{2018Barret}. 

This simulation demonstrates that \textit{ATHENA} can achieve about ten times more effective UFO detection with only one-tenth of the exposure time compared with \textit{XMM-Newton}. Its powerful capabilities allow to track the response of the UFO on the variability time-scale of the ionizing continuum and the dynamical time-scale of the innermost regions of the accretion disk. Consequently, \textit{ATHENA} holds significant promise in advancing our understanding of AGN outflows and identifying the UFO launching mechanisms.

\section{Conclusions}\label{sec:conclusion}
In this work, we present a systematic analysis of the available archival \textit{XMM-Newton} observations of six highly accreting NLS1 galaxies, selected by UFO detection and exposure times, through high-resolution time- and flux-resolved spectroscopy. In summary, the results of our analysis are the following: 
\begin{enumerate}[-]
    \item The powerful method of the photoionization model scan over the high-resolution RGS spectra reveals three previously unreported UFOs in the archival datasets of RE J1034+396, PG 1244+026, and I ZW 1. All of them, confirmed through MC simulations, exhibit a detection significance above $3\sigma$. Additionally, in RE J1034+396, two previously unknown WAs have been detected.
    \item Outflows (WAs and UFOs) in 5 out of 6 (83\%) AGN of our sample shows a multi-phase structure. We discover an energy-conserved outflow in I ZW 1 and a strong correlation between the velocity and ionization parameter of UFOs in PG 1211+143. 4 (66\%) AGN in our sample host UFOs and WAs sharing a similar ionization parameter, supporting the shocked outflow interpretation for multi-phase ionized winds in AGN.
    \item The UFO detection in our sample combined with the literature \citep{2018Parker} exhibits an inclination-dependent behavior, where most UFOs are detected in AGN with an inclination angle between $30\mbox{--}60^\circ$, most likely due to selection effects.
    \item We notice that 2 out of 7 (28\%) UFOs in our sample seem to respond to the ionizing luminosity and 3 (43\%) UFOs hint at a radiatively-driven nature, requiring further observations for confirmation.
    \item Combined with published works, we discover no correlations between UFO responses and AGN intrinsic properties except for a tentative ($\sim1.8\sigma$) anti-correlation between the radiation acceleration on winds and Eddington ratio. Further observations and an enlarged sample are required to confirm this trend.
    \item We estimate the location and energetics of UFOs and find the constraints on soft X-ray UFOs are too loose to conclude that soft X-ray UFOs could exert a significant feedback impact on the surrounding environment, requiring future observations and new methods, although they are very promising based on some reasonable assumptions. The hard X-ray UFO in I ZW 1 is the only case in our sample with conclusively sufficient power to affect the host galaxy.
\end{enumerate}

%
%
\section{Acknowledgements}
We thank the anonymous referee for useful suggestions. This research has made use of data obtained with the \textit{XMM-Newton}, an ESA science mission funded by ESA Member States and the USA (NASA). We thank Peter Kosec for insightful discussions on the interpretation of our results. Y. X. acknowledges support from the European Space Agency (ESA) archival research visitor program. C. P. and S. B. acknowledge support for PRIN MUR 2022 SEAWIND 2022Y2T94C (funded by NextGenerationEU) and INAF LG 2023 BLOSSOM. D.R. is supported by NASA through the Smithsonian Astrophysical Observatory (SAO) contract SV3-73016 to MIT for Support of the Chandra X-Ray Center (CXC) and Science Instruments and the Margaret Burbidge Prize at University of Chicago. E.K. acknowledges XRISM Participating Scientist Program for support under NASA grant 80NSSC20K0733. The research leading to these results has received funding from the European Union’s Horizon 2020 Programme under the AHEAD2020 project (grant agreement n. 871158).


\bibliographystyle{aa}
\bibliography{ref.bib}

\begin{appendix}
\section{Details of Individual targets and Spectral Analysis}\label{app:sec:obs+spectral}

Our sample comprises six AGN. In this appendix, we present an overview of basic information for each source, alongside details of the Gaussian line scan results and the photoionization modeling of warm absorbers (listed in Tab.\ref{app:tab:WAfits}) and UFOs (see Tab.\ref{tab:xabsfits}).

\begin{table*}[!htbp]
\renewcommand{\arraystretch}{1.3} 
\centering
\caption{Table of the best-fit WA parameters in extracted spectra of targets in this work. The uncertainties of parameters are estimated at the 90\% confidence level ($\Delta \mathrm{C-stat}=2.71$).
}
\begin{tabular}{c|cccccc}
\hline
\hline
 Sources & \multicolumn{6}{c}{WA}   \\
\hline
       & $\log\xi$ & $N_\mathrm{H}$  & $\sigma_\mathrm{v}$  & $v_\mathrm{LOS}$ & $\Delta\rm C\mbox{-}stat$ & single trial\\
       & $\mathrm{erg\,cm\,s^{-1}}$ &($10^{20}$\,cm$^{-2}$) &  (km/s) &  (km/s) &  & significance ($\sigma$)\\
\hline
1H 1934-063 &&&&&&\\
\cline{2-7}
2015 & $1.88^{+0.04}_{-0.05}$ &  $8^{+3}_{-2}$ & $<150$ & $-400^{+200}_{-300}$& 159 & >8 \\
2021 & $1.89^{+0.04}_{-0.04}$ & $8^{+3}_{-2}$ & $100^{+50}_{-50}$ & $-300^{+100}_{-100}$ & 215 & >8\\
F1 & $1.88^{+0.05}_{-0.06}$ & $10^{+5}_{-3}$ &  $<100$ &  $-250^{+100}_{-200}$  & 99 & >8 \\
F2 & $1.88^{+0.05}_{-0.03}$ & $8^{+3}_{-2}$  & $100^{+50}_{-50}$  & $-300^{+100}_{-100}$ & 142 & >8 \\
\hline
RE J1034+396 &&&&&\\
\cline{2-7}
\multirow{3}{*}{avg$^\star$} & $3.61^{+0.03}_{-0.03}$ &  $27^{+3}_{-3}$  & $300^{+100}_{-50}$ &  $-1100^{+100}_{-100}$ & 514& >8 \\
 & $0.0^{+0.2}_{-0.2}$ &  $1.1^{+0.4}_{-0.5}$ & $7200^{+1600}_{-2200}$ &  $-19800^{+1500}_{-2700}$ & 109 & >8\\
 & $1.89^{+0.11}_{-0.20}$ &  $0.65^{+0.19}_{-0.09}$ & $300^{+100}_{-100}$ &  $-350^{+100}_{-100}$  & 100 & >8\\
 \cline{2-7}
\multirow{3}{*}{F1$^\star$} &  $3.64^{+0.11}_{-0.20}$ & $32^{+10}_{-9}$ & $100^{+200}_{-50}$ & $-1400^{+200}_{-200}$ & 149 & >8\\
&  $-0.9^{+0.7}_{-0.8}$ &  $2.4^{+0.7}_{-1.1}$ & $10000^{\bullet}$ & $-30900^{+5100}_{-1800}$ & 36 & 5.1\\
&  $1.97^{+0.18}_{-0.21}$ & $1.6^{+1.0}_{-0.5}$ & $150^{+100}_{-100}$ & $-300^{+200}_{-200}$ & 53 & 6.5\\
\cline{2-7}
\multirow{3}{*}{F2$^\star$} & $3.59^{+0.07}_{-0.07}$ & $27^{+6}_{-4}$ &  $300^{+100}_{-100}$&  $-1000^{+200}_{-200}$ & 180 & >8 \\
&  $-0.2^{+0.1}_{-0.1}$ & $1.1^{+0.4}_{-0.3}$  & $2000^{+900}_{-1000}$ & $-21300^{+3000}_{-600}$ & 67 & 7.4\\
 &  $1.78^{+0.17}_{-0.27}$ & $0.6^{+0.3}_{-0.2}$ & $300^{+400}_{-100}$ & $-600^{+200}_{-300}$ & 23 & 3.8\\
 \cline{2-7}
\multirow{2}{*}{high$^\star$} & $3.46^{+0.09}_{-0.09}$ & $27^{+4}_{-4}$ & $550^{+300}_{-200}$ & $-700^{+200}_{-200}$ & 100 & >8\\
&  $0.0^{+0.3}_{-0.4}$ & $1.2^{+0.8}_{-1.0}$ & $10000^{\bullet}$ & $-21000^{+2400}_{-3900}$ & 20 & 3.5\\
\hline
PG 1211+143 &&&&&\\
\cline{2-7}
T7 & $2.40^{+0.18}_{-0.19}$ & $18^{+11}_{-9}$  & $<100$ & $-3000^{+900}_{-600}$ & 22 & 3.7\\
\hline
I ZW 1 &&&&&\\
\cline{2-7}
\multirow{2}{*}{avg$^\star$} & $-0.40^{+0.10}_{-0.09}$ &  $5.0^{+0.7}_{-0.7}$ & $200^{+50}_{-50}$ &  $-1800^{+100}_{-100}$ & 384  & >8 \\
 & $2.70^{+0.12}_{-0.11}$ &  $6^{+6}_{-3}$  & $<100$  &  $-2400^{+300}_{-300}$ & 41   & 5.6  \\
 \cline{2-7}
2005 & $-0.34^{+0.21}_{-0.31}$ &  $7^{+2}_{-2}$  & $150^{+100}_{-100}$  & $-1900^{+200}_{-300}$& 76 & 8.0 \\
\cline{2-7}
\multirow{2}{*}{F1$^\star$} & $-0.27^{+0.09}_{-0.09}$  & $3.5^{+0.8}_{-0.7}$ & $500^{+200}_{-300}$  &  $-1800^{+200}_{-200}$ & 111 & >8 \\
 & $2.44^{+0.30}_{-0.18}$ & $5^{+7}_{-2}$ & $100^{+200}_{-50}$ & $-1800^{+300}_{-300}$ & 28 & 4.4 \\
 \cline{2-7}
\multirow{2}{*}{F2$^\star$} & $-0.14^{+0.19}_{-0.09}$ & $4.8^{+1.0}_{-0.9}$  &  $100^{+100}_{-50}$ &  $-2000^{+200}_{-200}$ & 91 & >8 \\
 & $2.78^{+0.10}_{-0.18}$ & $30^{+30}_{-20}$ & $<100$  &  $-2400^{+900}_{-600}$ & 19 & 3.3  \\
 \cline{2-7}
2020 & $-1.00^{+0.09}_{-0.19}$ & $14^{+3}_{-3}$ & $200^{+100}_{-100}$ & $-1700^{+200}_{-200}$ & 172 & >8 \\
\hline
IRAS 17020+4544 &&&&&\\
\cline{2-7}
\multirow{3}{*}{avg$^\star$} & $-0.3^{+0.1}_{-0.1}$ &  $4.7^{+0.6}_{-0.7}$ & $750^{+700}_{-200}$ &  $50^{+200}_{-300}$ & 638 & >8\\
 & $1.9^{+0.1}_{-0.1}$ &  $23^{+4}_{-4}$ & $<100$&  $300^{+200}_{-100}$ & 160 & >8\\
 & $-2.8^{+0.2}_{-0.1}$ &  $6^{+1}_{-1}$ &  $700^{+200}_{-200}$ &  $-3900^{+300}_{-300}$ & 164 & >8\\
 \cline{2-7}
\multirow{3}{*}{F1$^\star$} & $-1.0^{+0.2}_{-0.2}$ &  $3.9^{+1.0}_{-0.6}$ & $1000^{+800}_{-100}$  &  $-6000^{+600}_{-600}$ & 248 & >8\\
 &  $2.0^{+0.1}_{-0.1}$ &  $16^{+4}_{-6}$ & $<100$ &  $250^{+300}_{-300}$ & 72 & 7.8\\
 & $-2.5^{+0.2}_{-0.2}$ &  $9^{+1}_{-1}$ & $750^{+200}_{-200}$ &  $-3900^{+300}_{-300}$ & 68 & 7.5 \\
 \cline{2-7}
\multirow{3}{*}{F2$^\star$} & $-0.2^{+0.3}_{-0.1}$ &  $5.0^{+2.1}_{-1.4}$ & $600^{+100}_{-100}$ &  $100^{+200}_{-200}$  & 222 & >8\\
 & $1.8^{+0.1}_{-0.1}$ &  $21^{+4}_{-4}$ & $100^{+50}_{-50}$ &  $200^{+300}_{-200}$ & 104 & >8\\
 &  $-2.7^{+0.3}_{-0.2}$ &  $5^{+2}_{-2}$ & $1000^{+700}_{-200}$ &  $-4200^{+600}_{-600}$& 44 & 5.8 \\
 \cline{2-7}
\multirow{3}{*}{F3$^\star$} & $0.0^{+0.2}_{-0.3}$ &  $4.7^{+1.0}_{-0.9}$ & $500^{+300}_{-300}$  &  $0^{+300}_{-300}$  & 205 & >8 \\
 & $1.9^{+0.1}_{-0.1}$ &  $29^{+10}_{-6}$ & $<100$  &  $350^{+200}_{-200}$  & 92 & >8\\
 & $-2.7^{+0.2}_{-0.3}$ &  $8^{+2}_{-2}$ & $700^{+300}_{-200}$ &  $-3900^{+300}_{-300}$ & 81 & >8 \\
 \cline{2-7}
\hline
\hline
\end{tabular}
\label{app:tab:WAfits}
\begin{flushleft}
{$^{\star}$ Multiple WAs are detected in this spectrum.

$^{\bullet}$ The parameter is fixed.}
\end{flushleft}
\end{table*}

\subsection{1H 1934-063}\label{app:subsec:1h1934-063}
1H 1934-063 is a nearby ($z=0.0102$) radio-quiet NLS1 galaxy, presenting variable X-ray fluxes \citep{2012Ponti}. The spin of the central SMBH was estimated at $a_\star<0.56$ \citep{2022Xu} or $a_\star>0.4$ \citep{2019Jiangdensity} depending on whether a high-density relativistic reflection model was adopted. In the 2015 observation of 1H 1934-063, a $\sim20$s time lag was detected between the disk reprocessing component and the primary continuum, indicating a compact corona with a height of $9\pm4\,R_\mathrm{g}$ ($R_\mathrm{g}\equiv GM_\mathrm{BH}/c^2$) in the lamppost geometry. In the RGS spectrum, distinct signatures of a WA and a moderately ionized UFO have been identified as well \citep{2022Xu}. Moreover, the tentative evidence for the reprocessing of the coronal photons onto the base of winds has also been observed. Furthermore, two \textit{XMM-Newton} observations were conducted on 1H 1934-063 in 2021 and the data analysis is presented here.

\begin{figure}[htbp]
    \centering
	\includegraphics[width=\columnwidth]{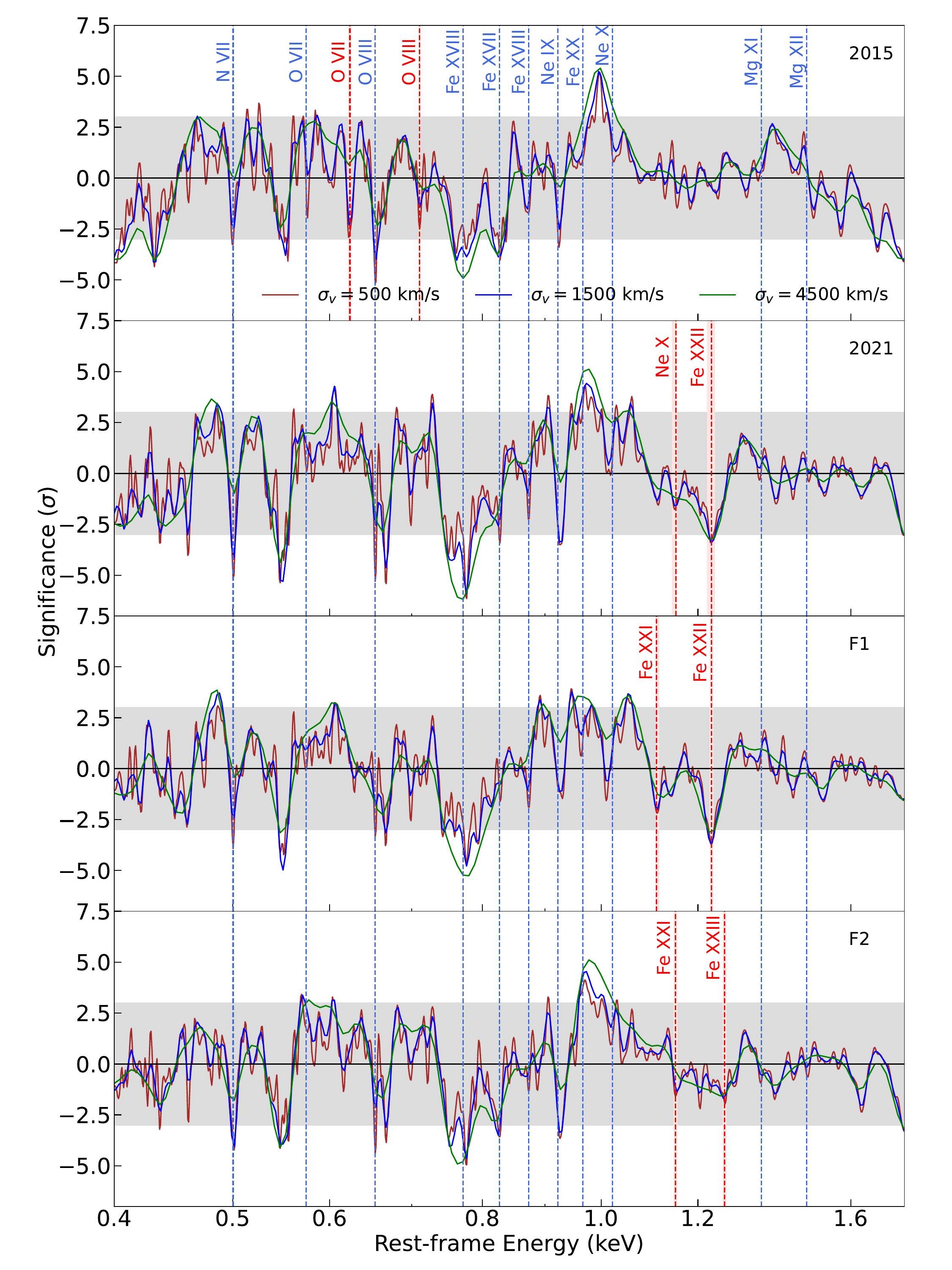}
    \caption{The single trial significance obtained from the Gaussian line scan with different line widths (500, 1500, 4500\,km/s) over the rest-frame spectra of 1H 1934-063 in the RGS band. The vertical dashed \textit{blue} lines represent the rest-frame energies of the known ion transitions as a reference, while the \textit{red} lines and the shaded region respectively correspond to the centroid and $1\sigma$ uncertainties of the prominent UFO absorption lines detected in the following photoionization modeling. The gray region marks the $3\sigma$ significance.
    }
    \label{app:fig:gaussian-1h1934-063}
\end{figure}

The results of the Gaussian line scan in the RGS band are shown in Fig.\ref{app:fig:gaussian-1h1934-063}, highlighting the consistent presence of the WA features, such as N \textsc{vii}, O \textsc{viii}, Fe \textsc{xviii}, Fe \textsc{xvii}, and Ne \textsc{ix} across all spectra. The prominent UFO absorption features transition from O \textsc{vii} and O \textsc{viii} to Fe complex lines, Fe \textsc{xxi-xxiii} between 2015 and 2021. In contrast, the properties of the WA remain constant for six years (see Tab.\ref{app:tab:WAfits}).

UFOs detected in the 2021 observations exhibit distinct properties from that in 2015 \citep{2022Xu} with higher ionization states, larger column densities, and faster speeds (see Tab.\ref{tab:xabsfits}). However, the observations of these two epochs share similar X-ray luminosity (see panel \textit{a} of Fig.\ref{fig:lc}), indicating that the difference is unrelated to the radiation field. The origin of the UFO change remains mysterious, perhaps resulting from different phases of UFOs or unknown prior variable accretion rates. In the flux-resolved spectra from consecutive 2021 observations, most UFO properties remain stable within their uncertainties, except for a slightly higher velocity at the brighter state.

\subsection{RE J1034+396}\label{app:subsec:rej1034+396}
RE J1034+396 is a nearby ($z=0.042$) NLS1 galaxy, firstly exhibiting significant X-ray quasi-periodic oscillation (QPO) within an AGN with the period of $2.7\times10^{-4}$\,Hz \citep[e.g.,][]{2008Gierli,2014Alston,2020Jin}. It has an extraordinarily steep soft X-ray in the spectrum \citep{1995Puchnarewicz}. The mass of the central SMBH was estimated within $10^{6}\mbox{--}10^{7}\,M_\odot$ \citep[summarized by ][]{2016Czerny}, with the most probable mass range of $(1\mbox{--}4)\times10^{6}\,M_\odot$ \citep[e.g.,][]{2008Gierli,2009Middleton,2018Chaudhury}. Over a span of more than a decade, \textit{XMM-Newton} has observed this source multiple times. A large observation campaign of this source was executed in 2020 and 2021 with a total amount of 855\,ks exposure time. In this work, we focus on the data from the 2020-2021 campaign, as well as three archival observations capturing RE J1034+396 at bright states.

\begin{figure}[htbp]
    \centering
	\includegraphics[width=\columnwidth]{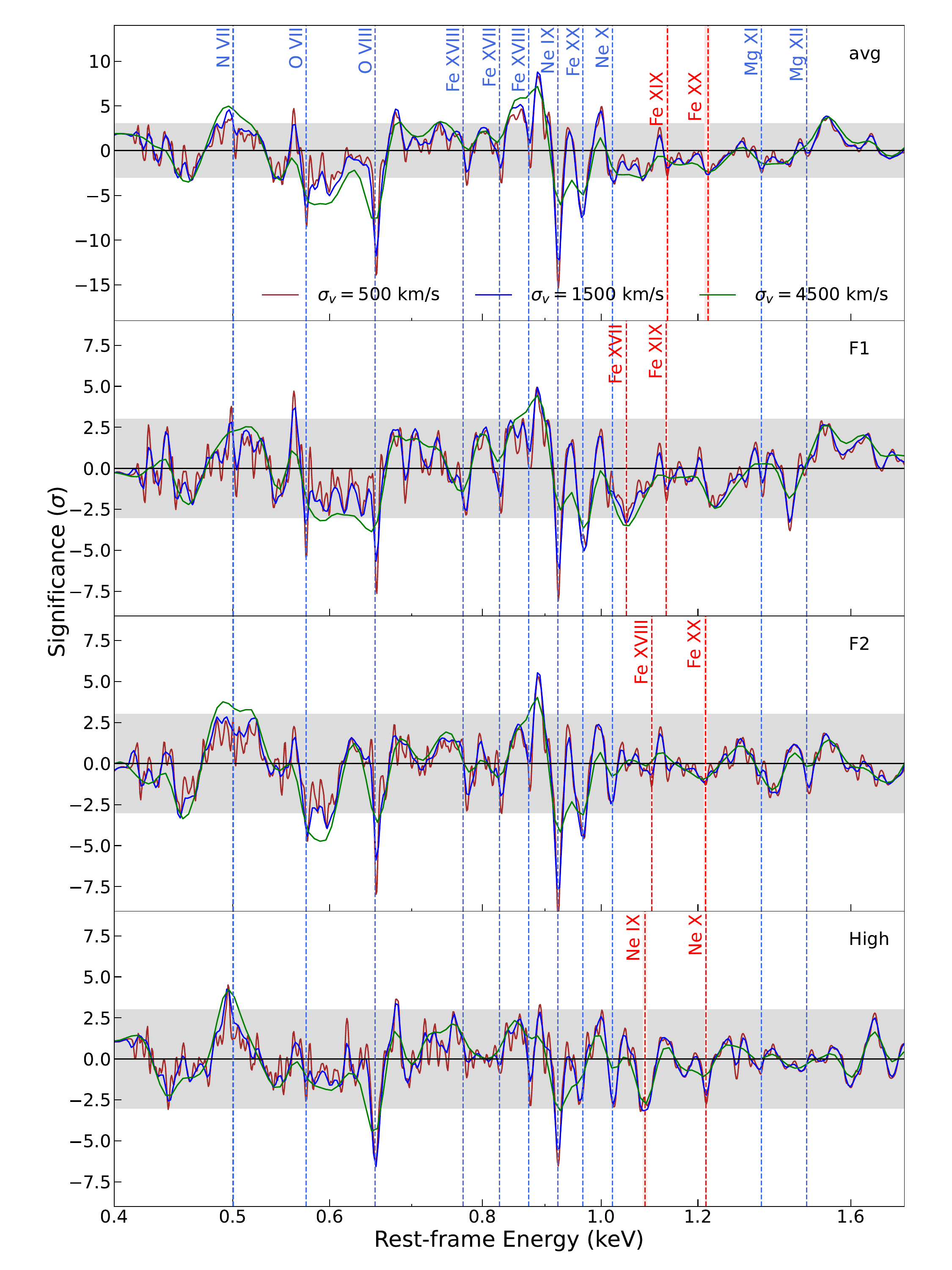}
    \caption{Similiar to Fig. \ref{app:fig:gaussian-1h1934-063}, but the scan is performed on the spectra of RE J1034+396.
    }
    \label{app:fig:gaussian-rej1034+396}
\end{figure}

We present the outcomes of the Gaussian line scan over the continuum model in Fig.\ref{app:fig:gaussian-rej1034+396}. The results for the first time unveil a series of strong absorption lines at their rest-frame energies, including Ne \textsc{ix}, Fe \textsc{xx}, O \textsc{vii}, and O \textsc{viii}. The spectra require two distinct WAs, each existing at different ionization states ($\log\xi\sim3.6$ and $\log\xi\sim1.9$). However, the cooler WA is superfluous for the `high' spectrum (see Tab.\ref{app:tab:WAfits}). A broad trough is evident on the blue side of the O \textsc{vii} line. We initially speculated its association with blueshifted O \textsc{vii}. However, our photoionization model scan suggests a neutral absorber ($\log\xi\sim0$) with an ultra-fast velocity. In the F1 and `high' spectra, the line width of this component is too broad to be constrained and thus fixed at 10000\,km/s. This particular neutral ionization state precludes its classification as a UFO. Consequently, we name it a `warm absorber', while its origin remains subject to future investigation. 

UFOs are degenerated to explain similar Fe complex absorption lines between $1\mbox{--}1.2$\,keV (Fig.\ref{app:fig:gaussian-rej1034+396}), indicating the same origin. The degenerated region can be observed in the panel \textit{b} of Fig.\ref{fig:absorption_scan} and the \textit{top} panel of Fig.\ref{fig:absorption_scan_MC}, where the best-fit UFO solutions for F1 and F2 spectra manifest at the high-speed end of the degenerate solutions while the `high’ spectrum solution is notably confined to a lower-speed domain. The statistical difference between these two distinct solutions (see Tab.\ref{tab:xabsfits}) within the `high' spectrum is $\Delta\mathrm{C}\mbox{--}\mathrm{stat}=16$ and it becomes even larger for UFOs in flux-resolved spectra, disfavoring the explanation of stochastic variability. Therefore, statistics differentiate these UFOs, resulting in correlations of the ionization parameter and column density with X-ray luminosity while the velocity does not exhibit the same behavior, of which origin is still unknown.

\subsection{PG 1244+026}\label{app:subsec:pg1244+026}
PG 1244+026, identified as a highly variable NLS1 galaxy \citep{2013Jin}, exhibits notable features in the spectrum, including a strong soft excess and a highly ionized Fe K emission line \citep{2006Crummy}. Within this source, a soft lag at high frequencies and hard lags at low frequencies were independently discovered by \citet{2014Alstonpg1244} and \citet{2014Kara}. It suggests the relativistic reflection off the inner accretion disk and the propagation of the accretion fluctuations. Over the span of 2011 to 2014, \textit{XMM-Newton} conducted six observations of PG 1244+026, with five of these observations occurring consecutively in 2014. Despite the observations, the literature lacks any reports of UFOs in PG 1244+026.

\begin{figure}[htbp]
    \centering
	\includegraphics[width=\columnwidth]{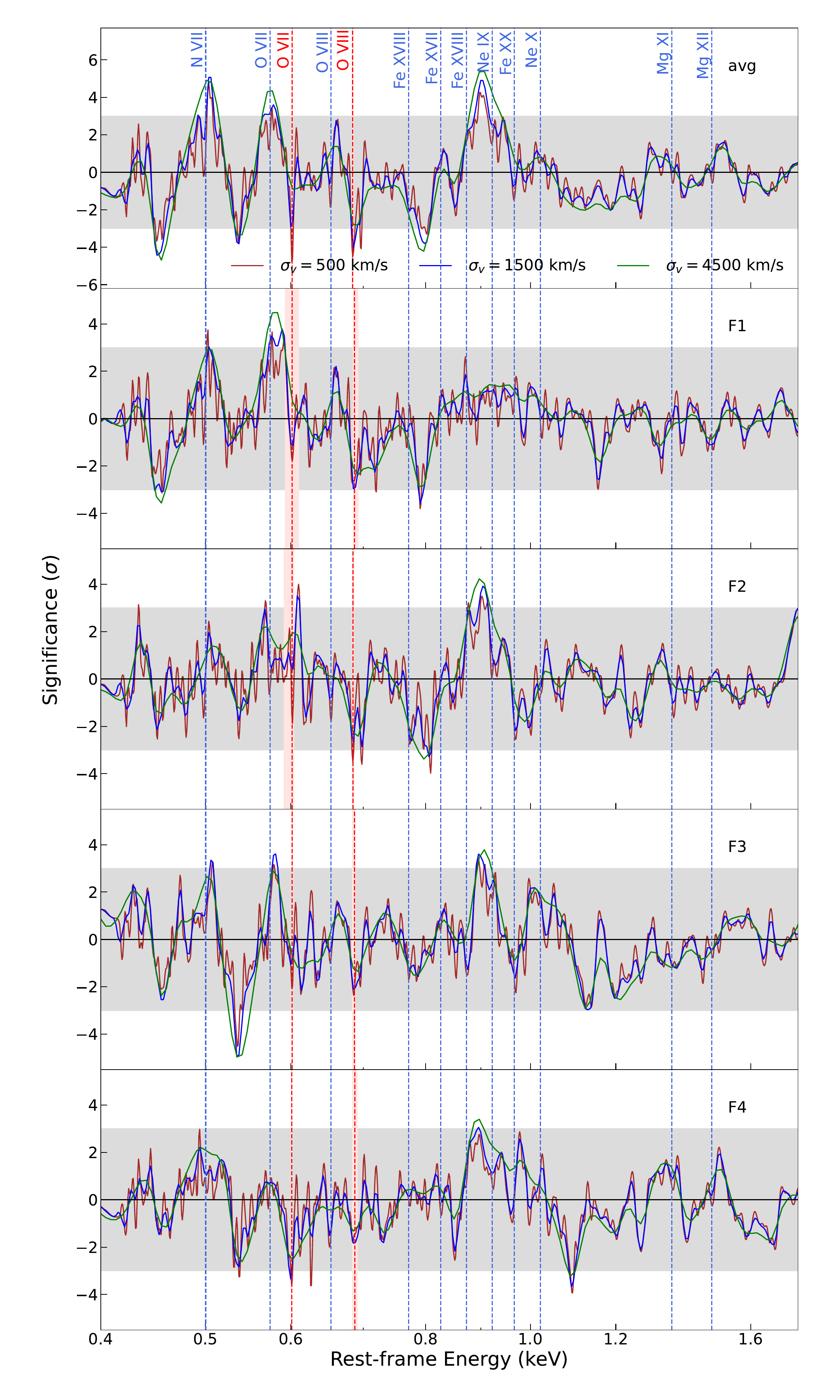}
    \caption{Similiar to Fig. \ref{app:fig:gaussian-1h1934-063}, but the scan is performed on the spectra of PG 1244+026.
    }
    \label{app:fig:gaussian-pg1244+026}
\end{figure}

The findings from the Gaussian line scan are presented in Fig.\ref{app:fig:gaussian-pg1244+026}. Within the time-averaged spectrum, a series of absorption lines appear as well as the N \textsc{vii}, O \textsc{vii} and $\sim0.9$\,keV emission lines. The identified UFO features encompass blueshifted O \textsc{vii} and O \textsc{viii}, featuring a stable velocity of $v_\mathrm{LOS}\sim-13500$\,km/s across all flux- and time-resolved spectra (time-resolved results are not shown). No warm absorber is detected in this system.

The UFO has stable velocity and ionization parameter within their uncertainties while simultaneously exhibiting an anticorrelation between the column density and X-ray luminosity (see Tab.\ref{tab:xabsfits}). However, we noticed that one observation (2011) occurred three years before the others (2014). Through the trial of the time-resolved spectroscopy, we found that the UFO properties in that observation resemble those found in the others, ruling out the explanation of different UFO phases. After removing the 2011 observation, time-resolved results lead to a decreasing trend of the column density at only $1\sigma$ from a constant. Therefore, the observed decreasing trend may result from long time-gap accretion variations, while the possibility of an intrinsic anticorrelation cannot be ruled out. Moreover, given its relatively low velocity $v_\mathrm{LOS}\sim-13500$\,km/s, the UFO may originate from a distant place from the SMBH, leading to a less responsive UFO to the source variation, which awaits future investigation.

\subsection{PG 1211+143}\label{app:subsec:pg1211+143}
PG 1211+143 is a well-studied and luminous NLS1/quasar at a redshift of $z=0.0809$, showing a typical X-ray luminosity of $\sim10^{44}\,\mathrm{erg/s}$ and a bright optical flux with a strong `Big Blue Bump'. This source is well-known for its spectral complexity and the presence of highly variable and multi-phase UFOs \citep[e.g.,][]{2003Pounds,2007Pounds,2009Pounds,2014Pounds,2016Pounds,2016Pounds2UFO,2018Reeves}. The velocity range of these UFOs spans from -18000\,km/s to -81000\,km/s \citep{2016Pounds,2018Danehkar}, revealing absorption lines across both soft and hard X-ray bands. Moreover, a corresponding UV counterpart to the UFO, outflowing at a comparable velocity, was documented by \citet{2018Kriss}. In addition, a soft time lag of $\sim500$s at $\sim10^{-4}$\,Hz \citep{2011deMarco} and a hard lag up to $\sim3$ks at $\sim10^{-5}$\,Hz \citep{2018Lobban} were reported in this source. The total \textit{XMM-Newton} exposure on this source culminates at $\sim900$\,ks, with a majority of observations conducted consecutively at varying flux levels. As a result, a time-resolved spectroscopic analysis is undertaken to investigate both temporal and flux variability.

\begin{figure*}[htbp]
    \centering
	\includegraphics[width=\columnwidth]{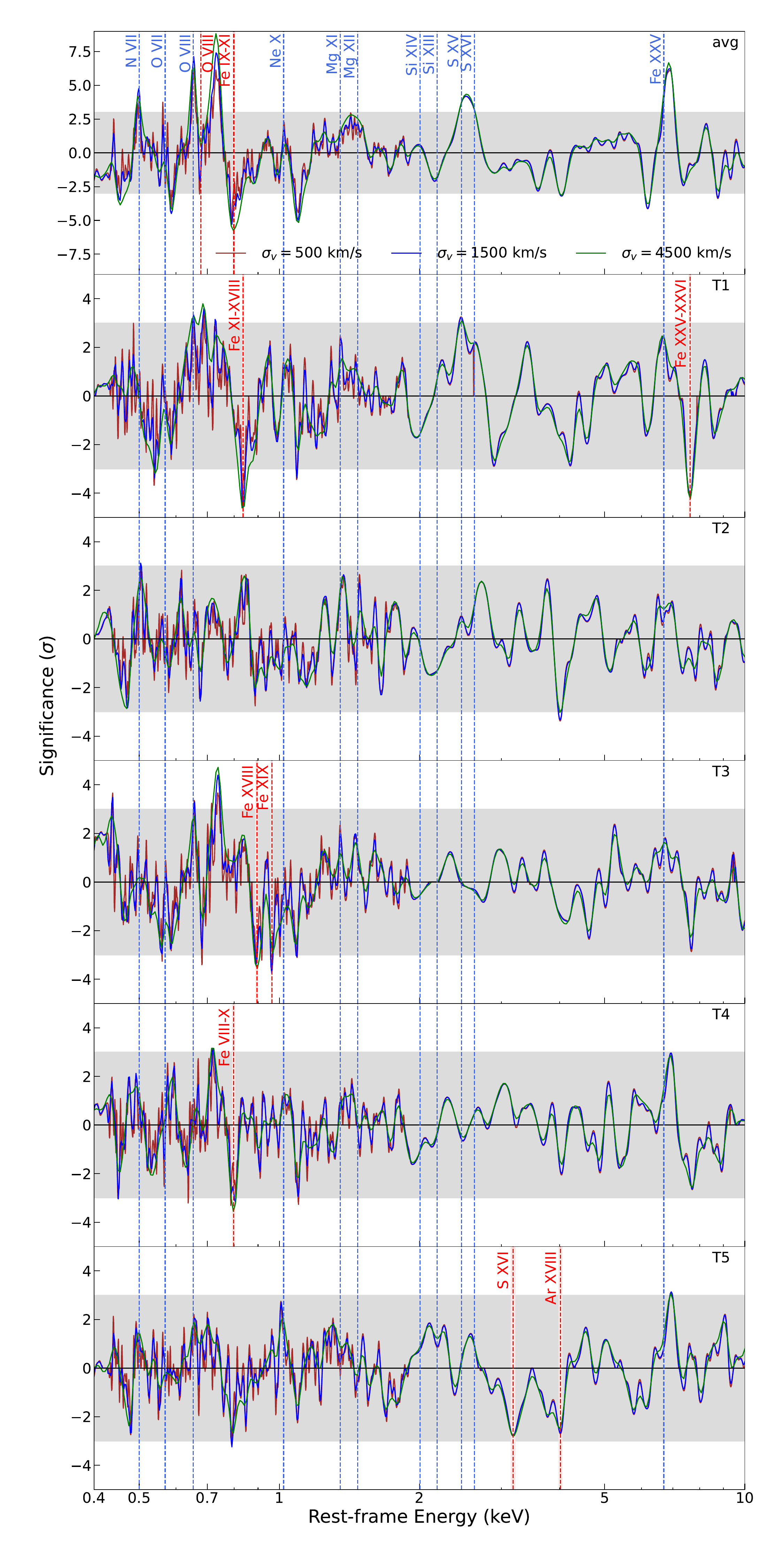}
	\includegraphics[width=\columnwidth]{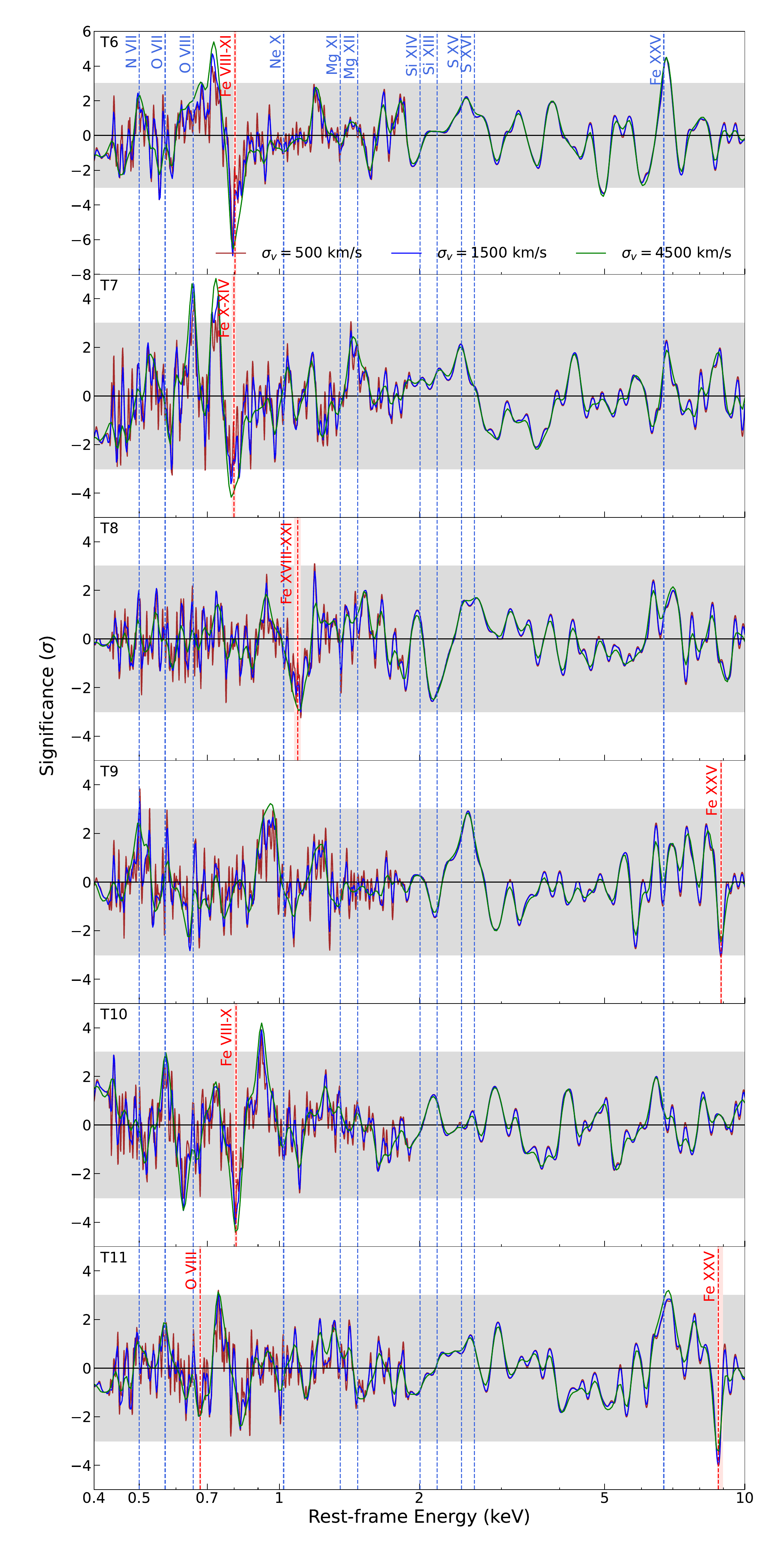}
    \caption{Similiar to Fig. \ref{app:fig:gaussian-1h1934-063}, but the scan is performed on the spectra of PG 1211+143.
    }
    \label{app:fig:gaussian-pg1211+143}
\end{figure*}

The Gaussian line scan results are shown in Fig.\ref{app:fig:gaussian-pg1211+143}, where the UFO features are marked by vertical red lines. The predominant UFO features are blueshifted O \textsc{viii} and Fe UTA lines in the soft X-ray regime, along with S \textsc{xvi}, Ar \textsc{xviii}, and Fe \textsc{xxv/xxvi} features in the hard X-ray band. The UFOs identified within our analysis are multi-phase and highly variable, in accordance with prior investigations \citep[e.g.,][]{2018Reeves,2016Pounds}.  According to our criterion for the outflow detection significance, we only find one fast warm absorber in the `T7' spectrum (see Tab.\ref{app:tab:WAfits}).

Consistent with the literature \citep[e.g.][]{2016Pounds2UFO,2018Reeves}, we found that UFOs are highly variable and multi-phase. The LOS velocity ranges from $-10000$ to $-76000$\,km/s and the ionization parameter $\log\xi$ spans between 1.5 and 5. The predominant UFO features are blueshifted Fe UTA lines in the soft X-ray regime, along with S \textsc{xvi}, Ar \textsc{xviii}, and Fe \textsc{xxv/xxvi} features in the hard X-ray band. We do not find any UFOs with a significance $\Delta\mathrm{C}\mbox{--}\mathrm{stat/d.o.f.}>16/4$ in the T2 spectrum, whereas a secondary UFO attains noteworthy significance in the T1 and T11 spectrum. The most robust UFO detection falls in the faintest T6 spectrum ($\Delta\mathrm{C}\mbox{--}\mathrm{stat/d.o.f.}=100/4$), attributed to the passage of an absorbing cloud \citep{2018Reeves}. Among UFOs in our results, the solution featuring $v_\mathrm{LOS}\sim-18000$\,km/s emerges as the most recurrent, while others manifest a notably stochastic behavior. Given the consecutive observations of 2014, such complexity cannot be attributed to the long time-gap accretion variations and was explained by some intrinsic disk instability \citep{2016Pounds2UFO}.

\subsection{I ZW 1}\label{app:subsec:izw1}
I ZW 1 is a nearby ($z\sim0.06$) NLS1 galaxy with a bright \citep[$\sim10^{44}\,\mathrm{erg/s}$,][]{2004Gallo} and variable X-ray luminosity \citep{2017Wilkins}. Its X-ray spectrum reveals the presence of a strong Fe K emission and a weak soft excess \citep{2007Gallo}. \citet{2017Wilkins} unveiled a time lag of $\sim160$s between the reflection-dominated energy and continuum-dominated band within $3\mbox{--}12\times10^{-4}$\,Hz. Notably, the signatures of the light bending and X-ray echoes around the event horizon were observed through the delayed and redshifted flare emission observed in a luminous event in 2020 \citep{2021Wilkins}. Furthermore, the soft X-ray spectrum of I ZW 1 reveals the presence of two WAs, exhibiting a long-term variable nature \citep{2007Costantini,2018Silva}. It was attributed to the presence of either a two-phase WA originating from a shared clump or absorbing gas existing in a non-equilibrium state. The exploration of the hard X-ray spectrum unveils the existence of a highly ionized ($\log\xi\sim4.9$) and high-speed ($v\sim-0.25c$) UFO during the 2005 and 2015 observations \citep{2019Reeves}. A subsequent observation in 2020, analyzed by \citet{2022Rogantini}, discovered a UFO, characterized by a similar velocity yet manifesting a lower ionization state ($\log\xi\sim3.8$).

\begin{figure}[htbp]
    \centering
	\includegraphics[width=\columnwidth, trim={20 60 20 10}]{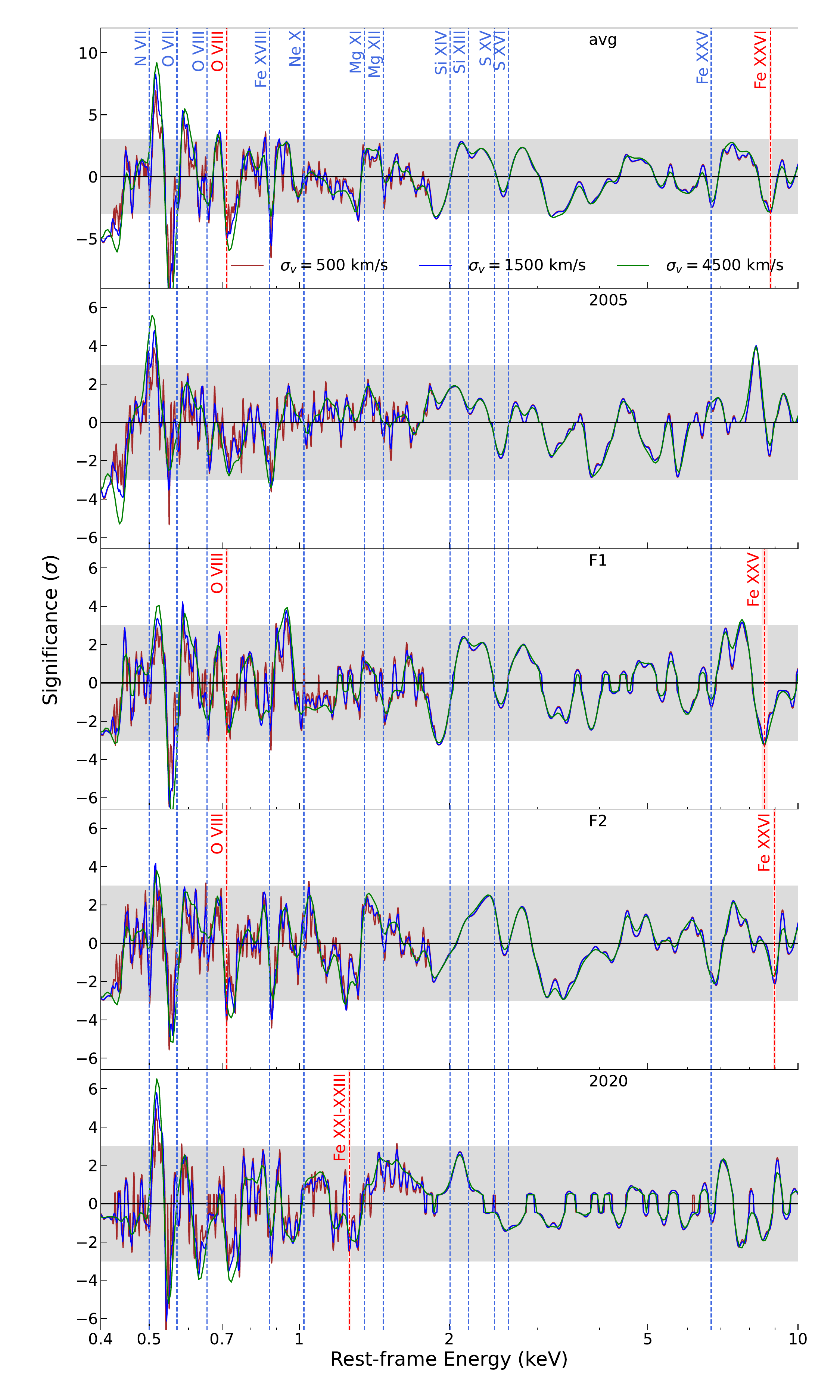}
    \caption{Similiar to Fig. \ref{app:fig:gaussian-1h1934-063}, but the scan is performed on the spectra of I ZW 1.
    }
    \label{app:fig:gaussian-izw1}
\end{figure}

We present the results of the Gaussian line scan in Fig.\ref{app:fig:gaussian-izw1}, where discernible absorption lines such as O \textsc{viii} and Fe \textsc{xviii} are clearly identified at their rest-frame positions. A prominent trough appears around 0.53\,keV across all spectra, which is associated with low-ionized oxygen. The UFO features are marked by the red vertical dashed lines in both soft (O \textsc{viii} and Fe \textsc{xxi-xxiii}) and hard (Fe \textsc{xxv-xxvi}) X-rays. Consistent with findings by \citep{2018Silva}, our analyses have detected two distinct slow absorbers, characterized by neutral ($\log\xi\sim-0.4$) and high ($\log\xi\sim2.7$) ionization states (see Tab.\ref{app:tab:WAfits}).  However, the high-ionization WA is not statistically necessary for the 2005 spectrum, according to our detection threshold. The limited photon counts of the 2005 spectrum cannot provide robust secondary absorber detection. While similar residuals around 0.7\,keV are observed in the 2020 spectrum, they have already been explained by the warm absorber featuring a lower ionization parameter ($\log\xi\sim-1$), identified as Fe \textsc{v-vi} lines. The requirement for this lower ionization state is attributed to a distinct absorption feature solely within the 2020 spectrum, located around 0.63\,keV. The absence of this feature within the 2015 spectra prevents the neutral warm absorber from entering the cooler regime, thus compelling the consideration of an alternative UFO to account for the unexplained 0.7\,keV feature.

We found a highly-ionized ($\log\xi>4.5$) and fast-moving UFO in 2015 observations through blueshifted Fe \textsc{xxv-xxvi} absorption lines, consistent with previous works \citep{2019Reeves}. This UFO in flux-resolved spectra remains stable. The UFO detected within the 2020 spectrum exhibits a comparable ionization parameter ($\log\xi\sim3.8$) to the one found in \citet{2022Rogantini}, accounting for residuals $\sim1.2$\,keV \citep{2022Wilkins}. In 2015 observations, our scan uncovers a previously unreported secondary UFO in soft X-rays (I ZW 1-2), particularly evident for the absorption feature around 0.7\,keV (Fig.\ref{app:fig:gaussian-izw1}). The column density and velocity of this UFO are stable in 2015 observations, while the ionization parameter seems to increase with the X-ray luminosity.

\subsection{IRAS 17020+4544}\label{app:subsec:iras17020+4544}
IRAS 17020+4544 emerges as a nearby ($z=0.0604$) radio-loud NLS1 galaxy, known to be characterized by a WA \citep{1997Leighly} and a compact radio emission \citep{2004Snellen}. The archival multi-wavelength campaign for IRAS 17020+4544 has revealed the presence of multi-phase UFOs ($v\sim24000\mbox{--}27000$\,km/s) in the X-ray spectra \citep{2015Longinotti,2018Sanfrutos}, a galaxy-scale molecular outflow \citep{2018Longinotti} in addition to a sub-relativistic jet on parsec scales in the radio band \citep{2017Giroletti}. Moreover, a UV counterpart of the X-ray UFO outflow has been detected through Ly$\alpha$ absorption, showing a velocity similar to its X-ray counterpart. Collectively, these discoveries suggest that AGN-driven hot gas gives rise to large-scale shocks into the interstellar medium.

\begin{figure}[htbp]
    \centering
	\includegraphics[width=\columnwidth]{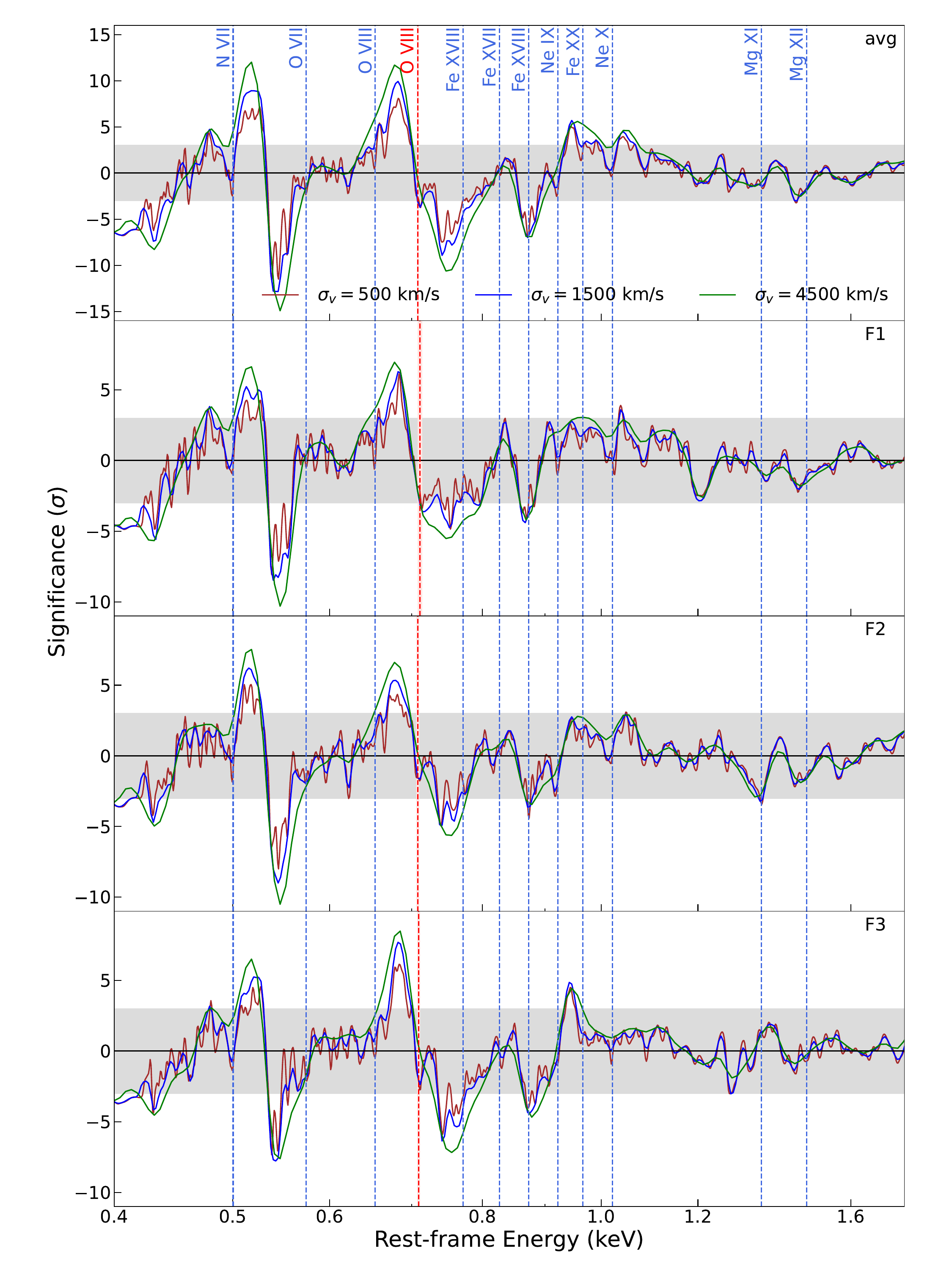}
    \caption{Similiar to Fig. \ref{app:fig:gaussian-1h1934-063}, but the scan is performed on the spectra of IRAS 17020+4544.
    }
    \label{app:fig:gaussian-iras17020+4544}
\end{figure}

The outcomes of the line scan are depicted in Fig.\ref{app:fig:gaussian-iras17020+4544}, exhibiting strong absorption and emission lines within the spectra. The prominent absorption features are located around 0.53, 0.75, and 0.88,keV. As reported in \citet{2018Sanfrutos}, our investigation identifies the presence of three distinct warm absorbers (WAs) in the spectra, each marked by substantial statistical significance (see Tab.\ref{app:tab:WAfits}). WA1 and WA3 emerge as absorbers characterized by low ionization states, mainly explaining the absorption feature at 0.53\,keV. On the other hand, WA2 is moderately ionized and models the feature around 0.75 and 0.88\,keV. 

We found a moderately ionized ($\log\xi\sim1.9$) UFO with a LOS velocity of $v_\mathrm{LOS}\sim-23100$\,km/s in the soft X-ray band. It accounts for the residual around 0.71\,keV (Fig.\ref{app:fig:gaussian-iras17020+4544}). This UFO is consistent with the findings of \citet{2018Sanfrutos}, while the remaining three UFOs in their work disappear in our analyses due to their tiny significance ($\Delta$C-stat/d.o.f.$<10/4$). The properties of such UFO are remarkably stable across flux-resolved spectra.

\section{UFO properties versus unabsorbed X-ray luminosity}\label{app:sec:correlation}

\begin{figure*}[htbp]
    \centering
    \includegraphics[width=\columnwidth]{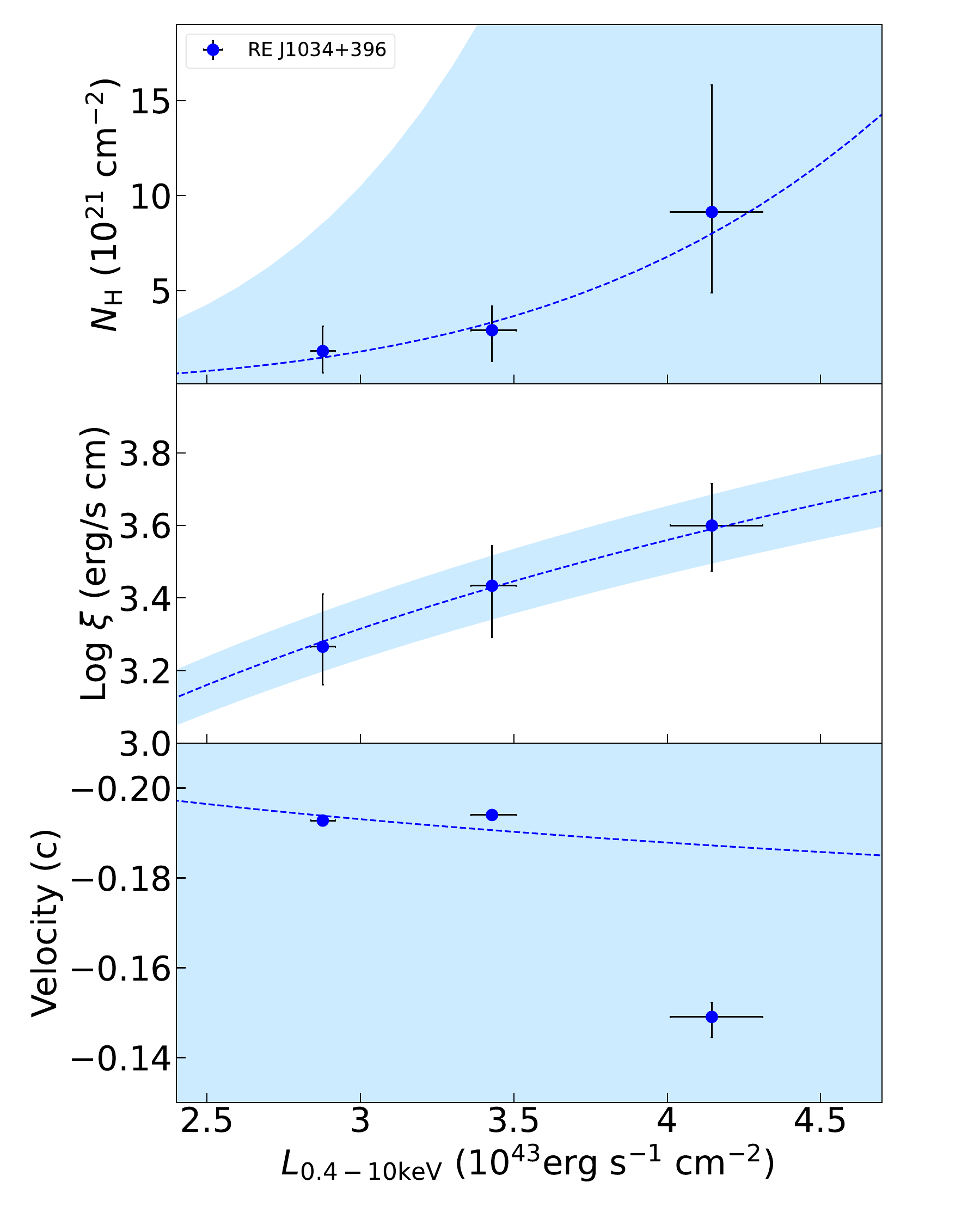}
    \includegraphics[width=\columnwidth]{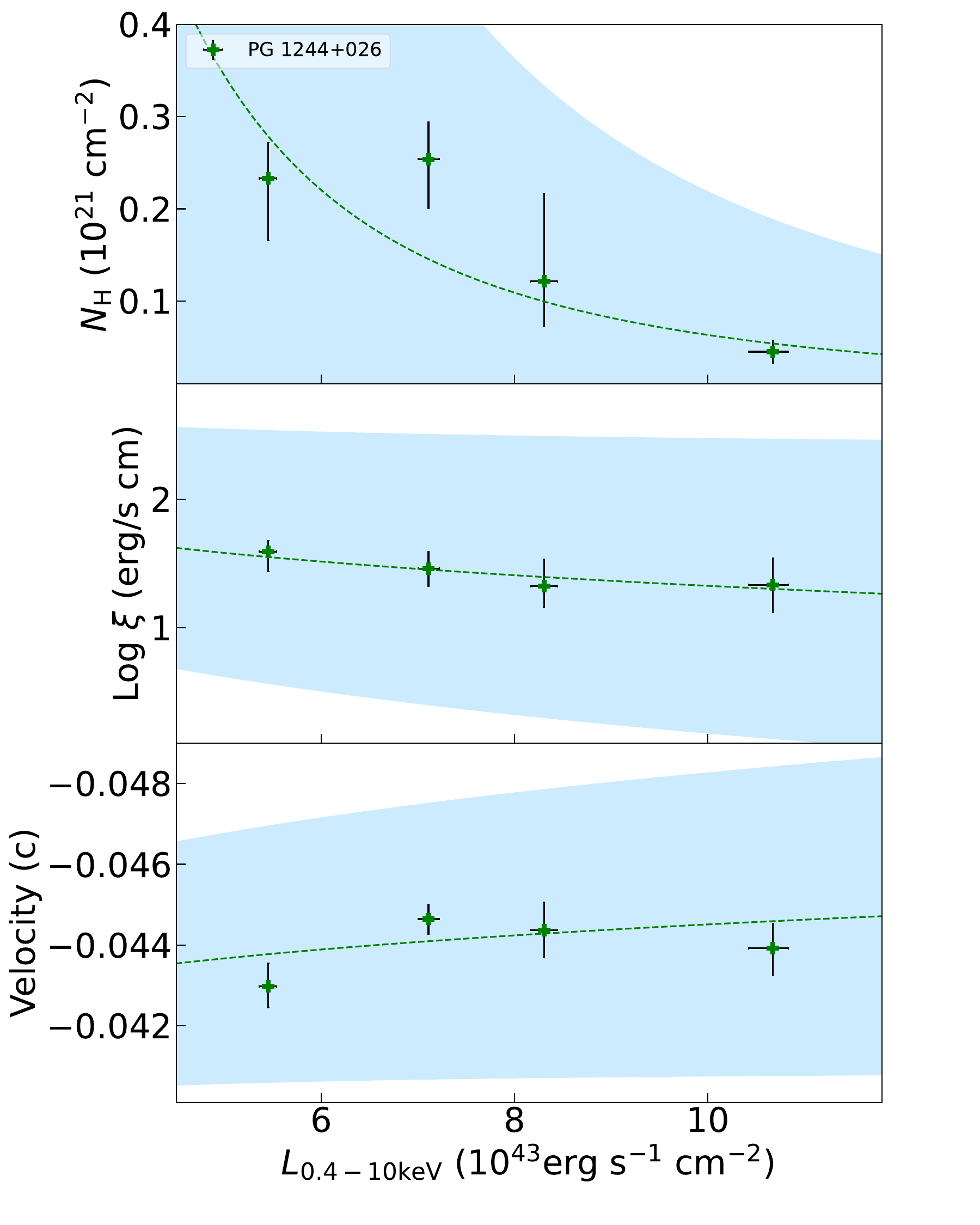}
    \includegraphics[width=\columnwidth]{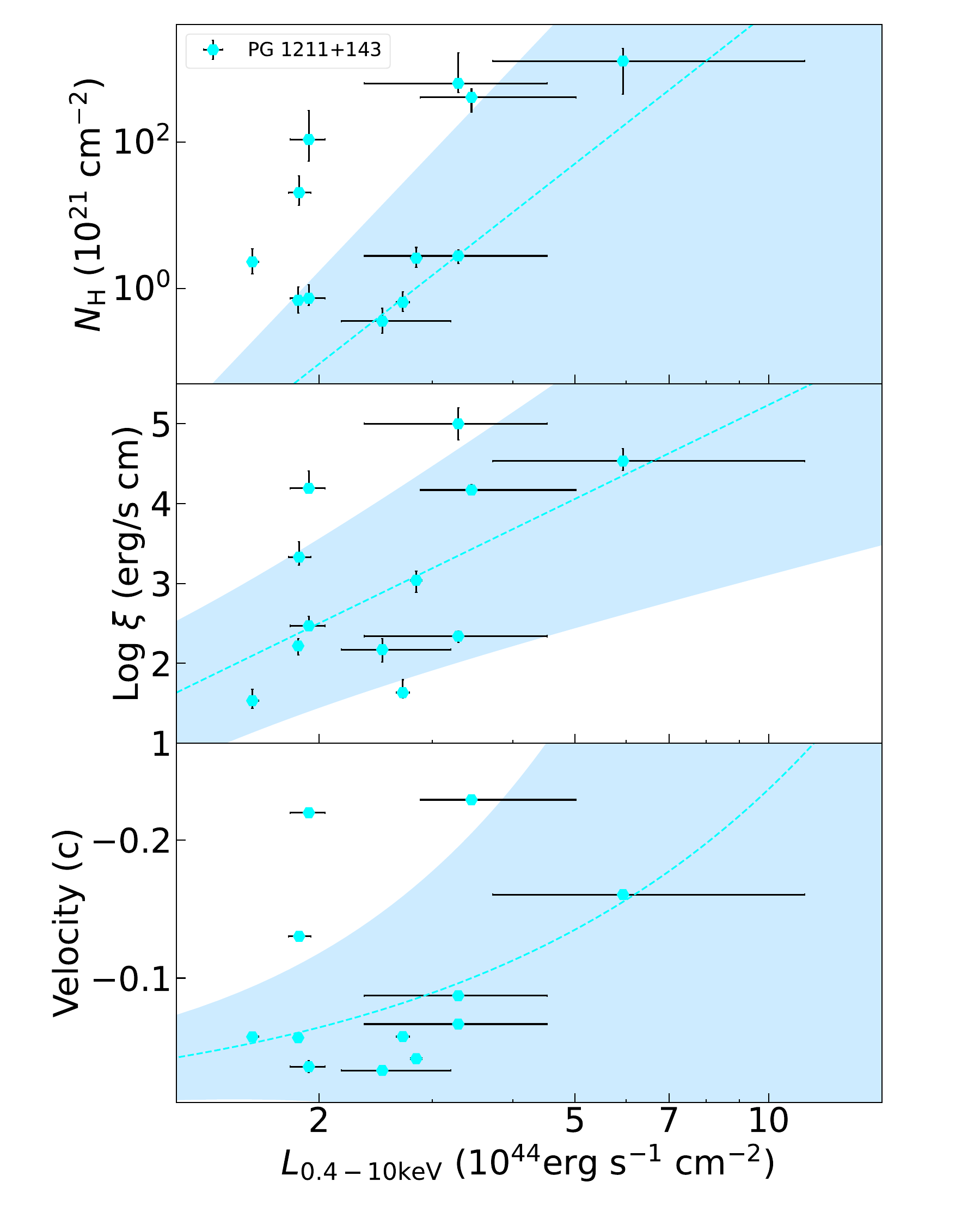}
    \includegraphics[width=\columnwidth]{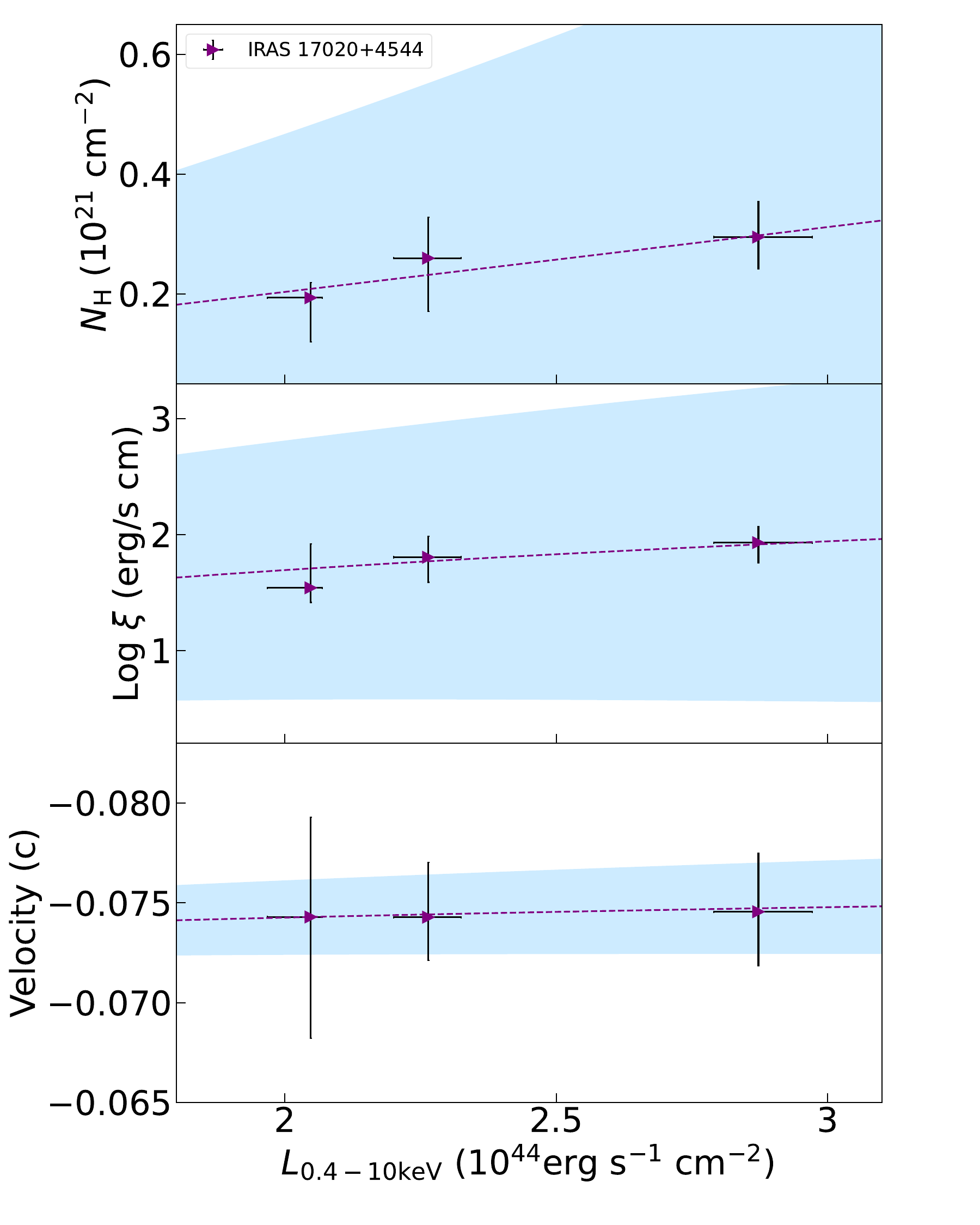}
    \caption{The column density, ionization parameter, and velocity of UFOs in our sample versus the unabsorbed luminosity between $0.4\mbox{--}10\,$keV. The fits with a power function are performed and depicted by \textit{dashed} lines with $1\sigma$ uncertainty shaded.
    }
    \label{app:fig:UFOs-lum}
\end{figure*}

The best-fitting column density, ionization parameter, and velocity of UFOs in our sample versus the unabsorbed luminosity between $0.4\mbox{--}10\,$keV are shown in Fig.\ref{app:fig:UFOs-lum}, where a powerlaw function is applied to fit using the \texttt{scipy.odr.ODR} package in Python, as depicted by the dashed lines. Given the variable and multi-phase nature of UFOs, discerning whether UFOs exhibiting distinct properties are manifestations of the same plasma is challenging. Consequently, to ensure that we are tracing the same absorber at different fluxes rather than comparing different UFOs in different epochs, in our fitting, we exclude UFOs that significantly differ from others and have long intervals from the rest within the same system, i.e. excluding UFOs in the `2015' spectrum of 1H 1934-063 and in the `2020' spectrum of I ZW 1 from the fits. In addition, UFOs with only two measurements are not included in our fits due to zero degrees of freedom, resulting in only 4 UFOs (in RE J1034+396, PG 1244+026, PG 1211+143, and IRAS 17020+4544) remaining in the fits. The second UFOs observed in T1 and T11 spectra of PG 1211+143 are marked by the same label as others and all UFOs in PG 1211+143 are fitted together because UFOs in this AGN have too many phases to identify which UFOs share the same origin. 

\section{The slope of the UFO properties versus $M_\mathrm{BH}$ and $L_\mathrm{ion}$}\label{app:sec:slope}
\begin{figure*}[htbp]
    \centering 
	\includegraphics[width=0.95\columnwidth, trim={0 0 80 0}]{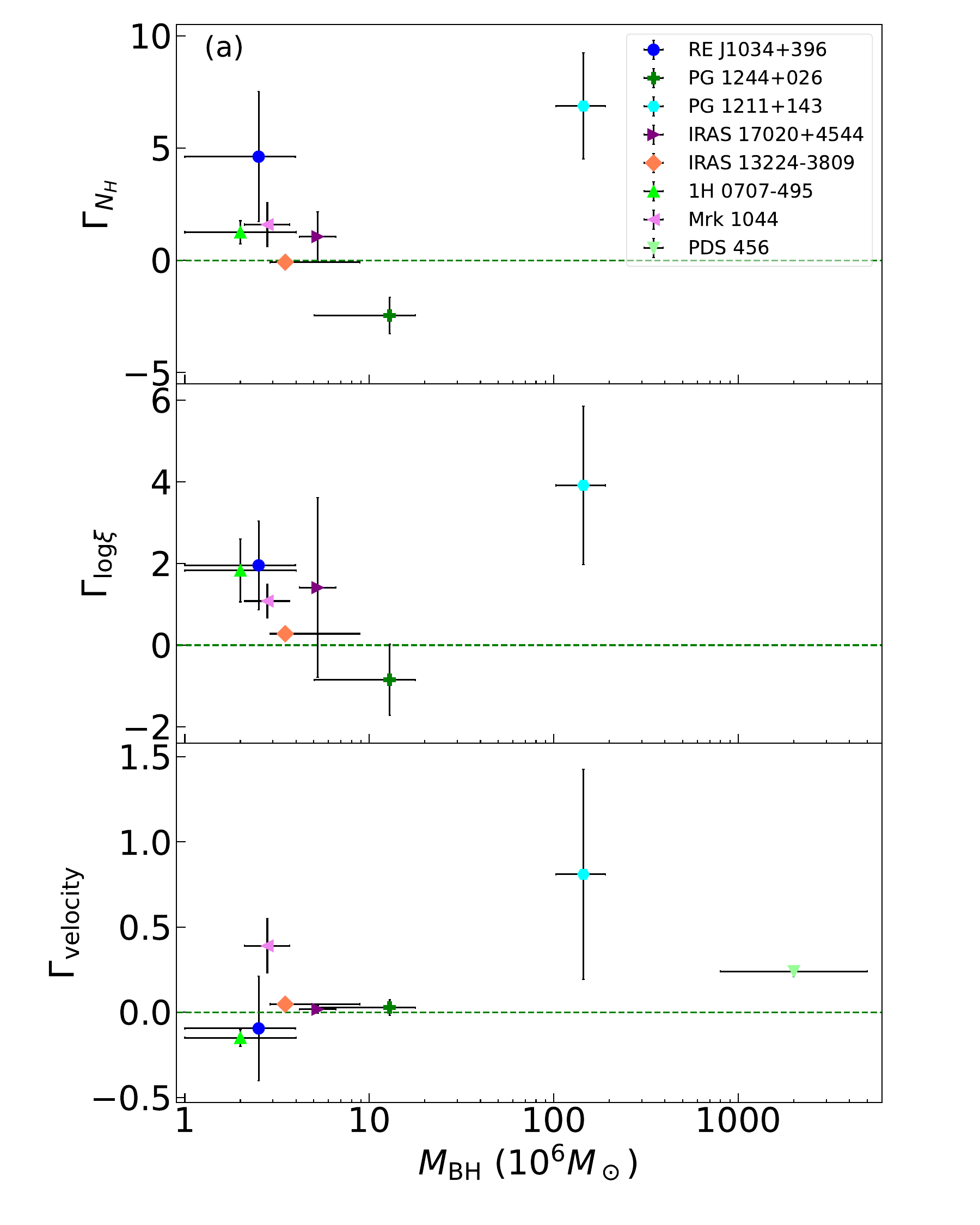}
	\includegraphics[width=0.95\columnwidth, trim={0 0 80 0}]{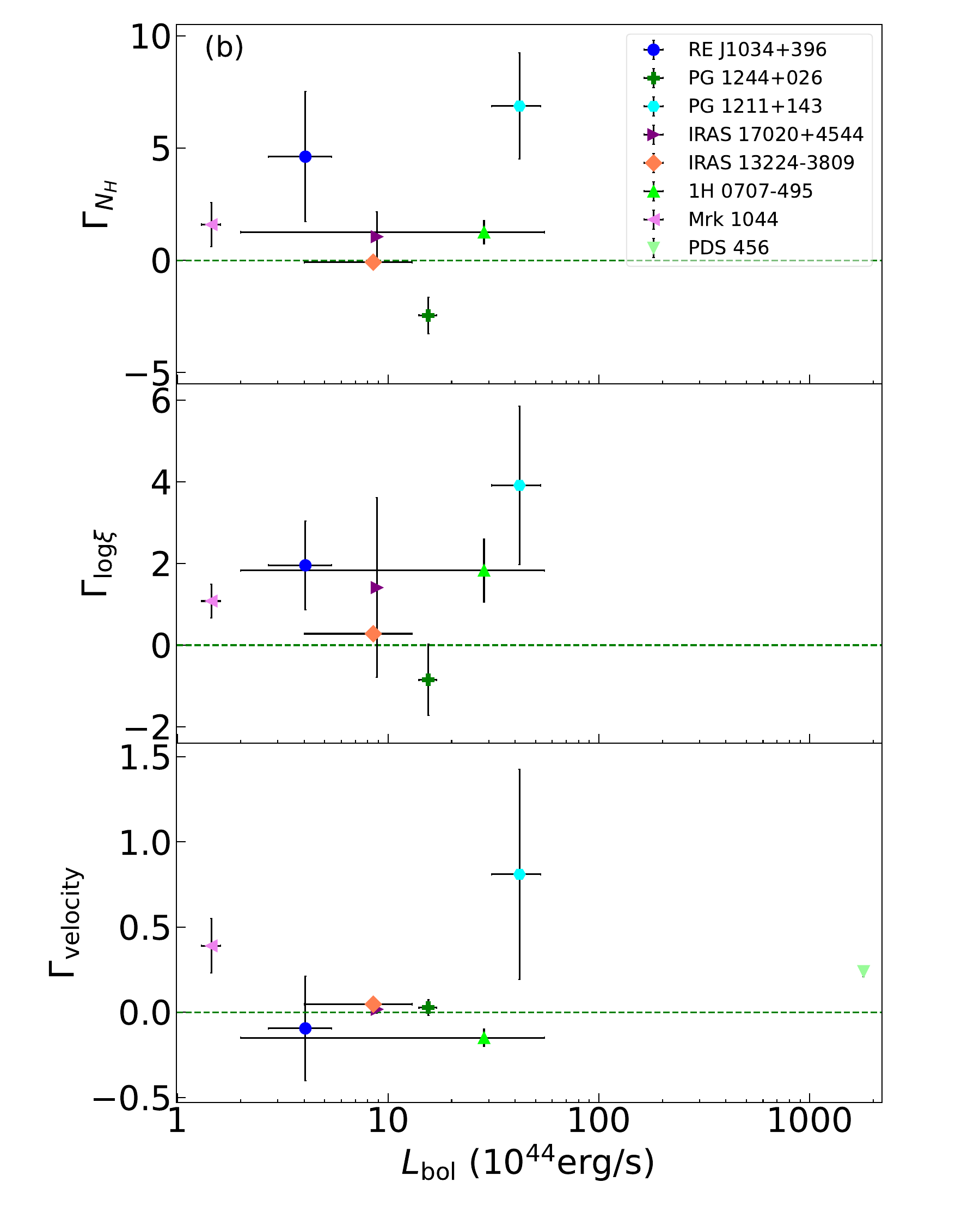}
	\includegraphics[width=0.95\columnwidth, trim={0 0 80 0}]{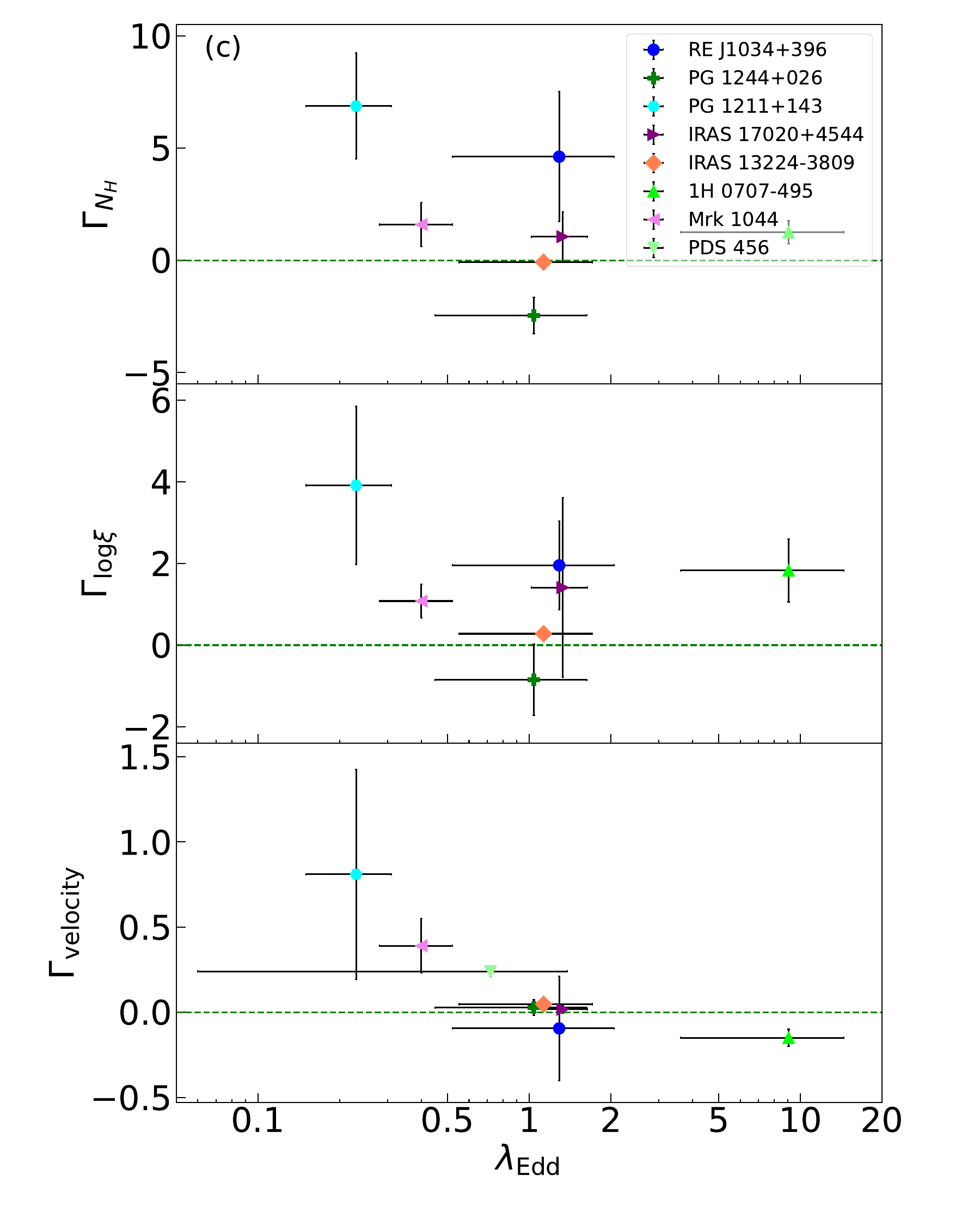}
	\includegraphics[width=0.95\columnwidth, trim={0 0 80 0}]{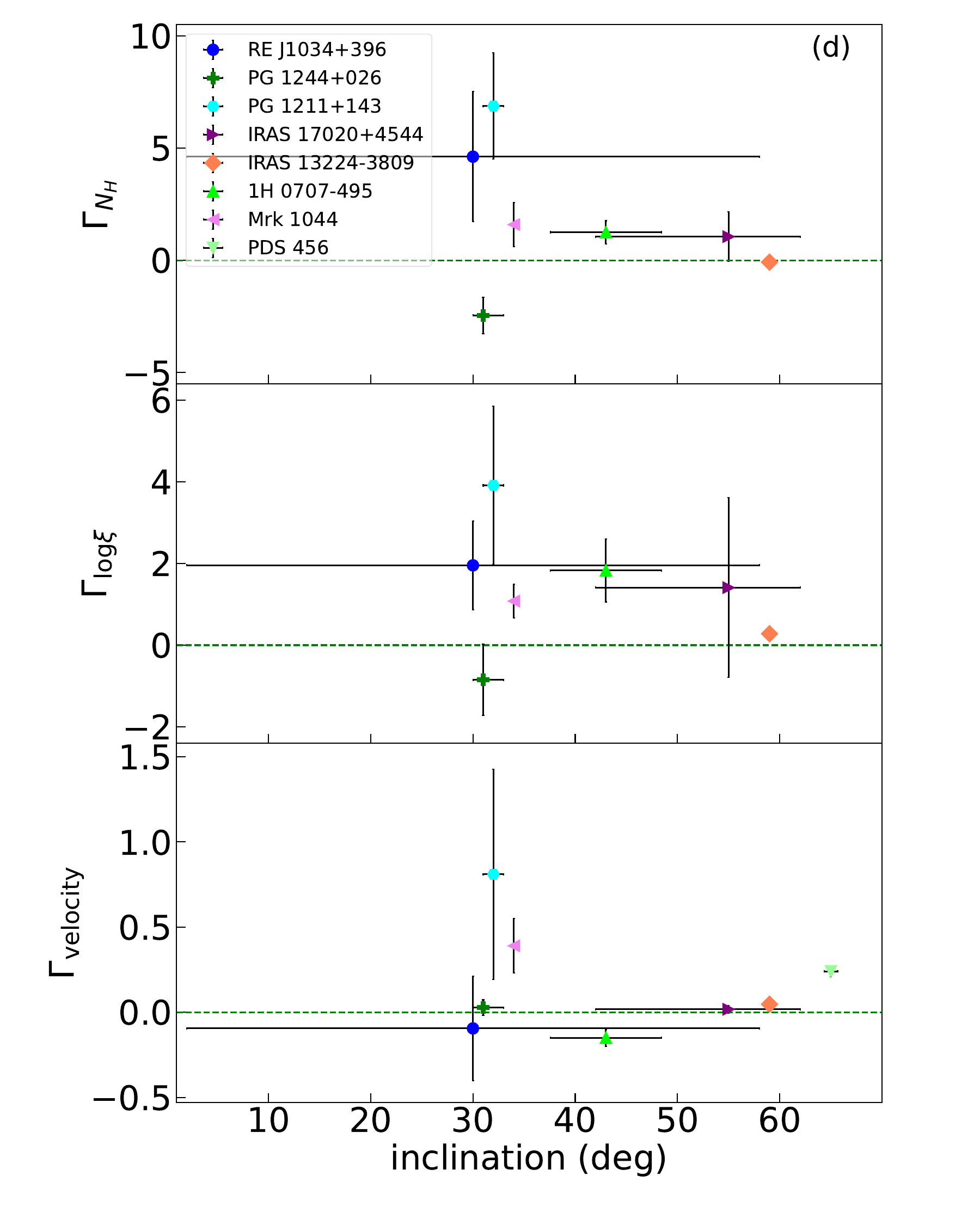}
    \caption{The evolution slopes of the UFO column density $\Gamma_{N_H}$ (\textit{top}), ionization state $\Gamma_\mathrm{\log\xi}$ (\textit{middle}) and velocity $\Gamma_v$ (\textit{bottom}) versus the black hole mass $M_\mathrm{BH}$ (\textit{a}), bolometric luminosity $L_\mathrm{bol}$ (\textit{b}), Eddington ratio $\lambda_\mathrm{Edd}$ (\textit{c}), and inclination angle (\textit{d}). The data come from the results listed in \cref{tab:obs,tab:fits}. The horizontal dashed lines denote the zero value of the slope.
    }
    \label{app:fig:slope}
\end{figure*}

The slopes of the UFO properties including the column density $\Gamma_{N_H}$, ionization state $\Gamma_\mathrm{\log\xi}$ and velocity $\Gamma_v$ are collected to compare with the intrinsic AGN properties, consisting of the black hole mass $M_\mathrm{BH}$, bolometric luminosity $L_\mathrm{bol}$, Eddington ratio $\lambda_\mathrm{Edd}$, and inclination angle. The results are depicted in Fig.\ref{app:fig:slope}. Our examination, facilitated by calculating the Pearson correlation coefficient, failed to reveal any significant correlations between the pairs of $\Gamma_{N_H}$/$\Gamma_\mathrm{\log\xi}$ and AGN properties by calculating the Pearson correlation coefficients. Only an anticorrelation is obviously observed between $\Gamma_v$ and $\lambda_\mathrm{Edd}$, discussed in Sec. \ref{subsec:evolution}. 

\end{appendix}
\end{document}